\documentclass[USenglish]{article}

\ifx\directlua\undefined\ifx\XeTeXcharclass\undefined
  \else\RequirePackage[no-math]{fontspec}[2017/03/31]\fi 
  \else\RequirePackage[no-math]{fontspec}[2017/03/31]\fi 
\usepackage[small]{dgruyter}
\usepackage{amssymb}
\usepackage{amsmath}
\usepackage{booktabs}
\usepackage{rotating}
\usepackage{lineno}
\usepackage{url}
\usepackage[version=4]{mhchem}
\usepackage[autostyle, english = american]{csquotes}
\usepackage{color}

\usepackage{silence}
\begin{document}


  \author*[1,2]{Elise Malmer Martinsen}
    \author*[3]{Andrew ~S. Voyles}
    \author[1,2]{Kevin Ching Wei Li}
    \author[4]{M.~Shamsuzzoha Basunia}
    \author[3,4]{Lee ~A. Bernstein}
    \author[1,2]{Hannah Lovise Okstad Ekeberg}
    \author[6]{Mazhar Hussain}
    \author[3]{Jonathan ~T. Morrell}
    \author[5]{Syed ~M. Qaim}
    \author[1,2]{Sunniva Siem}
    \author[7]{Md. Shuza Uddin}
    \author[6]{Haleema Zaneb}
  \runningauthor{...}
  \affil[1]{Department of Physics, University of Oslo, Oslo, NO-0316, Norway}
    \affil[2]{Norwegian Nuclear Research Centre, Oslo, Norway}
    \affil[3]{Department of Nuclear Engineering, University of California, Berkeley, 94720, CA, USA}
    \affil[4]{Lawrence Berkeley National Laboratory, Berkeley, 94720, CA, USA}
    \affil[5]{Institut für Neurowissenschaften und Medizin, INM-5: Nuklearchemie, Forschungszentrum Jülich GmbH, Jülich, 52425, Germany}
    \affil[6]{Physics Department, Government College University, Lahore, 54000, Pakistan}
    \affil[7]{Institute of Nuclear Science and Technology, Atomic Energy Research Establishment, Savar, Dhaka, Bangladesh}

            

            
            

            
  \title{Deuteron-induced reactions on natural Zr from threshold to $50$~MeV: production of $^{86\text{g}}$Y}
  \runningtitle{...}
  \abstract{Two stacks of thin Zr foils were irradiated with $30$ and $50$ MeV deuterons, respectively, using the Lawrence
Berkeley National Laboratory $88$-Inch Cyclotron, 
and $19$ excitation functions for $^{\text{nat}}$Zr($d$,$x$) reactions were measured over a beam energy range of $6.3$--$47.64$ MeV, where the independent cross sections for $^{\text{nat}}$Zr($d$,$x$)$^{88}$Nb and $^\text{{nat}}$Zr($d$,$x$)$^{86\text{m, g}}$Y were measured for the first time.
The well-characterized $^{\text{nat}}$Fe($d$,$x$)$^{56}$Co, $^{\text{nat}}$Ni($d$,$x$)$^{56}$Co, $^{\text{nat}}$Ni($d$,$x$)$^{58}$Co, $^{\text{nat}}$Ni($d$,$x$)$^{61}$Cu, $^{\text{nat}}$Ti($d$,$x$)$^{46}$Sc and $^{\text{nat}}$Ti($d$,$x$)$^{48}$V monitor reactions were used to determine the deuteron beam current throughout the stacks. 
All cross sections were determined using High Purity Germanium (HPGe) detector $\gamma$-ray spectroscopy. 
A variance minimization technique was employed to simultaneously constrain the deuteron beam currents with multiple monitor reactions, thus reducing systematic uncertainties.
An additional $16$ channels are reported for reactions on the nickel, titanium, and iron monitor foils, leading to a total of $35$ excitation functions, with $7$ reaction channels  reported for the first time in this work.
The measured excitation functions are compared to calculations provided by the reaction modeling codes $\textsc{TALYS}-2.0$, $\textsc{ALICE}-2020$, $\textsc{CoH}-3.5.3$ and $\textsc{EMPIRE}-3.2.3$, as well as the $\textsc{TENDL}-2023$ data library.
The degree of agreement between theory and experiments is discussed.
The possible production of the important PET radionuclide $^{86\text{g}}$Y via the $^{\text{nat}}$Zr($d$,$x$) route was critically examined.
The physical yields for $^{\text{nat}}$Zr($d$,$x$)$^{86}$Y and other yttrium isotopes produced were calculated and compared to other production pathways. Due to high-level of associated radionuclide impurities, this route cannot deliver $^{86\text{g}}$Y suitable for medical applications.}

  \received{...}
  \accepted{...}
  \journalname{...}
  \journalyear{...}
  \journalvolume{..}
  \journalissue{..}
  \startpage{1}
  \aop
  \DOI{...}

\maketitle

\section{Introduction}
\label{sec:Introduction}

Studies of excitation functions of charged-particle induced reactions are important for testing nuclear models as well as for practical applications. 
Over the last three decades, extensive experimental work has been performed worldwide on reaction products induced by light charged particles, mainly protons of energies up to about $30$ MeV, but also extending in some cases up to $100$ MeV and beyond \cite{exfor2024}. 
Regarding the test of nuclear models, except for the very light mass element region, the data are described fairly well by the available modern nuclear model codes, especially when the experimental database is strong. 
At intermediate energies, i.e. $E\geq 40$MeV, however, both experimental databases and nuclear model calculations need further improvements \cite{PhysRevC.103.034601, Fox2021MeasurementMeV, AmjedAslamHussainQaim+2021+525+537}. 
In recent years, interest in reactions induced by other charged particles has also been increasing, especially deuterons, although it is known that the modeling of the activation products formed in the interaction of deuterons with target nuclei is more challenging than with proton beams \cite{Rosch2017The90Y, Qaim2019TheranosticMethodologies, Gula2020State-of-the-artImaging, Uddin2022PositronStudies, PhysRevC.79.044610}. 
Here we present detailed cross section measurements on $^{\text{nat}}$Zr($d$,$x$)$^{86\text{m,g},87\text{m,g}, 88}$Y and several other reaction products up to a deuteron energy of $50$ MeV. 
The experimental data are compared with results of several nuclear model code calculations, and the results shed some light on the success of model calculations in reproducing the experimental data.

As regards practical applications, the measured cross-section data should be of considerable value in optimizing production conditions of radionuclides in the mass region $80$ to $95$, several of which are useful or potentially useful in nuclear medicine \cite{US_health_nuc_med}. 
Of particular interest to us in this study was the positron-emitting radionuclide $^{86\text{g}}$Y ($T_{1/2} = 14.7$ h) which has been gaining importance over the last two decades due to its ``theranostic application'', i.e., its use as a marker to determine the distribution of the injected radioactivity in a tumor-bearing patient quantitatively via positron emission tomography (PET), prior to medication with $^{90}$Y ($T_{1/2} = 2.7$ d), a $\beta^-$-emitting therapeutic radionuclide (for historical developments see \cite{Herzog1993MeasurementRadiotherapeutics, Rosch2017The90Y}). 
In the meantime, several other theranostic pairs have been developed \cite{Qaim2019TheranosticMethodologies}.
However, the  $^{86}$Y/$^{90}$Y pair still remains very interesting, mainly because of its stable trivalent chemistry, although $^{86\text{g}}$Y is not an ideal radionuclide for all PET investigations. 
The total $\beta^+$-emission intensity in its decay amounts only to $27.2 \pm 2.0 \%$ \cite{Uddin2022PositronStudies} and a large number of low and high energy $\gamma$-rays are emitted which adversely affect the resolution of PET scans. 
Nonetheless, through application of several corrections \cite{Herzog2008} it is possible to use this radionuclide in PET studies which can be utilized to monitor the uptake of $^{90}$Y in patients, enabling personalized dosing and enhancing treatment efficiency. 

The therapeutic radionuclide $^{90}$Y is commercially available through the $^{90}$Sr$\xrightarrow{\beta^-}$$^{90}$Y generator system. 
The positron emitter $^{86\text{g}}$Y has also been developed up to the stage of clinical scale production, but  further work is currently in progress.
Several reaction routes have been investigated for its production (for reviews see \cite{Rosch2017The90Y, Uddin2025An86Srd2n-reactions}).
In this work we have investigated a possible pathway for its production, namely $^{\text{nat}}$Zr($d$,$x$)$^{86\text{g}}$Y, for which the existing database for the reaction data is not strong. 
This work is part of a larger international project aiming to study multiple production pathways for $^{86\text{g}}$Y, including $^{86}$Sr($p$,$n$), $^{86}$Sr($d$,$2n$), and $^{\text{nat}}$Zr($d$,$x$) reactions. 
Results for the other production routes have already been recently published \cite{Uddin2025An86Srd2n-reactions, Uddin2020Accurate86Srpn-reaction, Uddin2022ExcitationY}. 
Compared to strontium, zirconium is less reactive and has a higher melting point, making it a more preferable target material. 
The $^{\text{nat}}$Zr($d$,$x$) production route may also serve as a proof-of-concept for future measurements using enriched Zr-targets.

\section{Experimental methods and materials}
The work described herein follows the methods utilized in our group's recent measurements for monitor reaction characterization of beam energy and
ﬂuence in stacked target irradiations with proton beams \cite{ Graves2016NuclearAl, Voyles2021Proton-inducedMeV, VOYLES201853, Kholil2025ExcitationTitanium, Uddin2024CrossProcesses, Uddin2024ExcitationSr, MorrellMeasurementActivation, Burahmah2023Pa232}, extending the methodology to deuteron irradiations.
Preliminary results for three of the measured channels from this experiment [$^{\text{nat}}$Zr($d$,$x$)$^{86,87,88}$Y] were previously included in the thesis work by Zaneb \cite{Zaneb_2018}.
This prior publication was based upon the same experimental data analyzed here, but using a far more simplistic approach to determining the beam current and energy in each foil, using a single monitor reaction.
In addition, preliminary results for all $^{\text{nat}}$Zr($d$,$x$) channels were reported in the master's thesis by Martinsen \cite{Martinsen_2024}. 
This paper will provide the final analysis and results of this measurement.

\subsection{Stacked-target design} 
In this experiment, two stacks of natural abundance targets were irradiated. 
The use of two stacks, each irradiated with a different incident beam energy, was employed to cover a wide, overlapping energy range.
This minimizes the systematic uncertainties of the degraded beam energies within each stack, while the overlap in energies between the two stacks enables a consistency check of the corresponding excitation functions.
The $30$ MeV stack consisted of five nickel ($99.9\%$ purity), five zirconium ($99.2\%$ purity) and five titanium foils ($99.6\%$ purity), as well as three $6060$ aluminum alloy foils. 
The $50$ MeV stack also consisted of five zirconium ($99.2\%$ purity) and five titanium foils ($99.6\%$ purity), but instead of nickel foils, the second stack contained five iron foils ($99.5\%$ purity).
In addition, this stack consisted of seven $6060$ aluminum alloy foils. 
The titanium, nickel and iron foils serve as monitor foils, while the zirconium foils are the foils of interest. 
The aluminum foils progressively degrade the beam to obtain a wide range of beam energies across the multiple target ``compartments'' (a group of one zirconium foil and its corresponding monitor foils). 
The nickel, zirconium, titanium and iron foils were cut into approximately $25$ mm by $25$ mm squares, while the aluminum foils had dimensions of about $67$ by $67$ mm. 
The length, width and thickness were measured using calipers and a digital micrometer, whilst the weights of the foils were measured using an analytical balance.
Before the irradiation was performed, the foils were cleaned with isopropanol to remove surface contamination. 
The areal density of each foil was calculated using the average mass of each foil, divided by the average area. 
The length, width, thickness, mass and areal density for each foil can be found in Tables \ref{Tab:foilcharacterization_30_MeV} and \ref{Tab:foilcharacterization_50_MeV} in Appendix \ref{sec:appendix_stack_design} for the stacks irradiated with $30\ $MeV and $50\ $MeV deuterons, respectively. 
The measured thicknesses of the foils were not used to calculate cross sections, but were only used as an additional test for target uniformity.

Stainless steel (316SS) foils are positioned at the start and end of the stacks for monitoring purposes. 
These foils can serve to track the spatial profile of the beam, as the activated foils may be used to develop radiochromic films (like the Gafchromic EBT3 used in this work) following irradiation. 
These films rely on a dose-proportional dye that develops when exposed to ionizing radiation, thus enabling relative quantification for the spatial profile of the beam at both the front and back of the stack. 

Following the characterization of the target foils, each nickel, zirconium, iron and titanium foil was sealed in Kapton tape and subsequently mounted on $6060$ aluminum alloy frames, each equipped with a $\approx 40$ mm diameter hollow aperture. 
The Kapton tape consists of two layers: $0.0254$ mm Kapton and $0.0432$ mm silicone adhesive.
This sealing process was implemented to reduce dispersible contamination generated from the foils during irradiation, and mitigate oxidation. 
For the irradiation, the foils were placed within a target holder made of $6060$ aluminum alloy, featuring an upstream hollow aperture to enable the unobstructed passage of the beam, as used in our previous measurements \cite{Uddin2020Accurate86Srpn-reaction}. 
To ensure stability of the target stack throughout irradiation, a spring mechanism was utilized to compress the frames in place.
The target stacks were irradiated  at the Lawrence Berkeley National Laboratory's 88-Inch Cyclotron \cite{KireeffCovo2018TheTesting}, for $20$ minutes at a nominal current of $125$\,nA for the 30 MeV stack, $20$ minutes at a nominal current of $100$\,nA for the 50 MeV stack.
The beam currents remained stable over the duration of each irradiation, measured using a current integrator on an electrically-isolated beamline. 

\subsection{Measurements of induced activities}
A single lead shielded, liquid nitrogen-cooled n-type ORTEC GMX-$50220$-S HPGe detector was used to measure the induced activities. 
The detector has a coaxial right cylinder geometry and the diameter and length are $64.9$ mm and $57.8$ mm, respectively. 
After end-of-bombardment (EoB) the irradiated targets were counted at fixed positions ($10$--$50$ cm). 
The measurements started approximately $20$ minutes after EoB and lasted for five weeks, where each spectrum was counted multiple times to reduce statistical uncertainty and to separate products with similar gamma transitions but different half-lives.  
Standard calibration sources were used for energy, efficiency and resolution calibration. 
The peak fitting was performed using Curie \cite{Curie}, and exported to a peak-data file containing information about the fitted energies, number of counts with uncertainties and the calculated number of decays with uncertainties. 

These activities are converted into independent and cumulative cross sections. 
The independent cross sections were measured when possible, only including direct production, while excluding decay feeding from other nuclei leading to the production of that nuclide.
For the first observed product nuclide in a mass chain the cumulative cross sections, $\sigma_c$, are generally reported. 
Independent cross sections, $\sigma_i$, are reported when it is possible to use spectrometry to distinguish between direct production and production through decay feeding of the nuclide, or in cases where no decay precursors exist.
Otherwise, the cumulative cross sections are reported.
Solutions to the Bateman equations are used to distinguish direct production from decay feeding  \cite{bateman1910solution}. 

\subsection{Determination of deuteron beam currents}
Thin foils of nickel, titanium and iron were used for beam current monitoring in each irradiation. 
In the stack irradiated with $30$ MeV deuterons, one such monitor foil of nickel and one of titanium were placed in each energy \enquote{compartment}, while monitor foils of iron and titanium were used in the stack irradiated with $50$ MeV deuterons.  
The IAEA-recommended monitor reactions $^{\text{nat}}$Fe($d$,$x$)$^{56}$Co, $^{\text{nat}}$Ni($d$,$x$)$^{56}$Co, $^{\text{nat}}$Ni($d$,$x$)$^{58}$Co, $^{\text{nat}}$Ni($d$,$x$)$^{61}$Cu, $^{\text{nat}}$Ti($d$,$x$)$^{46}$Sc and $^{\text{nat}}$Ti($d$,$x$)$^{48}$V were used to determine the deuteron beam current throughout the stacks \cite{Hermanne2018ReferenceReactions}. 
Figures showing our measured monitor reactions compared to the recommended IAEA values and previously measured data are included in the supplementary information to demonstrate reproduction of well-established monitor reactions. We are not reporting these as new cross section measurements.
Due to the non-trivial broadening of the deuteron-energy distributions as the beam penetrates the stack, an integral formula is used for the beam current calculations. 
The deuteron beam current as a function of deuteron energy, $\Phi (E_d)$ is given by:
\begin{equation}
    \Phi(E_d) = \frac{A_0}{N_T(1-e^{-\lambda\Delta t_{irr}})} \times \frac{1}{\frac{\int \sigma_{mon}(E_d) \frac{d\phi}{dE}dE}{\int \frac{d\phi}{dE}dE}}.
\end{equation}
where, for a particular monitor reaction, $A_0$ is the EoB activity for the monitor product, $N_T$ is the number of target nuclei per unit area (areal density), $\sigma_{\text{mon}}$ is the monitor reaction cross section provided by the IAEA evaluation, $\lambda$ is the decay constant of the radionuclide, $\Delta t_{\text{irr}}$ is the duration of the irradiation, and $\frac{d\phi}{dE}$ is the energy distribution of the deuterons in a single foil.

The energy loss of deuterons is dependent on a combination of both the areal densities of the targets the deuterons are traversing, as well as the associated stopping powers of the materials.
It has been shown \cite{VOYLES201853, Voyles2021Proton-inducedMeV} that the uncertainty in the deuteron energies is generally dominated by uncertainties in the stopping powers, as opposed to the monitor cross sections or areal densities.
In the present work, the deuteron energies and beam current were optimized by varying the stopping power of the deuterons, a process from our prior work referred to as variance minimization \cite{VOYLES201853, Voyles2021Proton-inducedMeV, morrell2020, Fox2021MeasurementMeV}.
In the analysis of this work, the stopping powers were globally varied through a surrogate variation of the effective areal densities by up to $\pm 20\%$.
This was simply because it was more numerically convenient, taking advantage of the charged particle transport calculations in Curie.
For the $30$ MeV stack, the variance minimization was performed using all five monitor reactions in compartment $4$, i.e., the compartment containing foil Zr04.
For the $50$ MeV stack, the three monitor reactions in compartment $5$ (containing foil Zr10) were used. 
When deciding on which compartment to use in the variance minimization, it is crucial to consider the slopes of the monitor cross section graphs for the different reactions. 
It is advantageous to use a compartment with an energy range where the slopes of the different monitor reaction cross sections are significantly different, and in particular where at least one monitor channel is decreasing while the others are increasing, or vice-versa. 
Likewise, a compartment containing the energetic threshold for a monitor channel provides a physical indication of calculating the wrong energy distribution, due to the negligible production rates near threshold.
Such cases generally lead to a clear minimum in disagreement between monitor reactions, providing the best constraints of beam energy in the stack.
Moreover, it is beneficial to use a compartment further back in the stack as the effects of the poorly characterized deuteron stopping power become progressively larger as the deuterons traverse through a greater volume of material. 
Compartment $5$ was chosen for variance minimization for the $50$ MeV stack as it is the last compartment in the stack and the monitor reaction cross sections have significantly different slopes. 
For the $30$ MeV stack, compartment $4$ was chosen as it has monitor reactions with a major difference in the slopes of the cross sections provided by IAEA \cite{Hermanne2018ReferenceReactions}, and it is the second to last compartment in the stack. 
The last compartment, compartment $5$, could not be reliably used for variance minimization in the $30$ MeV stack as the beam was unexpectedly stopped in this compartment. 
This fact alone provided a clear indication that stopping powers were underestimated in the initial stack designs for this measurement, as the beam was expected to exit compartment $5$ at approximately 4 MeV.
Both linear and constant models of the beam currents (as a function of beam energy) within one compartment were considered, however, the difference in results between these two approaches was negligible. 
In this work, a linear fit was employed to account for the physical reality that the beam current is degraded within a single stack.
The beam currents calculated using each monitor reaction 
show a clear improvement in consistency following variance minimization, particularly for the $30$ MeV stack.
Further details on this process, including figures before and after, can be seen in \cite{Martinsen_2024}.



Following the variance minimization, the correlated, uncertainty-weighted average beam currents, $\langle\Phi\rangle$, for the zirconium foils were determined using all the monitor reactions within the same compartment. 
In this work, the method to calculate the correlated uncertainty-weighted average is the same as in Ref. \cite{Voyles2021Proton-inducedMeV}, using the formula:
\begin{equation}
    \langle\Phi\rangle = \frac{\Sigma_{i,j}\Phi_j(V_{ij}^{-1})}{\Sigma_{i,j}(V_{ij}^{-1})},
\end{equation}
where $V_{ij}$ is the $i j$ element in the covariance matrix given by the ``sandwich estimator'':
\begin{equation}
    V_{ij} = \text{Cov}[\Phi_i, \Phi_j]=\sum_\beta\frac{\partial \Phi}{\partial\beta_i}\delta\beta_i\text{Corr}[\beta_i, \beta_j]\delta\beta_j\frac{\partial \Phi}{\partial\beta_j}.
\end{equation}
Here, $i$ and $j$ denote the monitor reactions in a given compartment and $\beta \in [A_0, \rho\Delta r, \lambda, \Delta t_{irr}, \int \sigma(E)\frac{d\phi}{dE}dE]$, where $\rho\Delta r$ is the areal density.
The exact correlations between these variables is dependent on each particular case.
Since these correlations are not exactly known, the nature of these correlations is assumed based on previous studies \cite{Voyles2021Proton-inducedMeV}.
$A_0$ is assumed to be $30\%$ correlated for all reactions, $\rho\Delta r$ is assumed to be $100\%$ correlated for reactions from the same monitor foil, $\lambda$ is assumed to be uncorrelated, $\Delta t_{irr}$ is assumed to be $100\%$ correlated and $\int \sigma(E)\frac{d\phi}{dE}dE$ is assumed to be $30\%$ correlated for reactions from the same monitor foil. 
This process was adopted to properly calculate the average beam current in each foil, as the various monitor foil measurements (especially for those within a single foil) are decidedly not independent measurements. 
A figure including the weighted average beam currents used for the cross-section measurements can be seen in \cite{Martinsen_2024}.
The weighted average beam currents were found to be within 2\% agreement of the current integrator measurements for the target assembly. 

\subsection{Calculation of measured cross sections}
The final cross sections for the observed ($d$,$x$) reactions were calculated using the quantified EoB activities and the variance-minimized deuteron beam currents. 
All cross sections reported in this work are flux averaged. 
For both the cumulative and independent activities quantified, cross sections were calculated as:
\begin{equation}
    \sigma = \frac{A_0}{\langle\Phi\rangle N_T(1-e^{-\lambda t_{irr}})},
\end{equation}
where $A_0$ is the EoB activity for the product nuclide, $\langle\Phi(E)\rangle$ is the flux-averaged beam current, $N_T$ is the target areal density, $\lambda$ is the decay constant of the radionuclide, and $\Delta t_{\text{irr}}$ is the duration of the irradiation.
The gamma-ray intensities and half-lives used in this work are tabulated in Tables \ref{tab:decay_data_Zr_part1} and \ref{tab:decay_data_mon_part1} in Appendix \ref{sec:decay_data}.
The uncertainties in the EoB activity ranged between $0.4$--$52.6\%$ for all reaction products, except for $^{\text{nat}}$Zr($d$,$x$)$^{95\text{m}}$Nb for $E=41.1$ MeV where the uncertainty in the EoB activity was $115\%$. 
The uncertainty in the beam current ranged between $2$--$5\%$, the uncertainty in the areal density ranged between $0.1$--$1.0\%$, the uncertainty in half-lives varies by product, but is generally on the order of $0.1\%$, and the uncertainty in the length of irradiation is $3$ seconds. 
Thus, the cross sections reported in this work have uncertainties ranging from $2.0$--$51.6\%$. 

\subsection{Nuclear model calculations}
All the measured excitation functions were compared to calculations provided by the standard reaction modeling codes $\textsc{TALYS}-2.0$, $\textsc{ALICE}-2020$, $\textsc{CoH}-3.5.3$ and $\textsc{EMPIRE}-3.2.3$, as well as the $\textsc{TENDL}-2023$ data library \cite{TALYS, ALICE, CoH, EMPIRE, TENDL}.
The codes were all explicitly run using their default parameters.
TENDL, on the other hand, was generated by its developers through a parameter optimization process. 
When reaction modeling codes are compared to cumulative cross sections, we have assumed the complete decay of all precursors.

\section{Results and discussion}
Using the final deuteron beam currents at each energy position, the excitation functions for $^{\text{88, 90, 92m, 95, 95m, 96}}$Nb, $^{\text{88, 89, 95}}$Zr, and ${^\text{86, 86m, 87, 87m, 88, 90m}}$Y were extracted for ($d$,$x$) reactions on $^{\text{nat}}$Zr foils up to $50$ MeV, presented in Table \ref{tab:xs_Zr}. 
For the three channels previously published in \cite{Zaneb_2018}, our results are consistent within uncertainty.
For the ($d$,$x$) reactions on $^{\text{nat}}$Ni, the extracted cross sections for $^{\text{57, 65}}$Ni, $^{\text{55, 57, 60}}$Co, and $^{\text{52, 54}}$Mn are presented in Table \ref{tab:xs_Ni}. 
Cross sections for $^{\text{47, 48}}$Sc were extracted for ($d$,$x$) reactions on $^{\text{nat}}$Ti, and are presented in Table \ref{tab:xs_Ti}. 
For ($d$,$x$) reactions on $^{\text{nat}}$Fe the extracted cross sections for $^{\text{55, 57, 58}}$Co, $^{52, 54}$Mn, and $^{\text{48}}$V are presented in Table \ref{tab:xs_Fe}. 
Excitation functions for $7$ reactions are reported for the first time in this work.
In the following section, we compare the measured cross sections to literature data retrieved from EXFOR \cite{Gonchar1993GO32, Tarkanyi2004TA08, Mercader1972ME22, Simeckova2021SI31, Vysotskij1991_exforO0380.1, Takacs2007EvaluatedNickel, Amjed2013Activation40MeV, Usman2016Measurements24MeV, Hermanne2013New50MeV, Ochiai2007DeuteronIFMIF, Avrigeanu2016Deuteron-inducedMeV, Zweit1991ExcitationTomography, Takacs1997ActivationPurpose, Lebeda2015ExperimentalReactions, Gagnon2010Experimental51Ti, Takacs2007EvaluatedTitanium, Khandaker2013Excitation24MeV, Takacs1997ExcitationBeams, Duchemin2015Cross34MeV, Khandaker2014ActivationTitanium, Hermanne2000ExperimentalTi, Kiraly2009Evaluated10MeV, Zhao1995ExcitationIron, Khandaker2013Activation24MeV, Dmitriev1969MethodsIsotope., Clark1969ExcitationIron, Avrigeanu2014LowIsotopes, zhenlan1984excitation, Jung1992CrossDeuterons, Nakao2006MeasurementsMaterials, Sudar1994ExcitationMeV, Takacs1996StudyTechnique} and results from the reaction modeling codes $\textsc{TALYS}-2.0$, $\textsc{ALICE}-2020$, $\textsc{CoH}-3.5.3$, and $\textsc{EMPIRE}-3.2.3$, as well as the $\textsc{TENDL}-2023$ database \cite{TALYS, ALICE, CoH, EMPIRE, TENDL}.
The measured data of particular interest for the production of $^{86}$Y through ($d$,$x$) reactions on $^{\text{nat}}$Zr are discussed in detail below. 
Figures of all other excitation functions for the $^{\text{nat}}$Zr($d$,$x$) reactions are given in Appendix \ref{sec:Zr_excitation_functions}, while the figures of all excitation functions measured from the monitor foils are given in the supplementary information \ref{sec:measured_excitation_functions}.

\begin{sidewaystable*}[htbp]
    \centering
    \renewcommand{\arraystretch}{1.1} 
    \caption{Measured cross sections for all the $^{\text{nat}}$Zr($d$,$x$) reaction products observed in this work. Cumulative cross sections are designated as $\sigma_c$, while independent cross sections are designated as $\sigma_i$. The Stack ID specifies which of the two irradiated stacks each measurement belongs to. Stack A is the stack irradiated with $50$ MeV deuterons, and stack B was irradiated with $30$ MeV deuterons. Uncertainties are denoted in the least significant digit, that is, $39.1(47)$ mb means $39.1 \pm 4.7$ mb.}
    \label{tab:xs_Zr}
    \begin{tabular}{*{10}{l}}
        \midrule
        & \multicolumn{9}{l}{Production cross section (mb)} \\
        \cmidrule(lr){2-10}
        $E_d$ (MeV) & $47.64(91)$ & $41.1(11)$ & $36.4(12)$ & $31.0(13)$ & $26.40(63)$ & $24.9(16)$ & $19.24(79)$ & $12.4(11)$ & $6.3(18)$  \\
        Stack ID & A & A & A & A & B & A & B & B & B  \\
        \toprule
        $^{86\text{g}}$Y ($\sigma_c$) & $39.1(47)$ & $46.0(48)$ & $33.1(37)$ & $13.0(18)$ & — & — & — & — & — \\
        $^{86\text{g}}$Y ($\sigma_i$) & $19.4(34)$ & $22.5(35)$ & $17.5(27)$ & $8.0(13)$ & — & — & — & — & — \\
        $^{86\text{m}}$Y ($\sigma_i$) & $19.7(32)$ & $23.5(34)$ & $15.6(26)$ & $5.0(12)$ & — & — & — & — & — \\
        $^{87}$Y ($\sigma_c$) & $39.4(29)$ & $26.9(21)$ & $24.4(31)$ & $31.8(34)$ & $28.9(12)$ & $29.4(32)$ & $10.01(53)$ & — & — \\
        $^{87}$Y ($\sigma_i$) & $7.3(12)$ & $6.1(10)$ & $6.0(15)$ & $7.1(15)$ & $7.0(27)$ & $6.8(16)$ & $3.1(16)$ & — & — \\
        $^{87\text{m}}$Y ($\sigma_c$) & $32.4(84)$ & $21.1(65)$ & $18.7(60)$ & $25.0(62)$ & $21.69(92)$ & $23.0(55)$ & $6.70(35)$ & — & — \\
        $^{88}$Y ($\sigma_c$) & $353(21)$ & $226(19)$ & $80(17)$ & $19.8(12)$ & $15.6(25)$ & $12.45(29)$ & $9.14(47)$ & $6.12(33)$ & $0.430(88)$ \\
        $^{88}$Y ($\sigma_i$) & $45.4(59)$ & $29.7(67)$ & $21.4(84)$ & $15.4(11)$ & — & $11.75(29)$ & — & — & — \\
        $^{88}$Zr ($\sigma_c$) & $308(20)$ & $196(18)$ & $59(14)$ & $4.37(32)$ & — & $0.696(19)$ & — & — & — \\
        $^{88}$Zr ($\sigma_i$) & $267(20)$ & $182(18)$ & — & — & — & — & — & — & — \\
        $^{88}$Nb ($\sigma_i$) & $41.5(39)$ & $14.3(19)$ & — & — & — & — & — & — & — \\
        $^{89}$Zr ($\sigma_c$) & $365(59)$ & $533(19)$ & $536(12)$ & $391(64)$ & $305(12)$ & $282(16)$ & $12.3(16)$ & $1.39(21)$ & $0.540(32)$ \\
        $^{90}$Nb ($\sigma_i$) & $134.1(59)$ & $192.6(58)$ & $193.5(76)$ & $221.9(92)$ & $294(10)$ & $348(12)$ & $420(11)$ & $187.5(37)$ & $0.395(18)$ \\
        $^{90\text{m}}$Y ($\sigma_c$) & $6.20(59)$ & $5.43(56)$ & $5.27(74)$ & — & $3.44(30)$ & $3.28(12)$ & — & $0.229(52)$ & —\\
        $^{92\text{m}}$Nb ($\sigma_i$) & $20.08(92)$ & $26.78(75)$ & $32.7(10)$ & $37.2(16)$ & $26.1(12)$ & $26.2(13)$ & $40.7(18)$ & $83.7(34)$ & $20.6(17)$ \\
        $^{95}$Zr ($\sigma_c$) & $12.61(94)$ & $14.4(22)$ & $14.52(98)$ & $16.20(82)$ & $15.71(88)$ & $17.29(60)$ & $19.04(67)$ & $31.11(88)$ & $24.5(12)$ \\
        $^{95}$Nb ($\sigma_i$) & $5.85(95)$ & $8.8(33)$ & $8.9(14)$ & $12.5(13)$ & $21.0(15)$ & $24.3(14)$ & $27.6(12)$ & $10.77(47)$ & $12.88(75)$ \\
        $^{95\text{m}}$Nb ($\sigma_i$) & $3.3(10)$ & $3.0(34)$ & $3.9(15)$ & $5.7(11)$ & $5.79(79)$ & $6.8(11)$ & $6.09(73)$ & $4.50(36)$ & $6.02(63)$ \\
        $^{96}$Nb ($\sigma_i$) & $1.61(21)$ & $1.92(36)$ & $2.29(58)$ & $2.80(14)$ & $3.27(15)$ & $3.98(15)$ & $7.20(28)$ & $23.12(58)$ & $6.74(33)$ \\
        \midrule
    \end{tabular}
\end{sidewaystable*}

\subsection{$^\text{{nat}}$Zr($d$,$x$)$^{86\text{m}}$Y (independent) and $^\text{{nat}}$Zr($d$,$x$)$^{86\text{g}}$Y (cumulative and independent)}

\begin{figure}[h!]
    \centering
    \includegraphics[width=\linewidth]{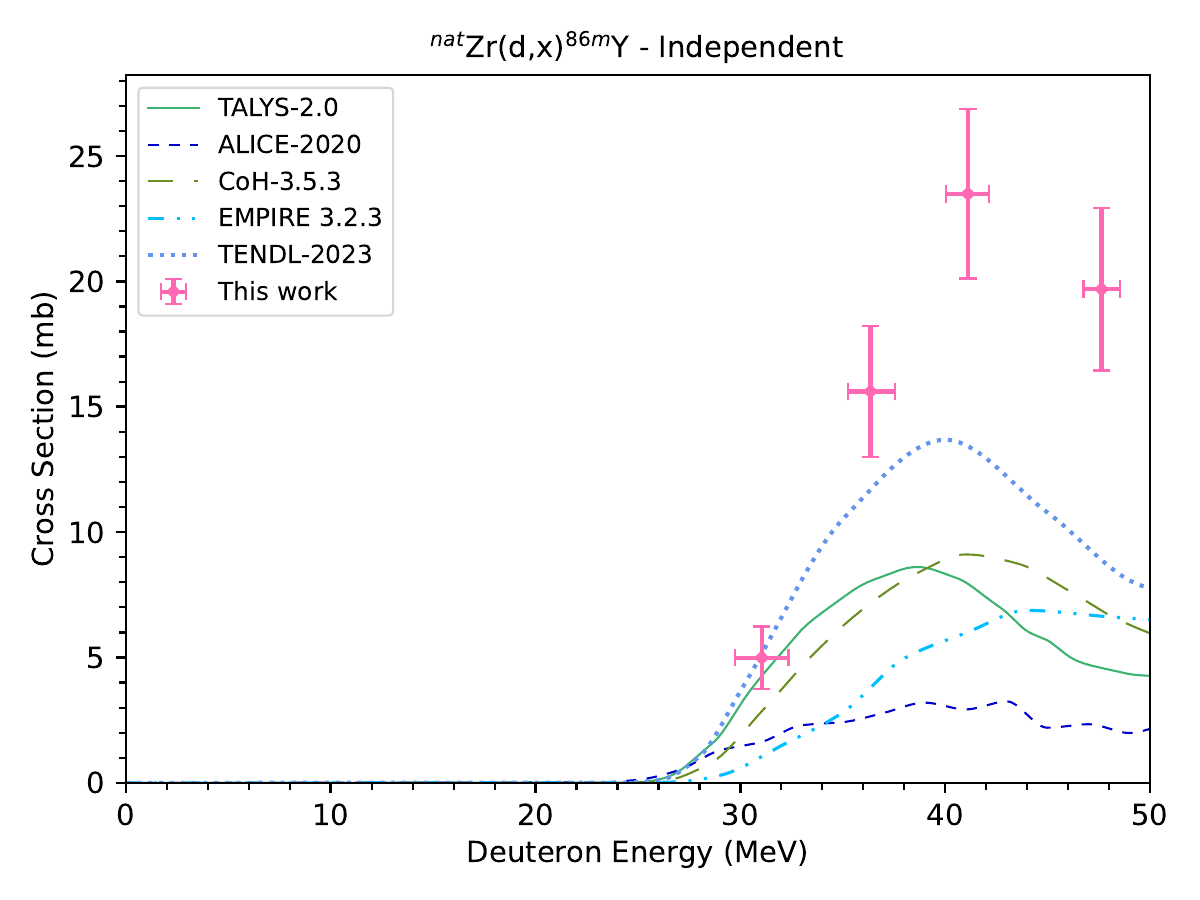}
    \caption{The excitation function for the independent production of $^\text{{nat}}$Zr($d$,$x$)$^{86\text{m}}$Y.}
    \label{fig:86Ym_ind}
\end{figure}

\begin{figure}[h!]
    \centering
    \includegraphics[width=\linewidth]{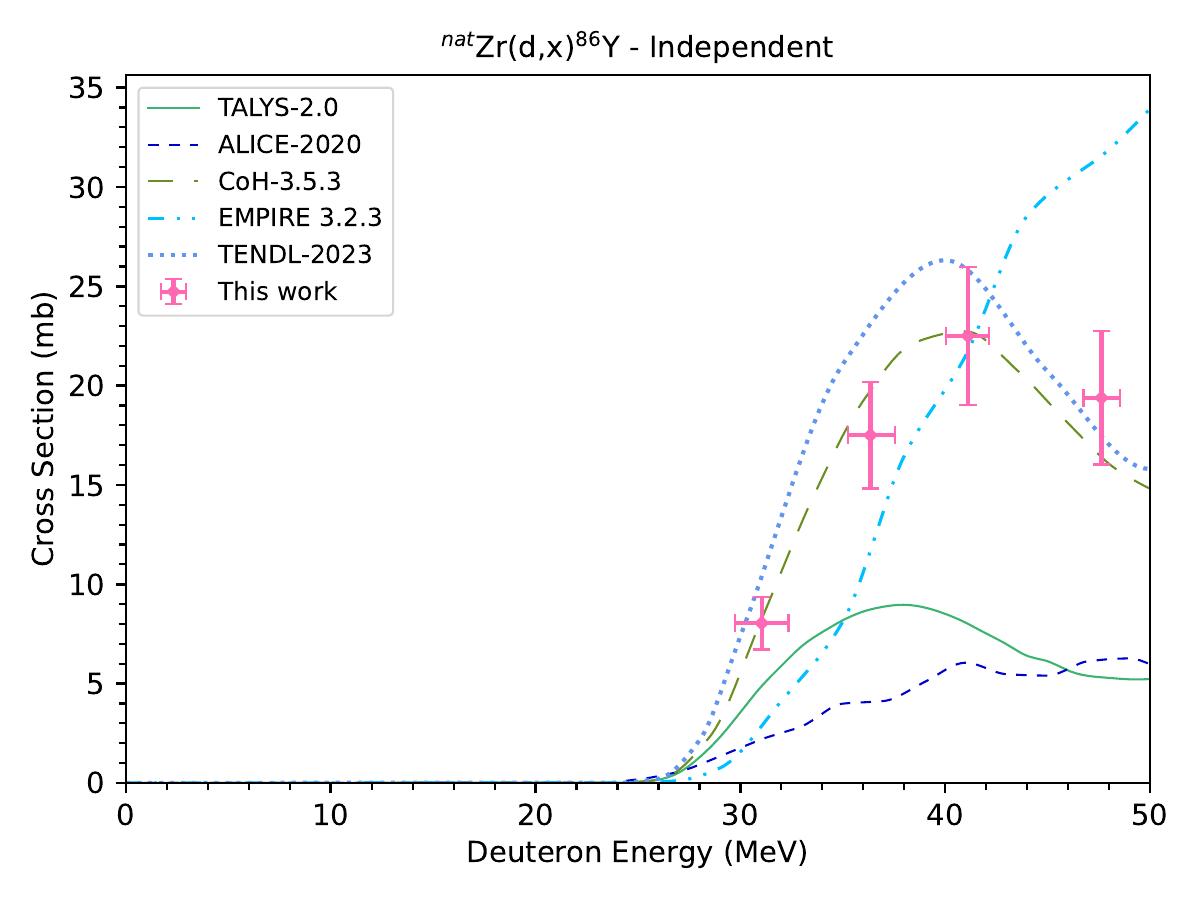}
    \caption{The excitation function for the independent production of $^\text{{nat}}$Zr($d$,$x$)$^{86\text{g}}$Y. No feeding from $^{86}$Zr to $^{86}$Y was observed in the energy region covered in this work.}
    \label{fig:86Y_ind}
\end{figure}

The independent cross sections for $^\text{{nat}}$Zr($d$,$x$)$^{86\text{m}}$Y ($t_{1/2}=47.4$ min) \cite{Negret2015Nuclear86} can be seen in Figure \ref{fig:86Ym_ind}. 
In addition, the cross sections for $^\text{{nat}}$Zr($d$,$x$)$^{86\text{g}}$Y excluding the feeding from the isomeric state are reported. 
This reaction channel includes feeding from $^{86}$Zr but not $^{86\text{m}}$Y, as the decay of $^{86}$Zr does not feed the isomeric state, $^{86\text{m}}$Y \cite{Negret2015Nuclear86}.
However, $^\text{{nat}}$Zr($d$,$x$)$^{86}$Zr is a weakly fed channel, with $Q = -36.8\ $MeV, and a maximum cross section of just $0.0173$ mb in the energies studied in this work, according to $\textsc{TENDL}-2023$.
Since $^{86}$Zr was not observed in any of the $\gamma$-ray spectra, with an MDA of $0.00477$ mb, we report this work as presenting the first reported independent measurement of $^\text{{nat}}$Zr($d$,$x$)$^{86\text{g}}$Y.
Independent cross sections for this reaction channel are presented in Figure \ref{fig:86Y_ind}.
The independent cross sections for $^\text{{nat}}$Zr($d$,$x$)$^{86\text{g}}$Y and  $^\text{{nat}}$Zr($d$,$x$)$^{86\text{m}}$Y were obtained by fitting to a two-step decay curve including $^{86\text{m}}$Y and $^{86\text{g}}$Y in the decay chain, and the relative uncertainties of the production rates range from $15.2$--$17.1\ \%$ for the ground state and $14.2$--$24.6\ \%$ for the isomeric state. 

In terms of reaction modeling, $\textsc{TALYS}$ and $\textsc{ALICE}$ both underestimate the magnitude of the excitation function for both the isomeric and ground states. 
$\textsc{CoH}$ provides an excellent reproduction of the cross sections reported for $^\text{{nat}}$Zr($d$,$x$)$^{86\text{g}}$Y in this work, but underestimates the magnitude for the $^\text{{nat}}$Zr($d$,$x$)$^{86\text{m}}$Y reaction, suggesting issues with the competition between $^{86\text{m}}$Y and $^{86\text{g}}$Y for partitioning of angular momentum in the compound nucleus. 


The cumulative cross sections for $^\text{{nat}}$Zr($d$,$x$)$^{86\text{g}}$Y, presented in Figure \ref{fig:86Y_cum}, are obtained by adding the cross sections for $^\text{{nat}}$Zr($d$,$x$)$^{86\text{m}}$Y, which are weighted by the decay mode branching ratio of the isomeric transition, to the cross sections for $^\text{{nat}}$Zr($d$,$x$)$^{86\text{g}}$Y. 
The cross sections reported in this work are in good agreement with the experimental data reported by Gonchar \textit{et al.} \cite{Gonchar1993GO32} and Tarkanyi \textit{et al.} \cite{Tarkanyi2004TA08}. 
The cross sections reported in this work also are consistent, within uncertainty, with the values previously reported by Zaneb \cite{Zaneb_2018}. 
However, the values reported in this work are systematically higher than the ones reported by Zaneb, despite the fact that the same experimental data are used in both analyses. 
This can be explained by the differing methods used in each work for determining the deuteron beam currents in the targets. 
$\textsc{TALYS}$ and $\textsc{ALICE}$ significantly underestimate the magnitudes of the cross sections. 
EMPIRE fails to reproduce both the shape and magnitude of all of the \ce{^{86}Y} channels as well.
Overall, the experimental data are best reproduced by $\textsc{CoH}$, which yields a particularly good agreement at lower energies, but underestimates the cross sections for higher energies.

\subsection{$^\text{{nat}}$Zr($d$,$x$)$^{87\text{m}}$Y (cumulative) and $^\text{{nat}}$Zr($d$,$x$)$^{87}$Y (independent and cumulative)}\label{sec:res_87_chain}
\begin{figure}[h!]
    \centering
    \includegraphics[width=\linewidth]{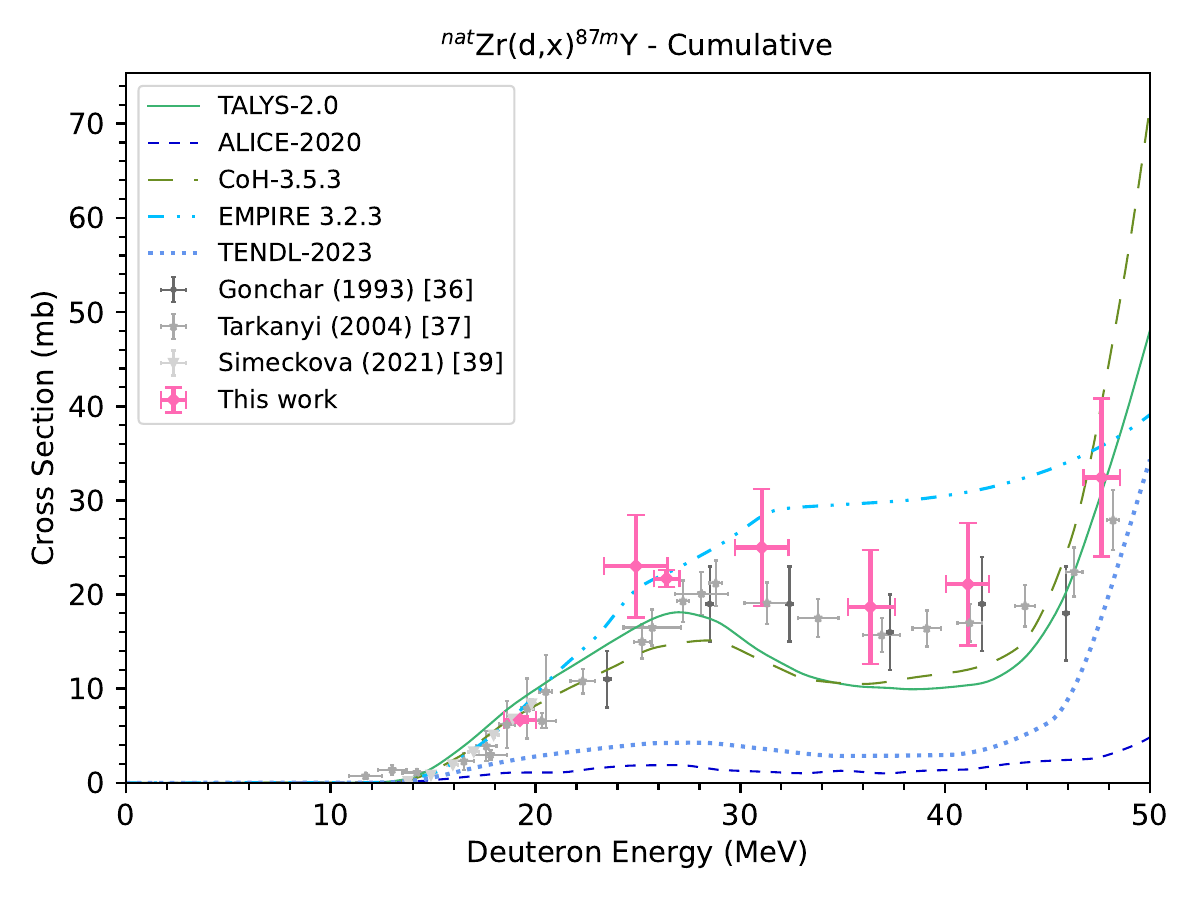}
    \caption{The excitation function of the cumulative production of $^\text{{nat}}$Zr($d$,$x$)$^{87\text{m}}$Y.}
    \label{fig:87Ym_cum}
\end{figure}
In this work, we observed the decays of $^{87}$Y and $^{87\text{m}}$Y which are fed by the decay of $^{87}$Zr. 
The production of radionuclides in a mass chain are therefore linked and in this work, a simultaneous, self-consistent analysis was performed for the decay curves of such chains of nuclei.
However, in this particular case of the $A=87$ isobars, the analysis was instead performed in two portions.
This is due to unresolved numerical issues when using Curie \cite{Curie}, leading to several approaches being employed to extract the cross sections of the $^\text{{nat}}$Zr($d$,$x$)$^{87\text{m}}$Y and $^\text{{nat}}$Zr($d$,$x$)$^{87}$Y reactions for the foils from the different stacks.
For the $^\text{{nat}}$Zr($d$,$x$)$^{87\text{m}}$Y cross sections, a three-step decay curve including $^{87}$Zr, $^{87\text{m}}$Y and $^{87}$Y was used for all the foils in the stack irradiated with $50\ $MeV deuterons. 
For the foils in the stack irradiated with $30\ $MeV deuterons, a two-step decay curve including $^{87\text{m}}$Y and $^{87}$Y was used to extract the cross sections for $^\text{{nat}}$Zr($d$,$x$)$^{87\text{m}}$Y.
When using the two-step decay curves, only spectra counted after $5$--$6$ half-lives of $^{87}$Zr were used. 
For the calculations of the cross sections for $^\text{{nat}}$Zr($d$,$x$)$^{87}$Y, a three-step decay curve including $^{87}$Zr, $^{87\text{m}}$Y and $^{87}$Y was used for all foils. 

\begin{figure}[h!]
    \centering
    \includegraphics[width=\linewidth]{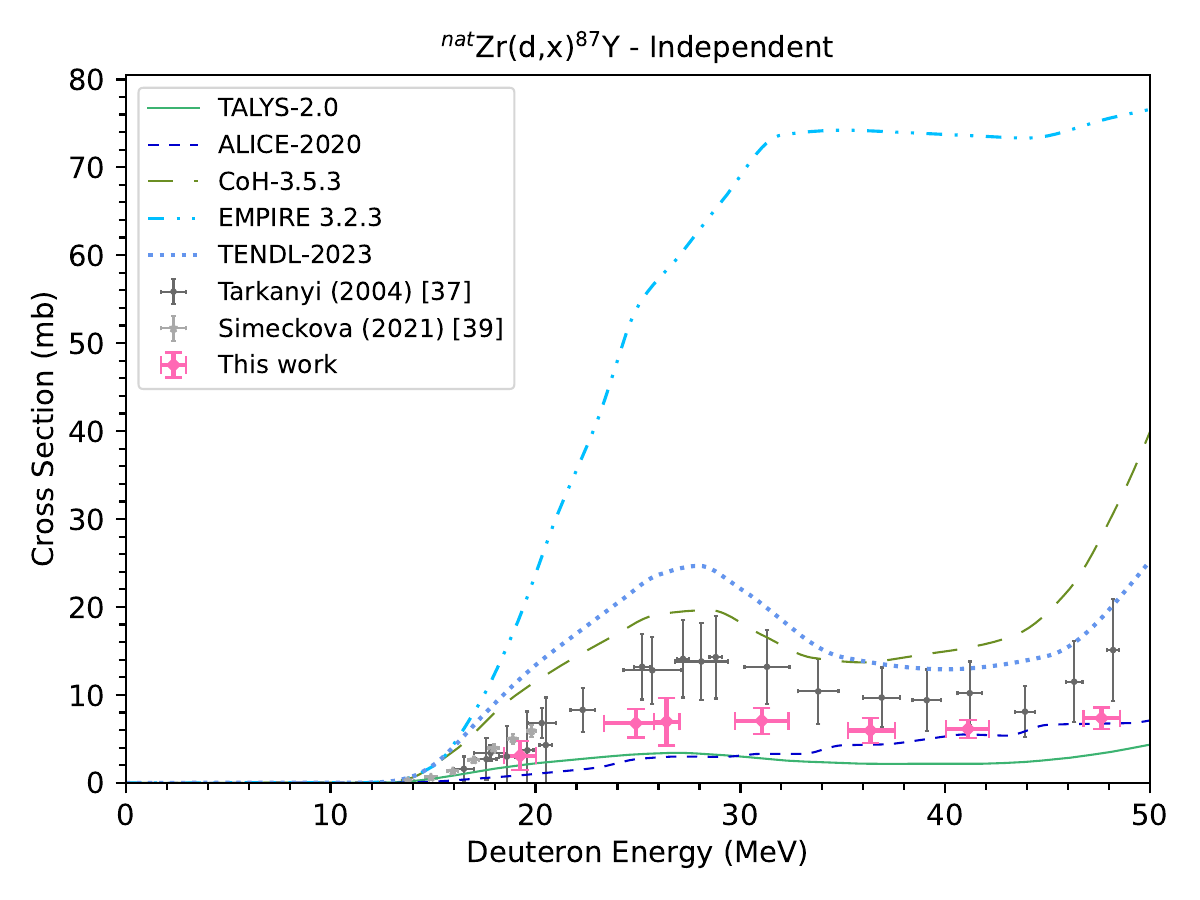}
    \caption{The excitation function of the independent production of $^\text{{nat}}$Zr($d$,$x$)$^{87}$Y.}
    \label{fig:87Y_ind}
\end{figure}

\begin{figure}[h!]
    \centering
    \includegraphics[width=\linewidth]{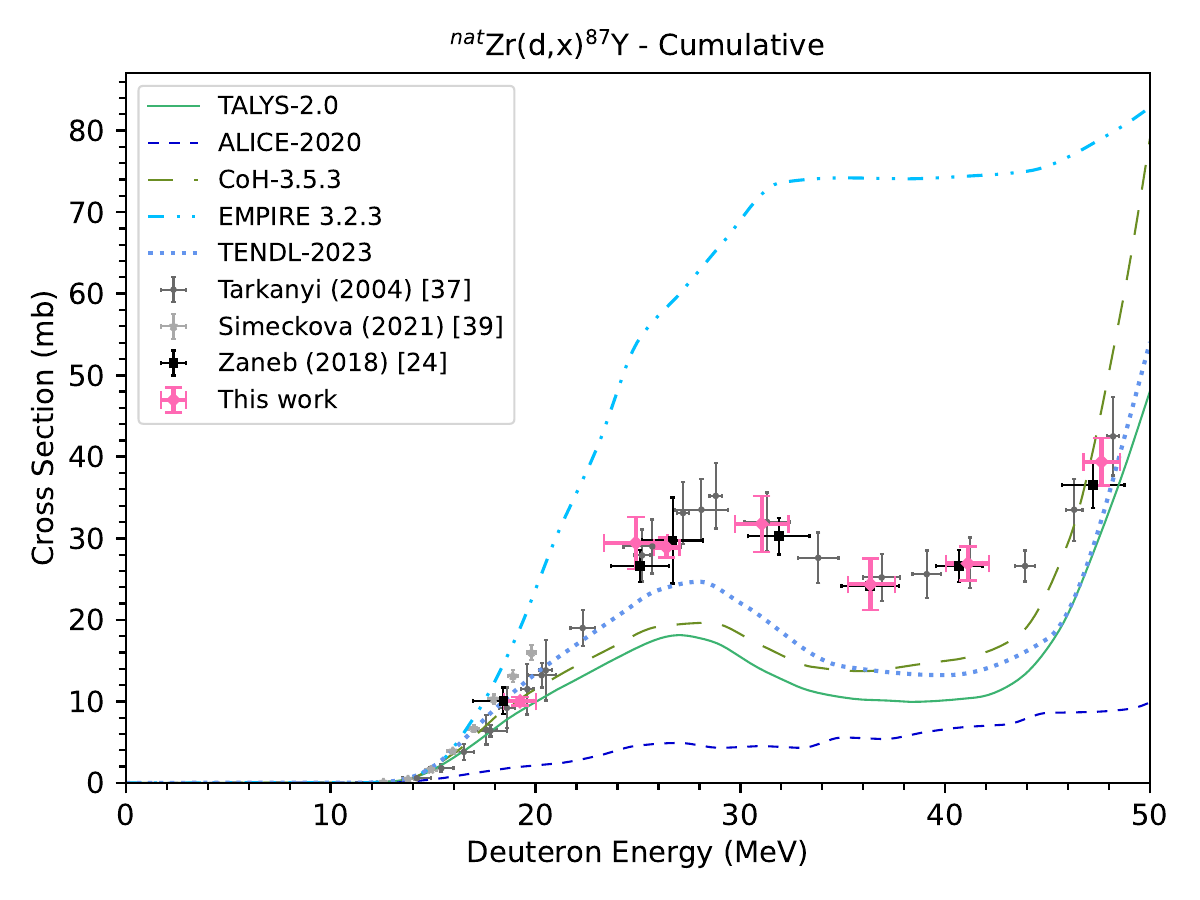}
    \caption{The excitation function of the cumulative production of $^\text{{nat}}$Zr($d$,$x$)$^{87}$Y.}
    \label{fig:87Y_cum}
\end{figure}

The measured cumulative excitation function for $^\text{{nat}}$Zr($d$,$x$)$^{87\text{m}}$Y ($t_{1/2} = 13.37$ h, IT $= 98.43\ \%$) \cite{Johnson2015Nuclear87} from this work is presented in Figure \ref{fig:87Ym_cum}.
The independent and cumulative cross sections for $^\text{{nat}}$Zr($d$,$x$)$^{87}$Y  ($t_{1/2} = 79.8$ h) \cite{Johnson2015Nuclear87} are reported in Figures \ref{fig:87Y_ind}, and \ref{fig:87Y_cum}, respectively. 

The two measurements of $^\text{{nat}}$Zr($d$,$x$)$^{87\text{m}}$Y for the $30\ $MeV stack have much lower relative uncertainties of $2.8\ \%$ and $4.7\ \%$, compared to the foils from the $50\ $MeV stack which range between $20.0$--$32.2\ \%$. 
The cross sections reported in this work (Figure \ref{fig:87Ym_cum}) are in good agreement with the data reported by Gonchar \textit{et al.} \cite{Gonchar1993GO32}, Tarkanyi \textit{et al.} \cite{Tarkanyi2004TA08} and Simeckova \textit{et al.} \cite{Simeckova2021SI31}. 
Both $\textsc{TALYS}$ and $\textsc{CoH}$ produce excitation functions which are in good agreement with the data, particularly near the compound peak around 27 MeV. 
The data suggest that the magnitude of the excitation function calculated by $\textsc{ALICE}$ and the data from the $\textsc{TENDL}$ database are underestimated for the measured range of energies, suggesting overestimation of the \ce{^{87}Y}, and confirmed in Figure \ref{fig:87Y_ind}.
$\textsc{EMPIRE}$ does not reproduce the shape of this channel, beyond approximately $25$ MeV.

The independent cross sections reported for $^\text{{nat}}$Zr($d$,$x$)$^{87}$Y can be seen in Figure \ref{fig:87Y_ind}. 
For the $30\ $MeV stack the relative uncertainties of the production rates are $38.7\ \%$ and $52.6\ \%$, while they range between $16.3$--$24.3\ \%$ for the $50\ $MeV stack. 
The independent cross sections for $^\text{{nat}}$Zr($d$,$x$)$^{87}$Y reported in this work (Figure \ref{fig:87Y_ind}) are systematically lower than the values reported by Tarkanyi \textit{et al.} and Simeckova \textit{et al.} \cite{Tarkanyi2004TA08, Simeckova2021SI31}, except for the cross section reported with the lowest deuteron energy.
However, some of the values reported in this work agree within $1\sigma$ uncertainty to the data reported by Tarkanyi \textit{et al.} \cite{Tarkanyi2004TA08}.
$\textsc{CoH}$ predicts a shape that is in good agreement with the experimental measurements reported by Tarkanyi \textit{et al.} \cite{Tarkanyi2004TA08}, but overestimates the magnitude. 
$\textsc{TALYS}$ predicts the shape and magnitude which is in best agreement with the data reported by Tarkanyi \textit{et al.} \cite{Tarkanyi2004TA08}.
$\textsc{ALICE}$ predicts a rough agreement in magnitude for the excitation functions within this energy range, however, there is a discrepancy for energies ranging from $25$--$30$ MeV.

The cumulative cross sections for $^\text{{nat}}$Zr($d$,$x$)$^{87}$Y can be seen in Figure \ref{fig:87Y_cum}. 
There is a mathematical inconsistency in the data presented for the production of the $87$ chain due to the high uncertainty for the production of the isomeric state, compared to the lower uncertainties for the cumulative production of the ground state. 
However, the data reported are consistent with the decay chain dynamics.
The uncertainties for the production of $^{87\text{m}}$Y are indeed higher than for the cumulative production of $^{87}$Y. 
This is due to the uncertainty in how much of the $^{87\text{m}}$Y activity is caused by feeding from $^{87}$Zr versus direct production of $^{87\text{m}}$Y.
Nevertheless, this uncertainty will not affect the cumulative production of $^{87}$Y. 
The measurements reported in this work (Figure \ref{fig:87Y_cum}) are in good agreement with the previously measured values reported by Tarkanyi \textit{et al.}, Simeckova \textit{et al.} and Zaneb \cite{Tarkanyi2004TA08, Simeckova2021SI31, Zaneb_2018}.
$\textsc{TALYS}$, $\textsc{CoH}$ and $\textsc{TENDL}$ predict a shape consistent with the experimental data, but underestimate the magnitude. 
$\textsc{ALICE}$ and $\textsc{EMPIRE}$ do not predict the correct magnitude or shape.

\subsection{$^\text{{nat}}$Zr($d$,$x$)$^{88}$Y (independent and cumulative)}

\begin{figure}[h!]
    \centering
    \includegraphics[width=\linewidth]{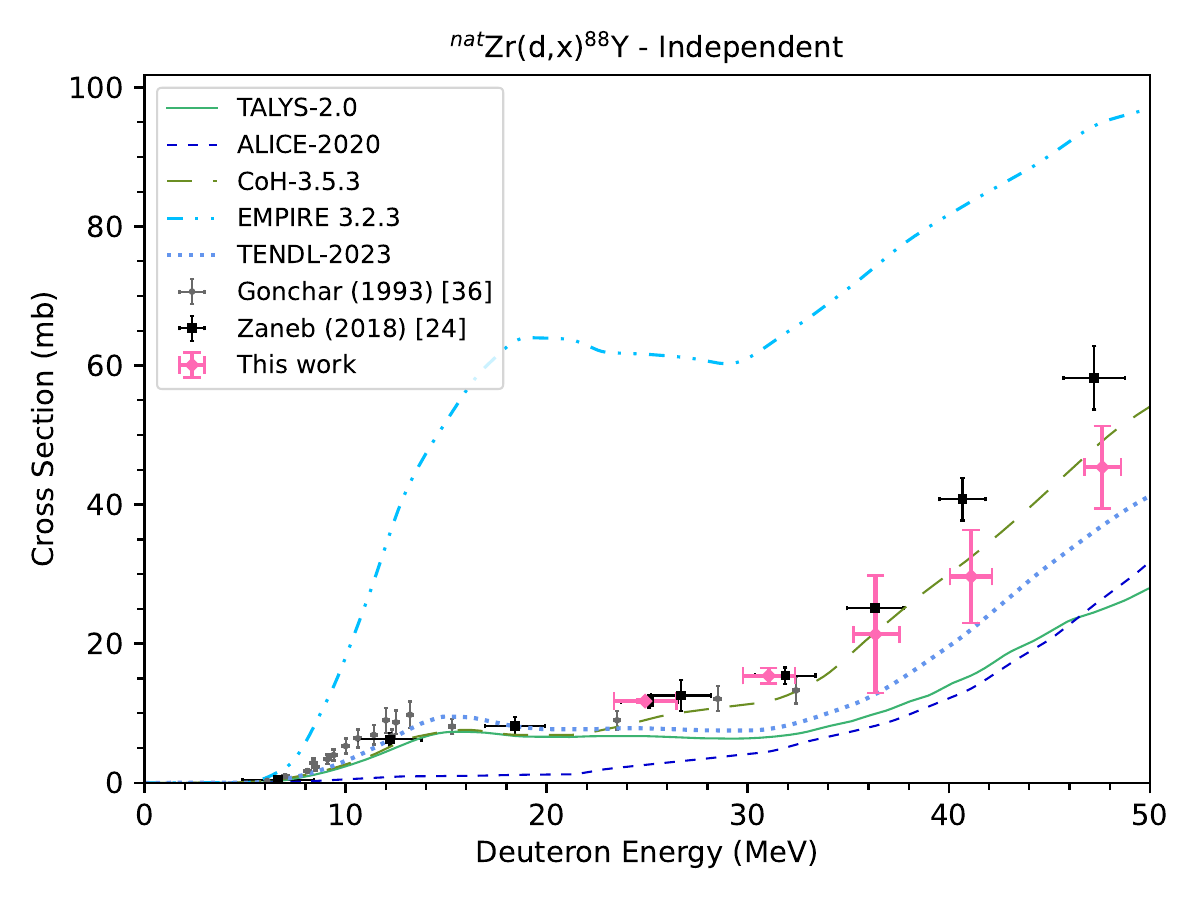}
    \caption{The excitation function of the independent production of $^\text{{nat}}$Zr($d$,$x$)$^{88}$Y. The data reported by Zaneb \cite{Zaneb_2018} are assumed to be independent cross sections.}
    \label{fig:88Y_ind}
\end{figure}

The independent cross sections for $^\text{{nat}}$Zr($d$,$x$)$^{88}$Y are reported, and can be seen in Figure \ref{fig:88Y_ind}. 
The fitted production rates exhibit a wide range of uncertainties, ranging from $0.4$--$39.5\ \%$.
The cross sections reported in this work are in good agreement with the experimental data reported by Gonchar \textit{et al.} \cite{Gonchar1993GO32}. 
The experimental data reported by Zaneb \cite{Zaneb_2018} are assumed to be independent cross sections, and agree with the data reported in this work for lower energies. 
For higher deuteron energies, the data reported by Zaneb \cite{Zaneb_2018} get progressively higher than the data reported in this work. 
All the codes produce excitation functions with shapes consistent with the experimental data. 
However, $\textsc{EMPIRE}$ overestimates the magnitude.  

\begin{figure}[h!]
    \centering
    \includegraphics[width=\linewidth]{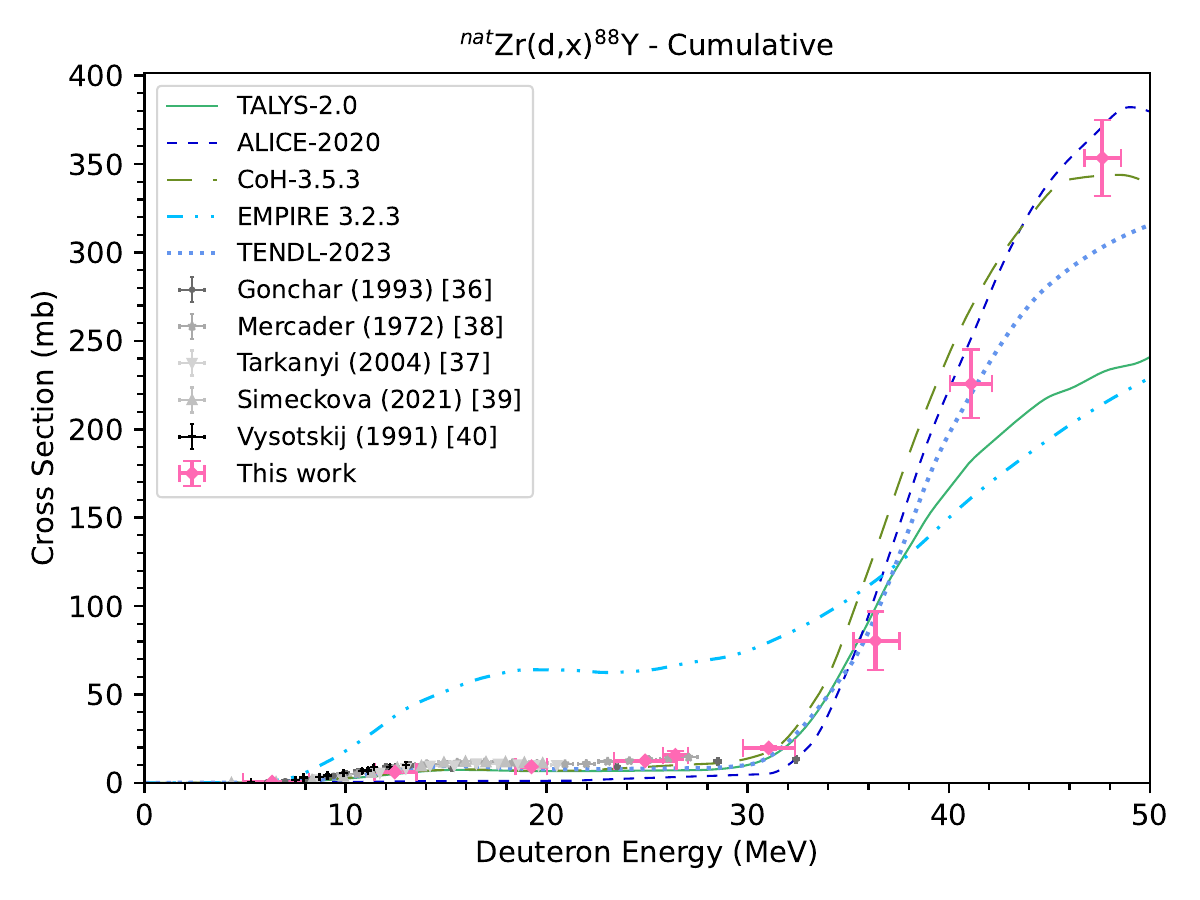}
    \caption{The excitation function of the cumulative production of $^\text{{nat}}$Zr($d$,$x$)$^{88}$Y.}
    \label{fig:88Y_cum}
\end{figure}

The cumulative cross sections for $^\text{{nat}}$Zr($d$,$x$)$^{88}$Y are reported and can be seen in Figure \ref{fig:88Y_cum}.
For foils in the stack irradiated with $30\ $MeV deuterons, the production rates of $^{88}$Y are obtained using a single-step decay curve, only including $^{88}$Y. 
For foils in the stack irradiated with $50\ $MeV, the cumulative cross sections for $^\text{{nat}}$Zr($d$,$x$)$^{88}$Y are obtained by adding the cumulative cross sections for $^\text{{nat}}$Zr($d$,$x$)$^{88}$Zr, weighted by the branching ratio, to the independent cross sections for $^\text{{nat}}$Zr($d$,$x$)$^{88}$Y. 
The reported cross sections are in good agreement with the experimental data reported by Mercader \textit{et al.} \cite{Mercader1972ME22}, Tarkanyi \textit{et al.} \cite{Tarkanyi2004TA08}, Simeckova \textit{et al.} \cite{Simeckova2021SI31} and Vysotskij \textit{et al.} \cite{Vysotskij1991_exforO0380.1}. 
However, there are no previous measurements to compare with for deuteron energies above $30\ $MeV. 
All the reaction modeling codes produce excitation functions with shapes consistent with the experimental data reported in this work. 
All codes provide excitation functions with the right shape.
However, only $\textsc{ALICE}$ and $\textsc{CoH}$ provide excitation functions with the right magnitudes. 
$\textsc{TENDL}$ highly underestimates the cumulative cross section for $^{88}$Y, while providing the right magnitude for the independent cross section. 
This implies that $\textsc{TENDL}$ is underestimating the feeding from $^{88}$Zr. 
$^{88}$Y is one of the main contaminants when producing $^{86\text{g}}$Y through $^\text{{nat}}$Zr($d$,$x$) reactions. 
Figures \ref{fig:88Y_ind} and \ref{fig:88Y_cum} show that the cumulative cross section is much higher than the direct production of $^{88}$Y.
Thus, the radiopurity can be improved significantly by chemically separating yttrium shortly after irradiation, and reducing the production of $^{88}$Y through feeding from $^{88}$Zr. 
While $^{88}$Y presently has no medical application, 
if production of $^{88}$Y is desired, production through $^\text{{nat}}$Zr($d$,$x$) reactions could be a viable production pathway. 

\subsection{$^\text{{nat}}$Zr($d$,$x$)$^{90\text{m}}$Y (cumulative)}

\begin{figure}[h!]
    \centering
    \includegraphics[width=\linewidth]{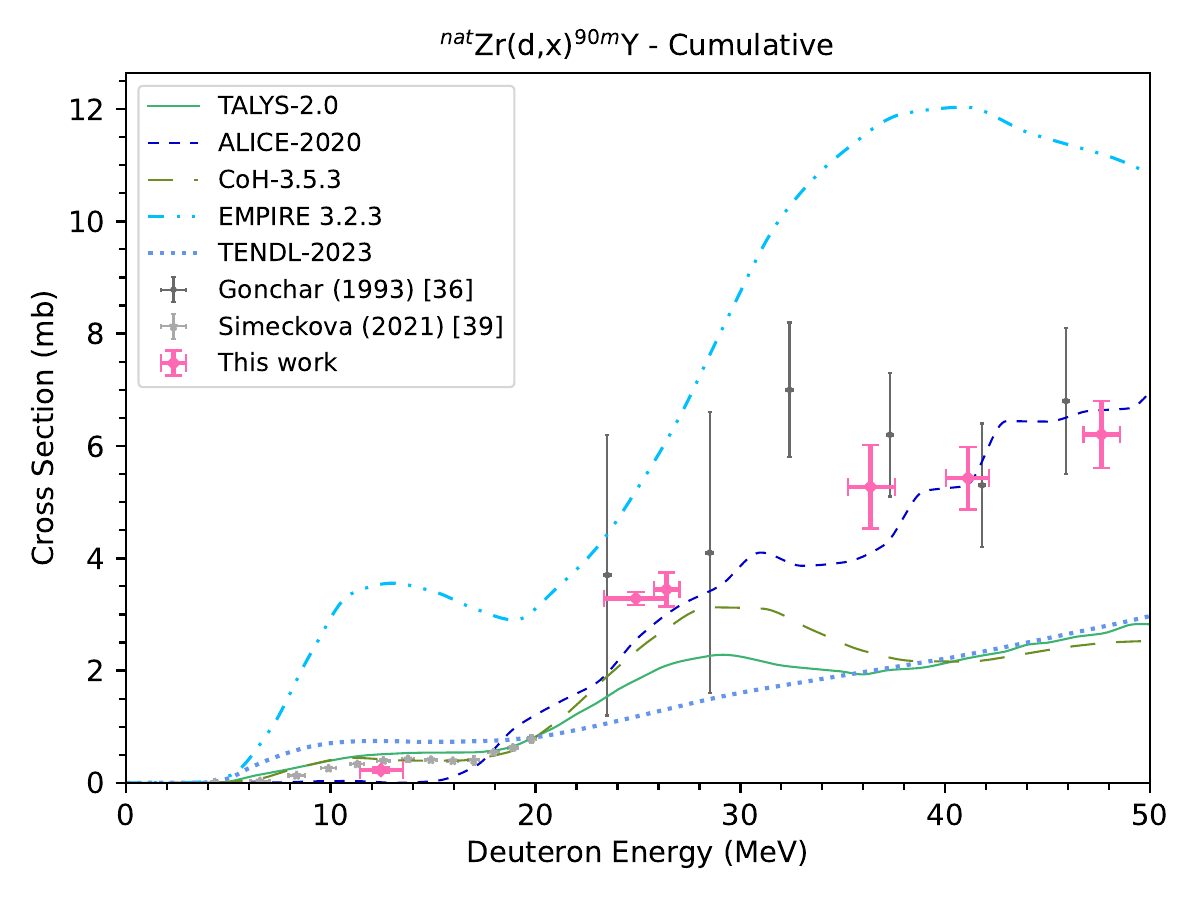}
    \caption{The excitation function of the cumulative production of $^\text{{nat}}$Zr($d$,$x$)$^{90\text{m}}$Y.}
    \label{fig:90Ym_cum}
\end{figure}

The cumulative cross sections for 
$^\text{{nat}}$Zr($d$,$x$)$^{90\text{m}}$Y were measured and can be seen in Figure \ref{fig:90Ym_cum}.
$^{90\text{m}}$Y was observed in almost all foils which were above the energetic threshold for this reaction.
Due to delays in counting the $30$ MeV stack foils, much of the $^{90\text{m}}$Y had already decayed when the first measurements were taken in some of these low-energy foils.
$^{90\text{m}}$Y was also observed in the approximately $31$ MeV foil, but some numerical errors were encountered when  calculating the cross section for this particular foil. 
As such, this data point has been omitted from the results of this work.
The production rates were obtained using a single-step decay curve, and have relative uncertainties ranging from $8.17$--$14.0\ \%$ for all foils except foils $E_d = 12.4$ MeV and $E_d = 24.9$ MeV which have a relative uncertainty of $22.7\ \%$ and $2.62\ \%$, respectively. 
The cross sections reported in this work are in good agreement with the data reported by Gonchar \textit{et al.} \cite{Gonchar1993GO32} and Simeckova \textit{et al.} \cite{Simeckova2021SI31}.
The codes provide excitation functions with approximately the right shapes, but $\textsc{ALICE}$ is the only code providing the right magnitude for higher energies. 
$\textsc{EMPIRE}$ overestimates the magnitude for lower energies as well as for higher energies. 

\subsection{Physical yields of yttrium isotopes}
\label{sec:physical_yields}

\begin{figure}[h!]
    \centering
    \includegraphics[width=\linewidth]{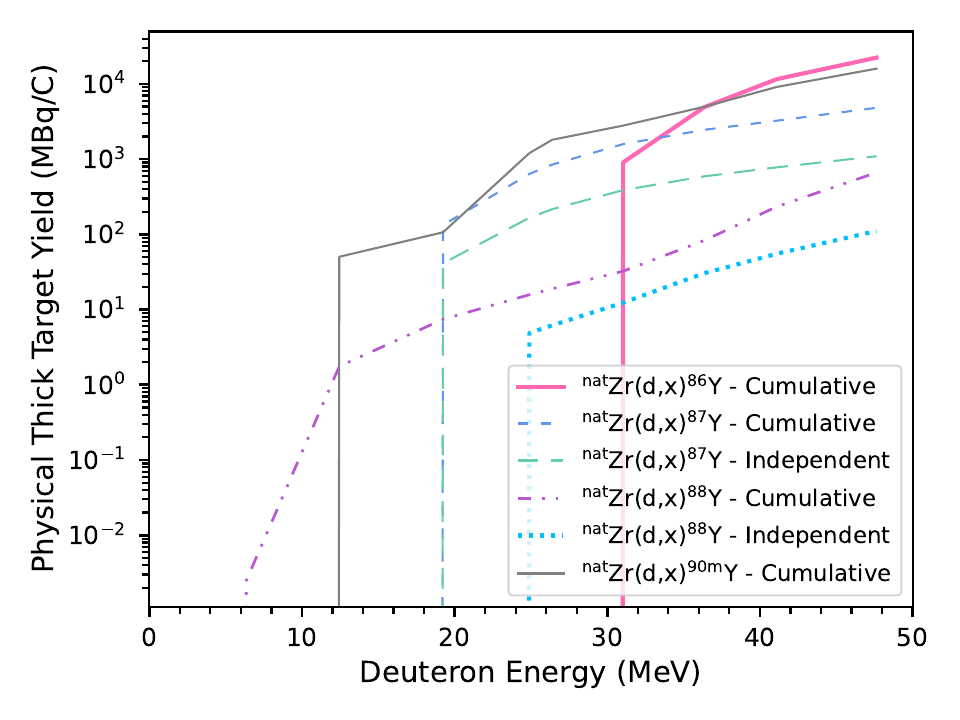}
    \caption{This figure shows the physical yields for the different yttrium isotopes produced in the $^\text{{nat}}$Zr($d$,$x$) reactions. $1\ $MBq/C$\ =0.0036$ MBq/$\mu$A$\cdot$h}
    \label{fig:yield_Zr}
\end{figure}

Figure \ref{fig:yield_Zr} shows the physical yields for all the different yttrium isotopes produced in the $^\text{{nat}}$Zr($d$,$x$) reactions as a function of deuteron beam energy. 
These yields were calculated using the recommendations in \cite{Otuka2015DefinitionsYields}, linearly interpolating between the cross sections reported in this work.
The production yield of the medically relevant $^{86\text{g}}$Y is observed to be higher than that of all other yttrium isotopes produced, with the main contaminants being $^{90\text{m}}$Y and $^{87}$Y.
Over the energy range of $E_d=50\rightarrow 40$ MeV, for example, the yield of $^{86\text{g}}$Y would be $47$ MBq for an irradiation at $1 \mu$A for $1$ hour, but the impurity level would be appreciable (approximately $35$\% $^{90\text{m}}$Y, $7$\% $^{87}$Y and $0.3$\% $^{88}$Y).
Whilst the $^{90\text{m}}$Y impurity would likely not be a major problem given its relatively short half-life, the $7$\% $^{87}$Y contamination likely renders this production regime unsuitable for medical applications.
This is perhaps somewhat expected given the previously measured cross sections for $^{87}$Y production in this energy range (which are in good agreement with the results of this work), which would require the production cross sections $^{86\text{g}}$Y to be approximately an order of magnitude higher than all considered predictions to reach the $<1$\% radiocontamination level that is typically desired.
Whilst more selective production of $^{86\text{g}}$Y may be possible at higher energies, in the measured energy range, the predicted radioisotopic impurity level of approximately $42\%$ is not suitable for use in human subjects.
However, this is to be expected, given the multiple stable isotopes of Zr present in natural abundance targets --- claiming this level of impurity as a viable production route for a medical application would be disingenuous.
Indeed, these measurements were performed as a first and valuable measurement of new nuclear data, and to evaluate the potential of Zr-based production pathways, using enriched Zr targets, as part of future work.

\begin{figure}[h!]
    \centering
    \includegraphics[width=\linewidth]{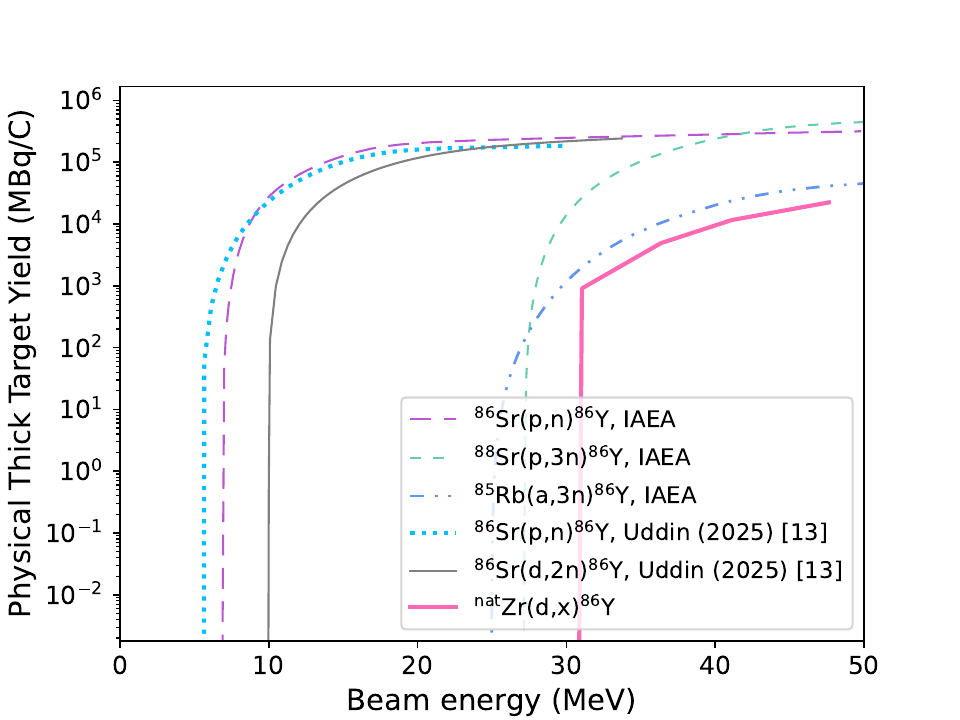}
    \caption{This figure shows the physical yields for the $^{86\text{g}}$Y produced through different production pathways. The data used for the $^{\text{nat}}$Zr($d$,$x$)$^{86\text{g}}$Y reaction are from this work. The data for the $^{86}$Sr($d$,$2n$)$^{86\text{g}}$Y reaction are from a recent publication under this collaboration \cite{Uddin2025An86Srd2n-reactions}, and for the $^{86}$Sr($p$,$n$)$^{86\text{g}}$Y reaction we have plotted both the data from the recent publication under this collaboration \cite{Uddin2025An86Srd2n-reactions} and data from the IAEA. All other data are from the IAEA \cite{InternationalAtomicEnergyAgencyIAEA2001ChargedProject, Tarkanyi2019RecommendedEmitters, Hermanne2021UpgradeIsotopes, Hermanne2023EvaluatedPET}. $1\ $MBq/C$\ =0.0036$ MBq/$\mu$A$\cdot$h}
    \label{fig:yield_routes}
\end{figure}

To estimate this potential, Figure \ref{fig:yield_routes} shows the physical yield for producing $^{86\text{g}}$Y, as a comparison between different production routes.
The data used for the $^{\text{nat}}$Zr($d$,$x$)$^{86\text{g}}$Y reaction are from this work. The data for the $^{86}$Sr($d$,$2n$)$^{86\text{g}}$Y reaction are from a recent publication under this collaboration \cite{Uddin2025An86Srd2n-reactions}, and for the $^{86}$Sr($p$,$n$)$^{86\text{g}}$Y reaction we have plotted both the data from the recent publication under this collaboration \cite{Uddin2025An86Srd2n-reactions} and data from the IAEA. All other data are from the IAEA \cite{InternationalAtomicEnergyAgencyIAEA2001ChargedProject, Tarkanyi2019RecommendedEmitters, Hermanne2021UpgradeIsotopes, Hermanne2023EvaluatedPET}.
We can see that production through $^{86}$Sr($p$,$n$) gives the highest yield for energies up to about $25$ MeV and is very suitable for production of $^{86\text{g}}$Y in the low energy range.
As seen in Ref. \cite{Uddin2025An86Srd2n-reactions}, the recent data from Uddin \emph{et al.} clearly shows the impact of having adopted older experimental data in the most recent IAEA evaluation for the $^{86}$Sr($p$,$n$)$^{86\text{g}}$Y reaction, which have since been superseded by newer measurements.
The data presented by Uddin \emph{et al.} in Ref. \cite{Uddin2025An86Srd2n-reactions} show an integral yield of $291$ MBq/$\mu$Ah for the production of $^{86\text{g}}$Y using $^{86}$Sr($p$,$x$) reactions with $E_p = 14 \rightarrow 7$ MeV, while the data from the IAEA gives an integral yield of $374$ MBq/$\mu$Ah for the same reaction and energy range. 
The yield calculated from the Uddin data is closer to the experimental yields obtained in many experiments. 
We therefore recommend that a targeted updated evaluation be performed by the IAEA in light of these new measurements.
Between $25$ and $30$ MeV the yield for the $^{86}$Sr($d$,$2n$) reaction becomes higher, though the greater radionuclidic impurity for this pathway makes the proton route the more attractive option for production of clinically-relevant $^{86\text{g}}$Y \cite{Uddin2025An86Srd2n-reactions}. 
Above $40$ MeV, the $^{88}$Sr($p$,$3n$) route gives the highest yield for production of $^{86\text{g}}$Y. 
Production through $^\text{{nat}}$Zr($d$,$x$)$^{86\text{g}}$Y gives the lowest yield among all the production pathways.
However, comparing yields obtained with enriched targets to those obtained with natural composed targets is somewhat unjustified, given the proof-of-concept nature of this measurement. 
Even assuming an enriched $^\text{{86}}$Zr target, the cross sections for $^{86\text{g}}$Y production would likely need to be an order of magnitude higher for this route to be considered viable, relative to the use of $^{88}$Sr targets. 
Further investigations need to be performed to measure and calculate the yields for $^{90}$Zr($d$,$\alpha 2n$)$^{86\text{g}}$Y, to compare against the established routes. 

\section{Conclusion} 
This work provides a complete measurement of the $^\text{nat}$Zr($d$,$x$) reactions using the stacked target activation method, with a special focus on the production of the medically relevant radionuclide $^{86\text{g}}$Y. 
We have measured $35$ different excitation functions, where $7$ were measured for the first time in this work.
Nineteen of the reported excitation functions are $^{\text{nat}}$Zr($d$,$x$) reactions, while the remaining $16$ excitation functions are reported from the monitor foils.
These results help provide a more complete picture of feasible production methods for $^{86\text{g}}$Y, indicating the $^\text{nat}$Zr($d$,$x$) pathway to likely be unfeasible in the measured energy range given the impurities and lower yields, relative to other established methods.
However, this is still an interesting production pathway to investigate further due to the convenience of using zirconium as a target material as opposed to strontium. 
Further investigations of the production of $^{86\text{g}}$Y using enriched $^{90}$Zr targets are recommended to test how this compares to enriched $^{86}$Sr and $^{88}$Sr.
In general, most of the reaction modeling codes predict either the right shape and/or amplitude, but show clear need for further improvement.
However, these results show that $\textsc{EMPIRE}$ and $\textsc{ALICE}$ don't reproduce the experimental data as well as the rest of the reaction modeling codes. 

\begin{acknowledgement}
  The authors would like to particularly acknowledge the assistance and support of Brien Ninemire, Scott Small, Tom Gimpel, and all the rest of the operations, research, and facilities staff of the LBNL 88-Inch Cyclotron. We also wish to thank Alexander Springer, who participated in these experiments. This research was supported by the U.S. Department of Energy Isotope Program, managed by the Office of Science for Isotope R\&D and Production, and carried out under Lawrence Berkeley National Laboratory (Contract No. DE-AC02-05CH11231).
  This research was supported by the Research Council of Norway through the Norwegian Nuclear Research Centre (project No. 341985) and through INTPART (project No. 310094).
  H. Zaneb thanks the Higher Education Commission (HEC) of Pakistan for granting her a scholarship to do part of her Ph.D. thesis work at LBNL, Berkeley, USA. 

\end{acknowledgement}


\appendix
\section*{Appendices}
\section{Stack design}
\label{sec:appendix_stack_design}

The length, width, thickness, mass and areal density for each foil can be found in Tables \ref{Tab:foilcharacterization_30_MeV} and \ref{Tab:foilcharacterization_50_MeV} for the stacks irradiated with $30\ $MeV and $50\ $MeV deuterons, respectively.  

\begin{table*}[h!]
    \centering
       \caption{The table shows the characteristics of each foil in the stack irradiated with $30$ MeV deuterons and the calculated areal densities. All lengths are measured in millimeter (mm) and masses are measured in gram (g). The areal densities are given in mg/cm$^2$. The foils are listed in the same order they were irradiated and the horizontal lines divide the foils into compartments.}
       \label{Tab:foilcharacterization_30_MeV}
       \small
    \begin{tabular}{lccccc}
        \toprule
        \text{Foil} & \text{Length} & \text{Width} & \text{Thickness} & \text{Mass} &  \text{Areal density}      \\
         & \text{(mm)} & \text{(mm)} & \text{(mm)} & \text{(g)}  & \text{(mg/cm$^2$)}      \\
        \midrule
        SS1  & & & & & $100.199(71)$ \\
        \midrule
        Ni01  & $25.283(15)$ & $25.005(6)$ & $0.030(1)$ & $0.1470(1)$ & $23.253(15)$  \\
        Zr01  & $24.725(91)$ & $24.993(33)$ & $0.025(1)$ & $0.0998(5)$ & $16.14(10)$  \\
        Ti01  & $24.178(86)$ & $25.190(50)$ & $0.026(1)$ & $0.0703(5)$ & $11.535(95)$  \\
        Al (E1)  & & & & &  $68.312(71)$\\
        \midrule
        Al (E2)  & &  & &  & $68.245(71)$  \\
        Ni02  & $25.030(34)$ & $25.248(26)$ & $0.030(1)$ & $0.1460(1)$ & $23.103(39)$  \\
        Zr02  & $24.853(32)$ & $25.550(70)$ & $0.026(1)$ & $0.1040(1)$ & $16.378(49)$  \\
        Ti02  & $24.405(41)$ & $24.520(65)$ & $0.026(1)$ & $0.0703(5)$ & $11.739(91)$  \\
        \midrule
        Al (E3)  & & & & & $68.349(71)$  \\
        Ni03  & $25.033(68)$ & $25.115(30)$ & $0.029(1)$ & $0.1450(1)$ & $23.064(69)$ \\
        Zr03  & $25.015(17)$ & $24.768(59)$ & $0.026(1)$ & $0.1000(1)$ & $16.140(1)$  \\
        Ti03  & $24.553(42)$ & $25.200(48)$ & $0.026(1)$ & $0.0698(5)$ & $11.273(86)$  \\
        \midrule
        Ni04  & $25.153(15)$ & $25.190(73)$ & $0.029(1)$ & $0.1470(1)$ & $23.201(69)$  \\
        Zr04  & $25.503(54)$ & $24.758(45)$ & $0.026(1)$ & $0.1023(5)$ & $16.195(91)$  \\
        Ti04  & $25.185(17)$ & $25.223(34)$ & $0.026(1)$ & $0.0705(6)$ & $11.099(92)$  \\
        \midrule
        Ni05  & $25.028(13)$ & $25.120(16)$ & $0.028(1)$ & $0.1430(1)$ & $22.746(19)$  \\
        Zr05  & $25.183(32)$ & $24.698(44)$ & $0.026(1)$ & $0.1020(1)$ & $16.400(36)$  \\
        Ti05  & $25.343(39)$ & $24.755(53)$ & $0.026(1)$ & $0.0710(1)$ & $11.317(30)$  \\
        \midrule
        SS2 & & & & & $100.865(71)$ \\
        \bottomrule
    \end{tabular}
\end{table*}

\begin{table*}[h!]
    \centering
       \caption{The table shows the characteristics of each foil in the stack irradiated with $50$ MeV deuterons and the calculated areal densities. All lengths are measured in millimeter (mm) and masses are measured in gram (g). The areal densities are given in mg/cm$^2$. The foils are listed in the same order they were irradiated and the horizontal lines divide the foils into compartments.}
       \label{Tab:foilcharacterization_50_MeV}
       \small
    \begin{tabular}{lccccc}
        \toprule
        \text{Foil} & \text{Length} & \text{Width} & \text{Thickness} & \text{Mass} &  \text{Areal density}      \\
         & \text{(mm)} & \text{(mm)} & \text{(mm)} & \text{(g)}  & \text{(mg/cm$^2$)}      \\
        \midrule
        SS1  & & & & & $100.199(71)$ \\
        \midrule
        Fe01  & $25.095(39)$ & $25.228(22)$ & $0.027(1)$ & $0.1270(1)$ & $20.061(36)$  \\
        Zr06  & $24.813(34)$ & $24.66(26)$ & $0.026(1)$ & $0.0990(1)$ & $16.18(17)$  \\
        Ti06  & $25.355(44)$ & $25.390(65)$ & $0.026(1)$ & $0.0708(5)$ & $10.990(85)$  \\
        Al (C1)  & & & & &  $261.480(71)$\\
        \midrule
        Fe02  & $24.918(36)$ & $25.480(14)$ & $0.027(1)$ & $0.1280(1)$ & $20.161(31)$  \\
        Zr07  & $25.198(5)$ & $24.790(27)$ & $0.026(1)$ & $0.1000(1)$ & $16.009(18)$  \\
        Ti08  & $25.11(14)$ & $25.430(62)$ & $0.026(1)$ & $0.0713(5)$ & $11.16(10)$  \\
        Al (E4)  & & & & & $68.290(71)$  \\
        \midrule
        Al (E5)  & & & & & $68.237(71)$  \\
        Fe03  & $24.70(13)$ & $25.475(33)$ & $0.027(1)$ & $0.1270(1)$ & $20.19(11)$ \\
        Zr08  & $24.740(39)$ & $25.378(15)$ & $0.026(1)$ & $0.1020(1)$ & $16.246(27)$  \\
        Ti09  & $24.995(24)$ & $25.198(71)$ & $0.026(1)$ & $0.0708(5)$ & $11.234(86)$  \\
        Al (E6)  & & & & & $68.215(71)$  \\
        \midrule
        Al (E7)  & & & & & $68.185(71)$  \\
        Fe04  & $25.180(48)$ & $25.290(41)$ & $0.027(1)$ & $0.1280(1)$ & $20.100(50)$  \\
        Zr09  & $24.788(36)$ & $25.36(10)$ & $0.026(1)$ & $0.1020(5)$ & $16.225(69)$  \\
        Ti10  & $25.71(19)$ & $25.53(48)$ & $0.026(1)$ & $0.0738(5)$ & $11.24(24)$  \\
        Al (E8)  & & & & & $68.155(71)$  \\
        \midrule
        Al (E9)  & & & & & $68.177(71)$  \\
        Fe05  & $25.165(13)$ & $25.593(85)$ & $0.027(1)$ & $0.1295(6)$ & $20.11(11)$  \\
        Zr10  & $24.855(33)$ & $25.523(15)$ & $0.025(1)$ & $0.1030(1)$ & $16.237(24)$  \\
        Ti11  & $25.500(26)$ & $24.240(22)$ & $0.026(1)$ & $0.0680(1)$ & $11.001(15)$  \\
        \midrule
        SS2 & & & & & $100.865(71)$ \\
        \bottomrule
    \end{tabular}
\end{table*}

\section{Decay data}
\label{sec:decay_data}

The half-lives and gamma-ray intensities presented in Tables 
\ref{tab:decay_data_Zr_part1} and \ref{tab:decay_data_mon_part1}
were used for all cross-section calculations in this work, and have been taken from the most recent edition of Nuclear Data Sheets for each mass chain \cite{Wu2000Nuclear46, Burrows2007Nuclear47, Chen2022NuclearA=48, Dong2015Nuclear52, Dong2014Nuclear54, Junde2008Nuclear55, Junde2011Nuclear56, Bhat1998Nuclear57, Nesaraja2010Nuclear58, Browne2013Nuclear60, Zuber2015Nuclear61, Browne2010Nuclear65, Negret2015Nuclear86, Johnson2015Nuclear87, McCutchan2014Nuclear88, Singh2013Nuclear89, Basu2020Nuclear90, Baglin2012Nuclear92, Basu2010Nuclear95, Abriola2008Nuclear96}.

\begin{table*}[htbp]
    \centering
    \renewcommand{\arraystretch}{1.0} 
    \caption{Decay data for gamma-rays analyzed in the reporting of cross sections for each product in $^\text{{nat}}$Zr($d$,$x$) reactions. Uncertainties are denoted in the least significant digit, that is, $14.74(2)$ mb means $14.74 \pm 0.02$ mb.}
    \label{tab:decay_data_Zr_part1}
    \setlength{\tabcolsep}{30pt}  
    \begin{tabular}{llll}
\toprule
Nuclide & Half-life & $E_{\gamma}$ (keV) & $I{\gamma}$ ($\%$) \\
\midrule
$^{86}$Y \cite{Negret2015Nuclear86} & 14.74(2) h &  443.13 & 16.90(5) \\
 &  & 627.72 & 32.6(10) \\
 &  & 645.87 & 9.2(11) \\
 &  & 703.33 & 15.4(4) \\
 &  & 777.37 & 22.4(6) \\
 &  & 1076.63 & 82.5 (41) \\
 &  & 1153.05 & 30.5(9) \\
 &  & 1854.38 & 17.2(5) \\
 &  & 1920.72 & 20.8(7) \\
$^{86\text{m}}$Y \cite{Negret2015Nuclear86} & 47.4(4) min & 208.10 & 93.8 (47) \\
$^{87}$Y \cite{Johnson2015Nuclear87} & 79.8(3) h & 388.53 & 82.2(41) \\
 &  & 484.81 & 89.8(9) \\
$^{87\text{m}}$Y \cite{Johnson2015Nuclear87} & 13.37(3) h & 380.79 & 78.0 (39) \\
$^{88}$Y \cite{McCutchan2014Nuclear88} & 106.627(21) d & 898.04 & 93.7(3) \\
 &  & 1836.06 & 99.2(3) \\
$^{88}$Zr \cite{McCutchan2014Nuclear88} & 83.4(3) d & 392.87 & 97.3 (47) \\
$^{88}$Nb \cite{McCutchan2014Nuclear88} & 14.55(11) min & 77.00 & 22.4(11) \\
 &  & 271.80 & 30.1(15) \\
 &  & 399.40 & 31.8(17) \\
 &  & 671.20 & 64(3) \\
 &  & 1057.10 & 100(6) \\
 &  & 1082.60 & 103(6) \\
$^{89}$Zr \cite{Singh2013Nuclear89} & 78.41(12) h & 909.15 & 99.0 (50) \\
$^{90}$Nb \cite{Basu2020Nuclear90} & 14.60(5) h & 141.18 & 66.8(7) \\
 &  & 1129.22 & 92.7(5) \\
 &  & 2186.24 & 17.96(17) \\
 &  & 2318.96 & 82.0(3) \\
$^{90\text{m}}$Y \cite{Basu2020Nuclear90} & 3.19(6) h & 202.53 & 97.0(5) \\
 &  & 479.51 & 90.5(3) \\
$^{92\text{m}}$Nb \cite{Baglin2012Nuclear92} & 10.15(2) d & 934.44 & 99.2 (50) \\
 $^{95}$Zr \cite{Basu2010Nuclear95} & 64.032(6) d & 724.19 & 44.27(22) \\
 &  & 756.73 & 54.38(22) \\
$^{95}$Nb \cite{Basu2010Nuclear95} & 34.991(6) d & 765.80 & 99.808(7) \\
$^{95\text{m}}$Nb \cite{Basu2010Nuclear95} & 3.61 d & 235.69 & 24.8(8) \\
$^{96}$Nb \cite{Abriola2008Nuclear96} & 23.35(5) h & 460.04 & 26.62(19) \\
 &  & 480.70 & 5.84(5) \\
 &  & 568.87 & 58.0(3) \\
 &  & 719.56 & 6.85(9) \\
 &  & 778.22 & 96.45(22) \\
 &  & 810.33 & 11.09(10) \\
 &  & 849.93 & 20.45(19) \\
 &  & 1091.35 & 48.5(15) \\
 &  & 1200.23 & 19.97(10) \\
\bottomrule
\end{tabular}
\end{table*}

\begin{table*}[htbp]
    \centering
    \renewcommand{\arraystretch}{1.0} 
    \caption{Decay data for gamma-rays analyzed in the reporting of cross sections for each product in $^\text{{nat}}$Ni($d$,$x$), $^\text{{nat}}$Ti($d$,$x$) and $^\text{{nat}}$Fe($d$,$x$) reactions. Uncertainties are denoted in the least significant digit, that is, $83.79(4)$ mb means $83.79 \pm 0.04$ mb.}
    \label{tab:decay_data_mon_part1}
    \setlength{\tabcolsep}{30pt}  
\begin{tabular}{llll}
\toprule
Nuclide & Half-life & $E_{\gamma}$ (keV) & $I{\gamma}$ ($\%$) \\
\midrule
$^{46}$Sc \cite{Wu2000Nuclear46} & 83.79(4) d & 889.28 & 99.9840(10) \\
 &  & 1120.55 & 99.9870(10) \\
$^{47}$Sc \cite{Burrows2007Nuclear47} & 3.3492(6) d & 159.38 & 68.3(4) \\
$^{48}$Sc \cite{Chen2022NuclearA=48} & 43.71(9) h & 175.36 & 7.47(18) \\
 &  & 983.53 & 100 (5) \\
 &  & 1037.52 & 97.5(20) \\
 &  & 1312.12 & 100 (5) \\
$^{48}$V \cite{Chen2022NuclearA=48} & 15.974(3) d & 944.13 & 7.870(7) \\
 &  & 983.52 & 99.98(4) \\
 &  & 1312.11 & 98.2(3) \\
$^{52}$Mn \cite{Dong2015Nuclear52} & 5.591(3) d & 744.23 & 90.0(12) \\
 &  & 935.54 & 94.5(13) \\
 &  & 1333.65 & 5.07(7) \\
 &  & 1434.09 & 100.0(14) \\
$^{54}$Mn \cite{Dong2014Nuclear54} & 312.20(20) d & 834.85 & 99.9760(10) \\
$^{55}$Co \cite{Junde2008Nuclear55} & 17.53(3) h & 477.20 & 20.2(17) \\
 &  & 931.10 & 75 (4) \\
 &  & 1316.60 & 7.1(3) \\
 &  & 1408.50 & 16.9(8) \\
$^{56}$Co \cite{Junde2011Nuclear56} & 77.236(26) d & 846.77 & 99.9(50) \\
 &  & 1037.84 & 14.05(4) \\
 &  & 1238.29 & 66.46(12) \\
 &  & 1771.36 & 15.41(6) \\
 &  & 2034.79 & 7.77(3) \\
 &  & 2598.50 & 16.97(4) \\
$^{57}$Co \cite{Bhat1998Nuclear57} & 271.74(6) d & 122.06 & 85.60(17) \\
 &  & 136.47 & 10.68(8) \\
 $^{57}$Ni \cite{Bhat1998Nuclear57} & 35.60(6) h & 127.16 & 16.7(5) \\
 &  & 1377.63 & 81.7(24) \\
 &  & 1757.55 & 5.75(20) \\
 &  & 1919.52 & 12.3(4) \\
$^{58}$Co \cite{Nesaraja2010Nuclear58} & 70.86(6) d & 810.76 & 99.5 (50) \\
$^{60}$Co \cite{Browne2013Nuclear60} & 1925.28(14) d & 1173.23 & 99.85(3) \\
 &  & 1332.49 & 99.9826(6) \\
$^{61}$Cu \cite{Zuber2015Nuclear61} & 3.336(10) h & 282.96 & 12.20 (61) \\
 &  & 656.01 & 10.80 (54) \\
$^{65}$Ni \cite{Browne2010Nuclear65} & 2.51719(26) h & 1115.53 & 15.43(13) \\
 &  & 1481.84 & 23.6 (12) \\
\bottomrule
\end{tabular}
\end{table*}

\section{Tabulated cross sections from monitor foils}
For the ($d$,$x$) reactions on $^{\text{nat}}$Ni, the extracted cross sections for $^{\text{57, 65}}$Ni, $^{\text{55, 57, 60}}$Co, and $^{\text{52, 54}}$Mn are presented in Table \ref{tab:xs_Ni}. 
Cross sections for $^{\text{47, 48}}$Sc were extracted for ($d$,$x$) reactions on $^{\text{nat}}$Ti, and are presented in Table \ref{tab:xs_Ti}. 
For ($d$,$x$) reactions on $^{\text{nat}}$Fe the extracted cross sections for $^{\text{55, 57, 58}}$Co, $^{52, 54}$Mn, and $^{\text{48}}$V are presented in Table \ref{tab:xs_Fe}. 

\begin{table*}[htbp]
    \centering
    \renewcommand{\arraystretch}{1.1} 
    \caption{Measured cross sections for all the $^{\text{nat}}$Ni($d$,$x$) reaction products observed in this work. The notations are the same as those in Table \ref{tab:xs_Zr}.
    }
    \label{tab:xs_Ni}
    \begin{tabular}{*{5}{l}}
        \midrule
        & \multicolumn{4}{l}{Production cross section (mb)} \\
        \cmidrule(lr){2-5}
        $E_d$ (MeV) & $27.35(64)$ & $20.45(76)$ & $14.1(10)$ & $8.8(14)$ \\
        Stack ID & B & B & B & B \\
        \toprule
        $^{52}$Mn ($\sigma_c$) & $2.68(36)$ & $0.764(79)$ & — & — \\
        $^{54}$Mn ($\sigma_i$) & $10.2(14)$ & — & — & — \\
        $^{55}$Co ($\sigma_c$) & $17.6(16)$ & $19.0(20)$ & $5.23(66)$ & — \\
        $^{57}$Co ($\sigma_i$) & $329(29)$ & $107.2(93)$ & $16.68(68)$ &  — \\
        $^{57}$Co ($\sigma_c$) &  — &  — &  — & $1.08(42)$\\
        $^{57}$Ni ($\sigma_c$) & $35.9(33)$ & $7.24(62)$ & $1.64(13)$ & — \\
        $^{60}$Co ($\sigma_c$) & $13.6(26)$ & — & $3.35(90)$ & $1.27(37)$ \\
        $^{65}$Ni ($\sigma_i$) & $0.466(58)$ & $0.900(89)$ & $1.332(62)$ & $2.29(12)$ \\
        \midrule
    \end{tabular}
\end{table*}

\begin{table*}[htbp]
    \centering
    \renewcommand{\arraystretch}{1.1} 
    \caption{Measured cross sections for all the $^{\text{nat}}$Ti($d$,$x$) reaction products observed in this work. The notations are the same as those in Table \ref{tab:xs_Zr}.
    }
    \label{tab:xs_Ti}
    \begin{tabular}{*{10}{l}}
        \midrule
        & \multicolumn{9}{l}{Production cross section (mb)} \\
        \cmidrule(lr){2-10}
        $E_d$ (MeV) & $47.10(95)$ & $40.5(11)$ & $35.7(12)$ & $30.3(14)$  & $25.54(65)$ & $24.0(16)$ & $18.13(80)$ & $10.9(12)$ & $3.9(22)$ \\
        Stack ID & A & A & A & A & B & A & B & B & B \\
        \toprule
        $^{47}$Sc ($\sigma_c$) & $52.7(20)$ & $48(17)$ & $40(14)$ & $27.8(96)$ & $14.64(33)$ & $13.1(46)$ & $5.86(83)$ & $2.02(79)$ & — \\
        $^{48}$Sc ($\sigma_c$) & $10.4(24)$ & $10.1(20)$ & $9.9(17)$ & $10.1(20)$ & — & — & — & — & — \\

        \midrule
    \end{tabular}
\end{table*}

\begin{table*}[htbp]
    \centering
    \renewcommand{\arraystretch}{1.1} 
    \caption{Measured cross sections for all the $^{\text{nat}}$Fe($d$,$x$) reaction products observed in this work. The notations are the same as those in Table \ref{tab:xs_Zr}.
    }
    \label{tab:xs_Fe}
    \begin{tabular}{*{6}{l}}
        \midrule
        & \multicolumn{5}{l}{Production cross section (mb)} \\
        \cmidrule(lr){2-6}
        $E_d$ (MeV) & $48.23(92)$ & $41.8(11)$ & $37.1(12)$ & $31.9(13)$ & $25.9(15)$ \\
        Stack ID & A & A & A & A & A \\
        \toprule
        $^{48}$V ($\sigma_c$) & $2.40(47)$ & $0.334(62)$ & $0.342(98)$ & $0.58(13)$ & $0.498(61)$ \\
        $^{52}$Mn ($\sigma_c$) & $31.3(64)$ & $38.5(75)$ & $42.5(73)$ & $33.2(57)$ & $4.77(62)$ \\
        $^{54}$Mn ($\sigma_i$) & $176(35)$ & $113(22)$ & $60(11)$ & $25.8(43)$ & $28.9(43)$ \\
        $^{55}$Co ($\sigma_i$) & $20.9(61)$ & $25.8(71)$ & $29.5(77)$ & $45(11)$ & $20.8(41)$ \\
        $^{57}$Co ($\sigma_i$) & $19.0(82)$ & $20.7(89)$ & $23.1(98)$ & $32(13)$ & $46(19)$ \\
        $^{58}$Co ($\sigma_i$) & — & — & $1.36(34)$ & $1.62(43)$ & $2.04(45)$ \\
        
        \midrule
    \end{tabular}
\end{table*}

\section{Measured excitation functions for $^\text{nat}$Zr($d$,$x$) reactions}
\label{sec:Zr_excitation_functions}
The measured cross sections presented in this section are compared to literature data \cite{Gonchar1993GO32, Tarkanyi2004TA08, Mercader1972ME22, Simeckova2021SI31, Vysotskij1991_exforO0380.1}, the $\textsc{TENDL}-2023$ data library \cite{TENDL} and the reaction modeling codes $\textsc{TALYS}-2.0$ \cite{TALYS}, $\textsc{ALICE}-2020$ \cite{ALICE}, $\textsc{CoH}-3.5.3$ \cite{CoH} and $\textsc{EMPIRE}-3.2.3$ \cite{EMPIRE}.

\begin{figure}[h!]
    \centering
    \includegraphics[width=\linewidth]{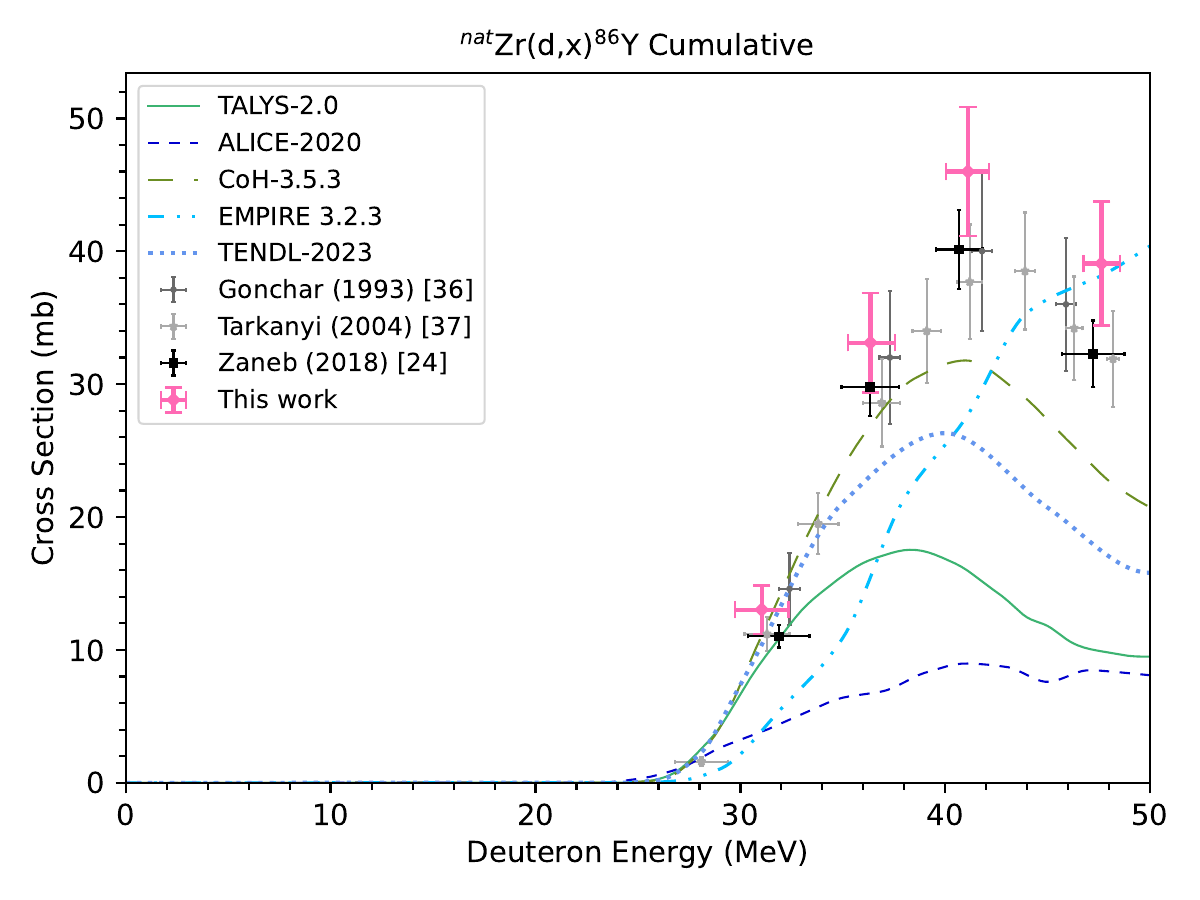}
    \caption{The excitation function for the cumulative production of $^\text{{nat}}$Zr($d$,$x$)$^{86}$Y ground state.}
    \label{fig:86Y_cum}
\end{figure}

\begin{figure}[h!]
    \centering
    \includegraphics[width=\linewidth]{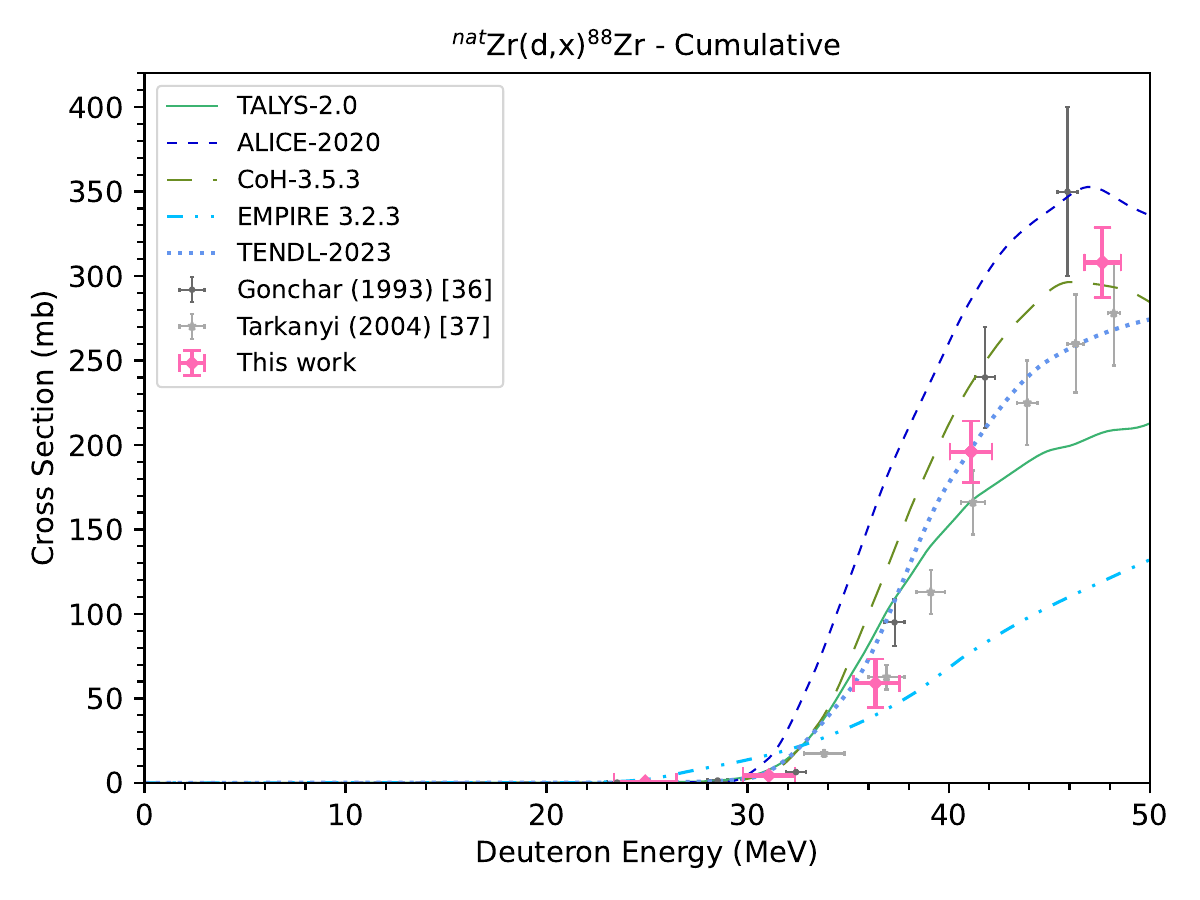}
    \caption{The excitation function of the cumulative production of $^\text{{nat}}$Zr($d$,$x$)$^{88}$Zr.}
\end{figure}

\begin{figure}[h!]
    \centering
    \includegraphics[width=\linewidth]{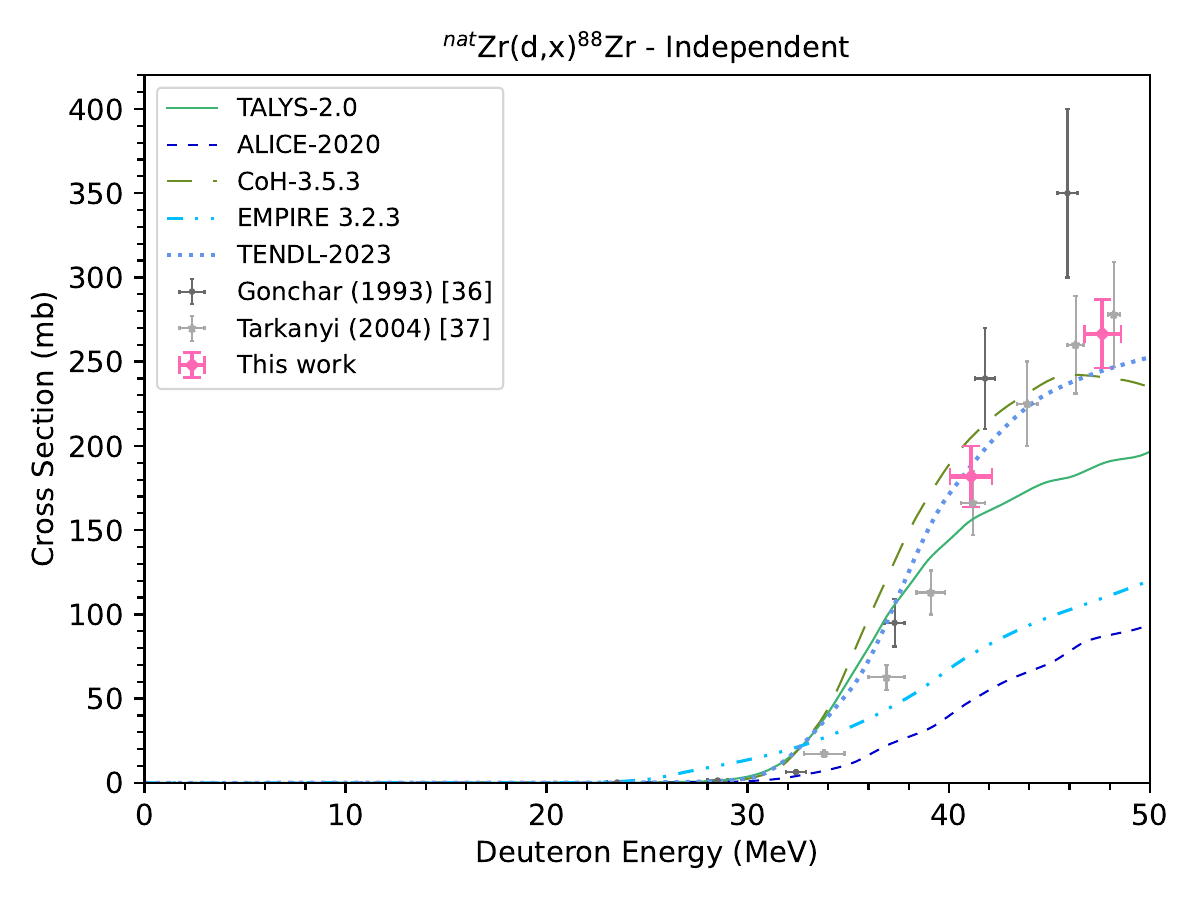}
    \caption{The excitation function of the independent production of $^\text{{nat}}$Zr($d$,$x$)$^{88}$Zr.}
\end{figure}

\begin{figure}[h!]
    \centering
    \includegraphics[width=\linewidth]{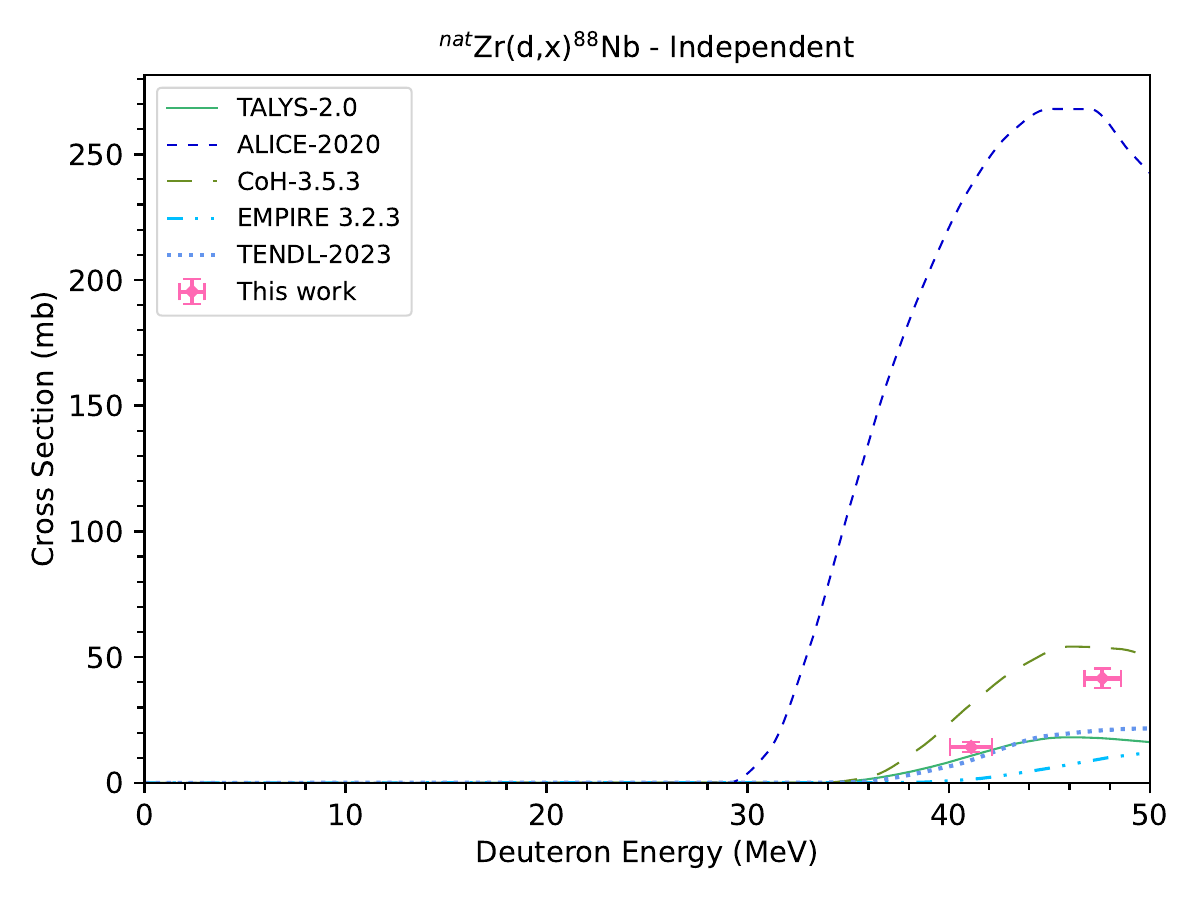}
    \caption{The excitation function of the independent production of $^\text{{nat}}$Zr($d$,$x$)$^{88}$Nb.}
\end{figure}

\begin{figure}[h!]
    \centering
    \includegraphics[width=\linewidth]{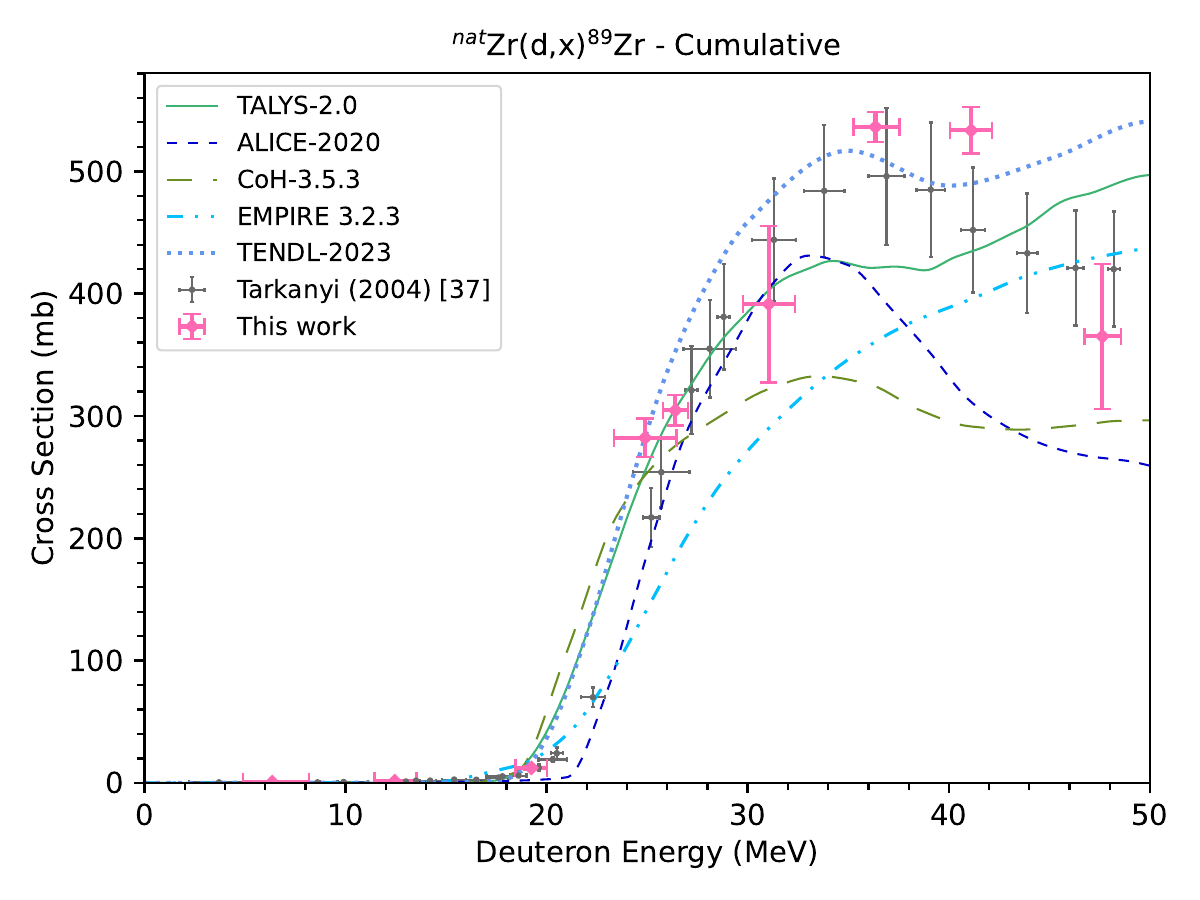}
    \caption{The excitation function of the cumulative production of $^\text{{nat}}$Zr($d$,$x$)$^{89}$Zr.}
\end{figure}

\begin{figure}[h!]
    \centering
    \includegraphics[width=\linewidth]{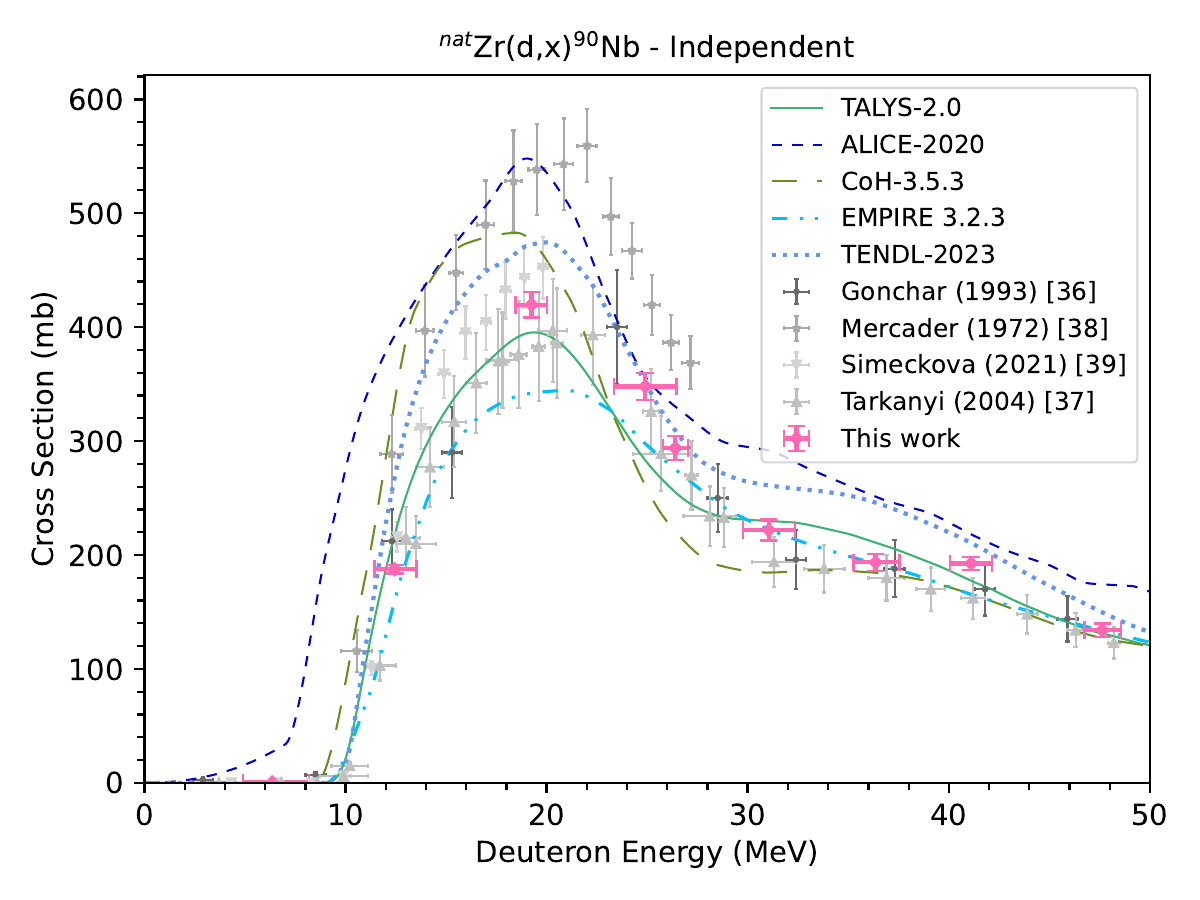}
    \caption{The excitation function of the independent production of $^\text{{nat}}$Zr($d$,$x$)$^{90}$Nb.}
\end{figure}

\begin{figure}[h!]
    \centering
    \includegraphics[width=\linewidth]{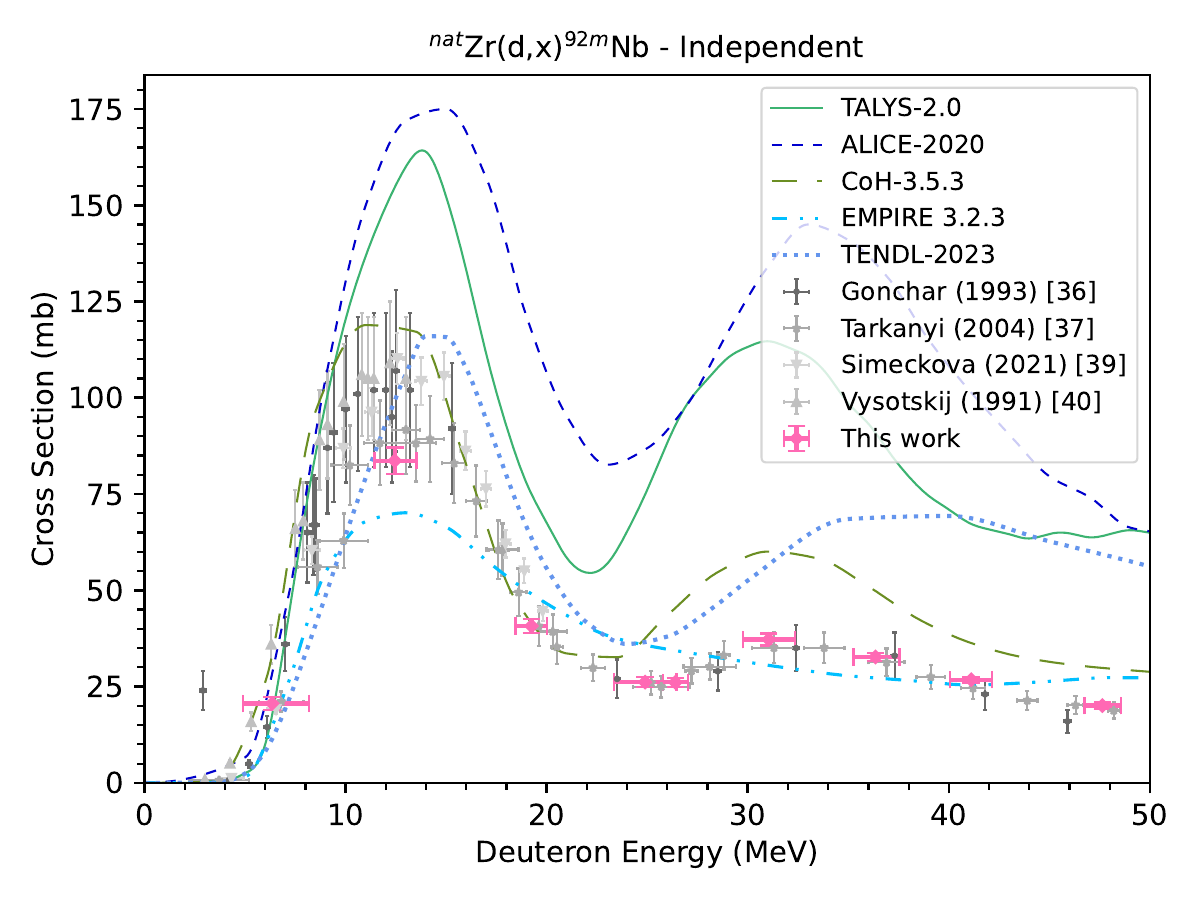}
    \caption{The excitation function of the independent production of $^\text{{nat}}$Zr($d$,$x$)$^{92m}$Nb.}
\end{figure}

\begin{figure}[h!]
    \centering
    \includegraphics[width=\linewidth]{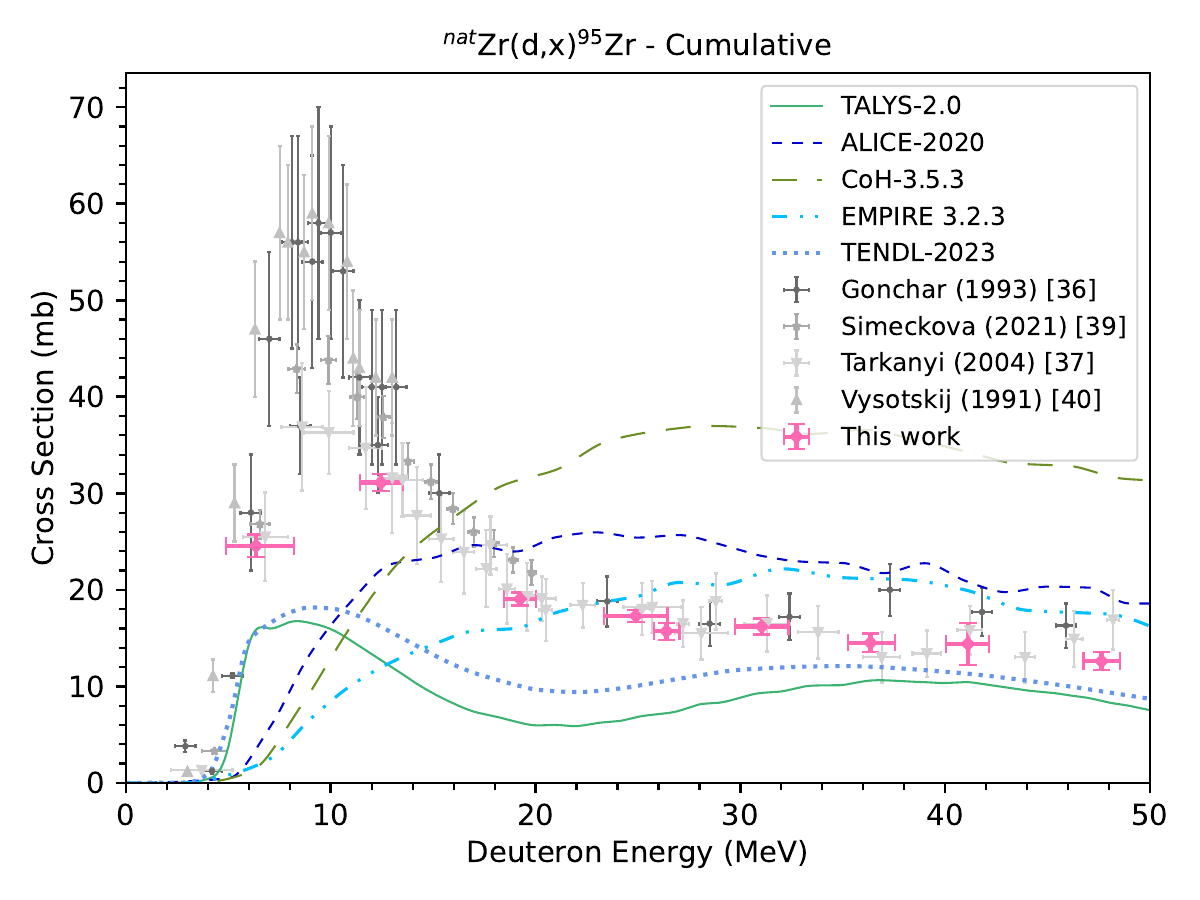}
    \caption{The excitation function of the cumulative production of $^\text{{nat}}$Zr($d$,$x$)$^{95}$Zr.}
\end{figure}

\begin{figure}[h!]
    \centering
    \includegraphics[width=\linewidth]{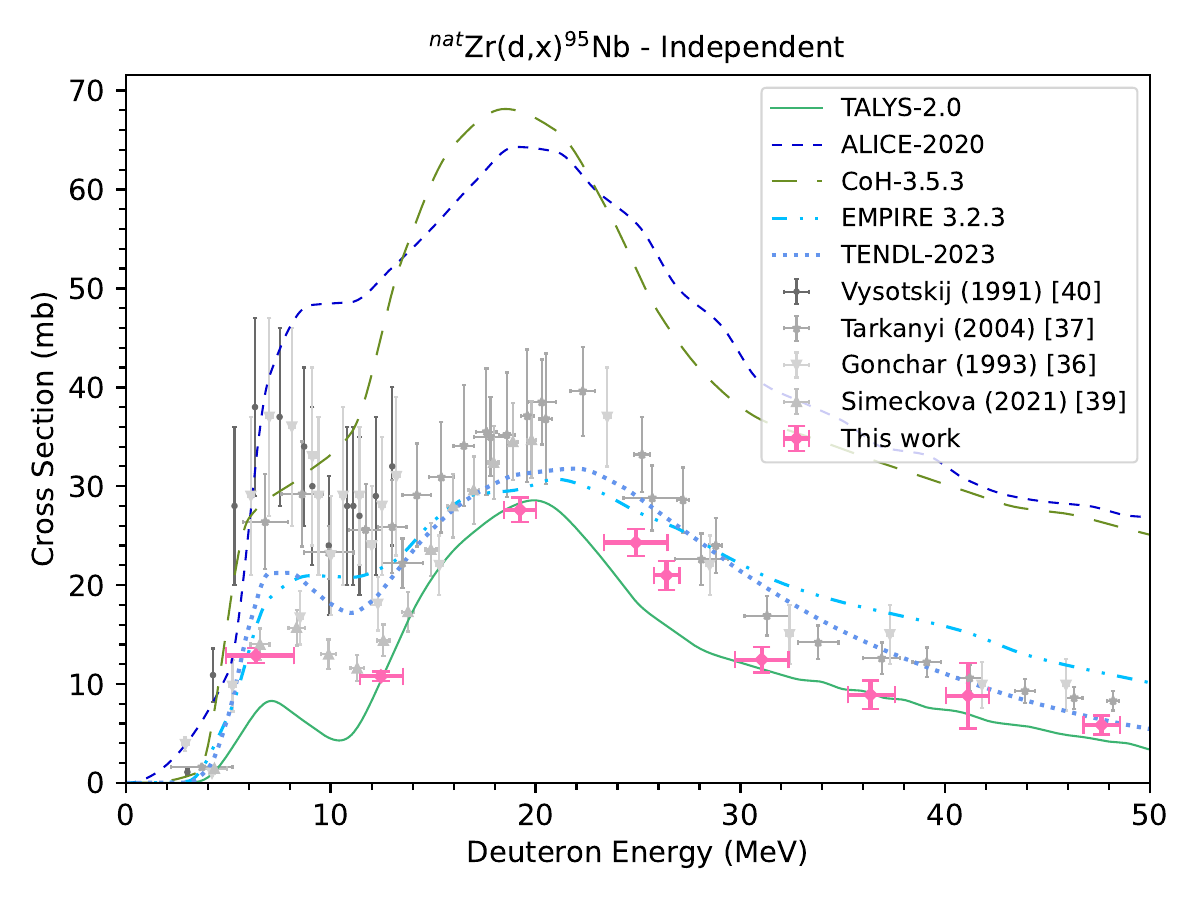}
    \caption{The excitation function of the independent production of $^\text{{nat}}$Zr($d$,$x$)$^{95}$Nb.}
\end{figure}

\begin{figure}[h!]
    \centering
    \includegraphics[width=\linewidth]{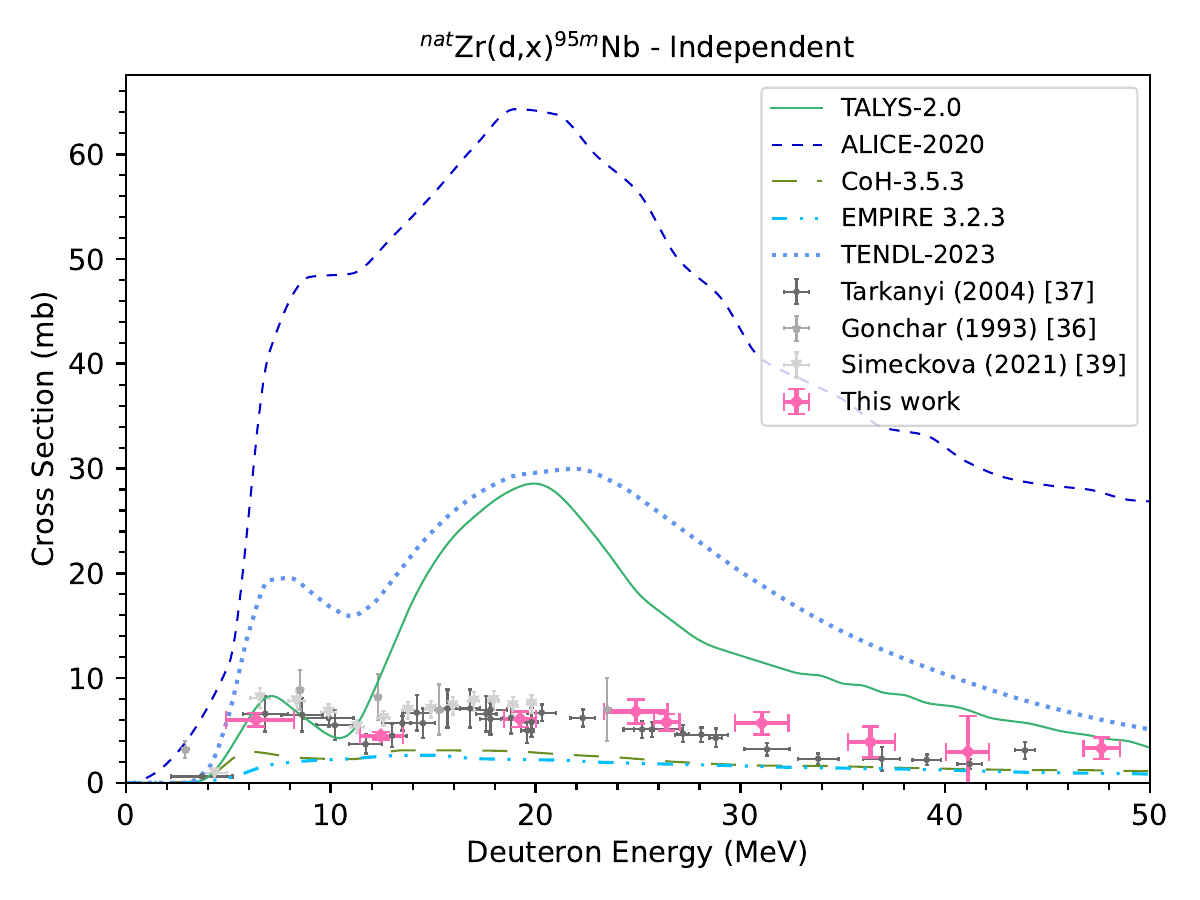}
    \caption{The excitation function of the independent production of $^\text{{nat}}$Zr($d$,$x$)$^{95m}$Nb.}
\end{figure}

\begin{figure}[h!]
    \centering
    \includegraphics[width=\linewidth]{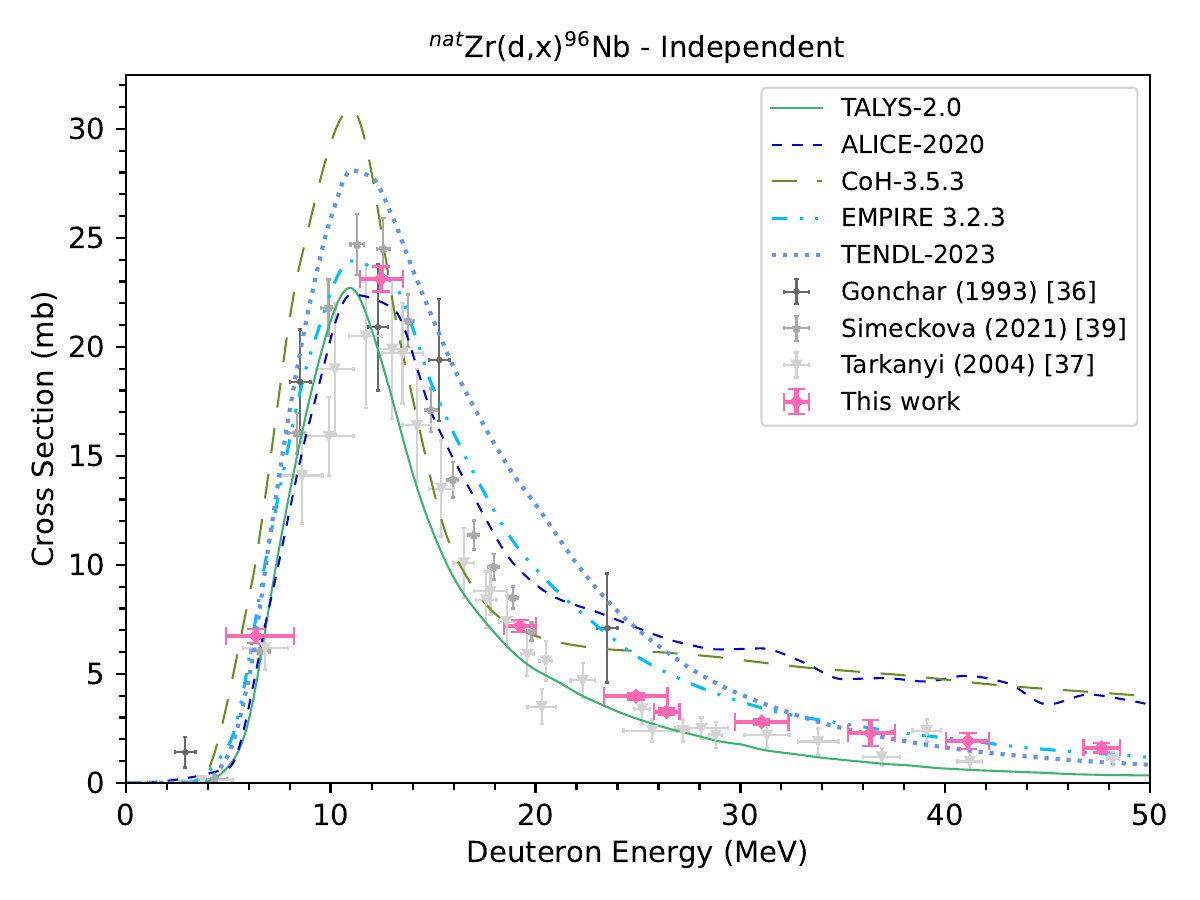}
    \caption{The excitation function of the independent production of $^\text{{nat}}$Zr($d$,$x$)$^{96}$Nb.}
\end{figure}



\setcounter{section}{18}
\section{Supplementary information}
\subsection{Measured excitation functions}
\label{sec:measured_excitation_functions}

\subsubsection{$^\text{nat}$Ni($d$,$x$) reactions}
The measured cross sections presented in this section are compared to literature data \cite{Takacs2007EvaluatedNickel, Amjed2013Activation40MeV, Usman2016Measurements24MeV, Hermanne2013New50MeV, Ochiai2007DeuteronIFMIF, Avrigeanu2016Deuteron-inducedMeV, Zweit1991ExcitationTomography, Takacs1997ActivationPurpose}, the $\textsc{TENDL}-2023$ data library \cite{TENDL} and the reaction modeling codes $\textsc{TALYS}-2.0$ \cite{TALYS}, $\textsc{ALICE}-2020$ \cite{ALICE}, $\textsc{CoH}-3.5.3$ \cite{CoH} and $\textsc{EMPIRE}-3.2.3$ \cite{EMPIRE}.

\begin{figure}[h!]
    \centering
    \includegraphics[width=\linewidth]{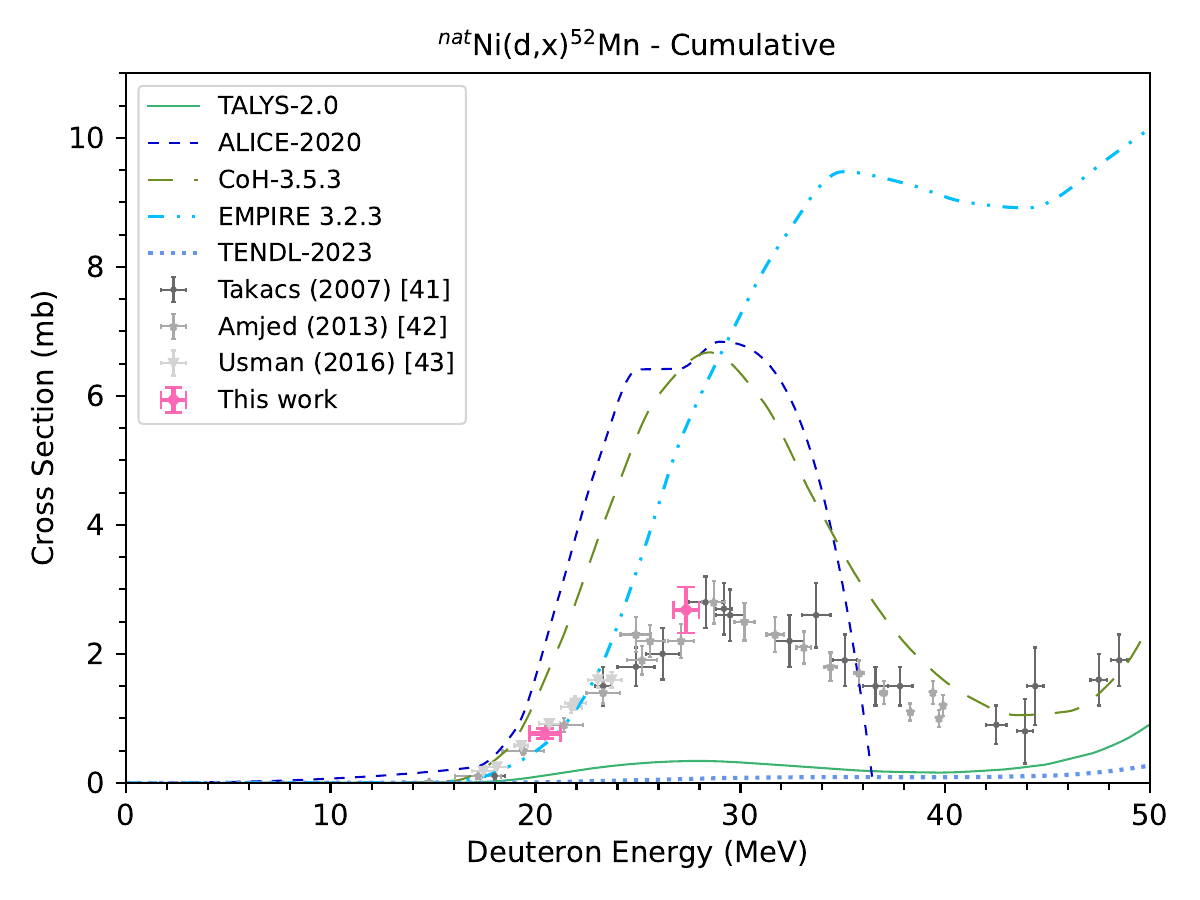}
    \caption{The excitation function of the cumulative production of $^\text{{nat}}$Ni($d$,$x$)$^{52}$Mn.}
\end{figure}

\begin{figure}[h!]
    \centering
    \includegraphics[width=\linewidth]{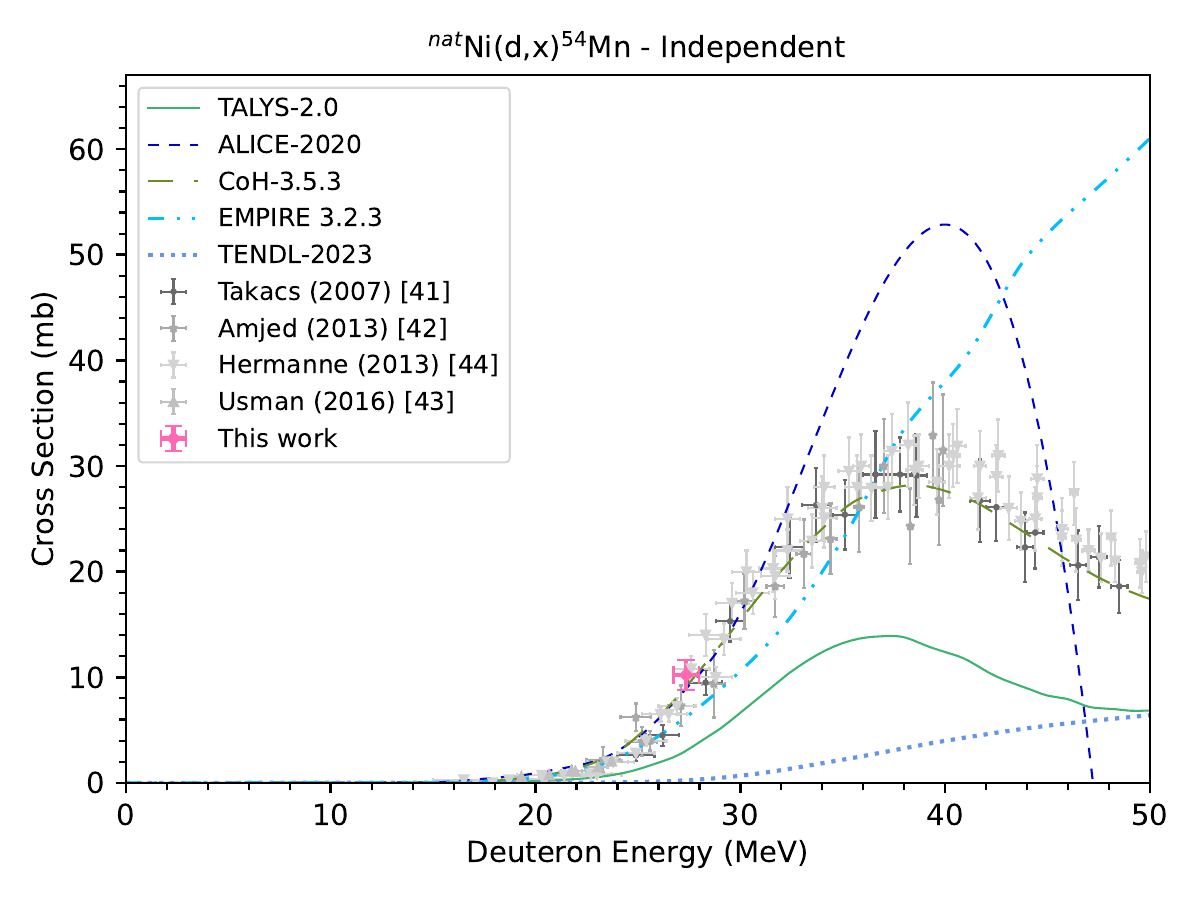}
    \caption{The excitation function of the independent production of $^\text{{nat}}$Ni($d$,$x$)$^{54}$Mn.}
\end{figure}

\begin{figure}[h!]
    \centering
    \includegraphics[width=\linewidth]{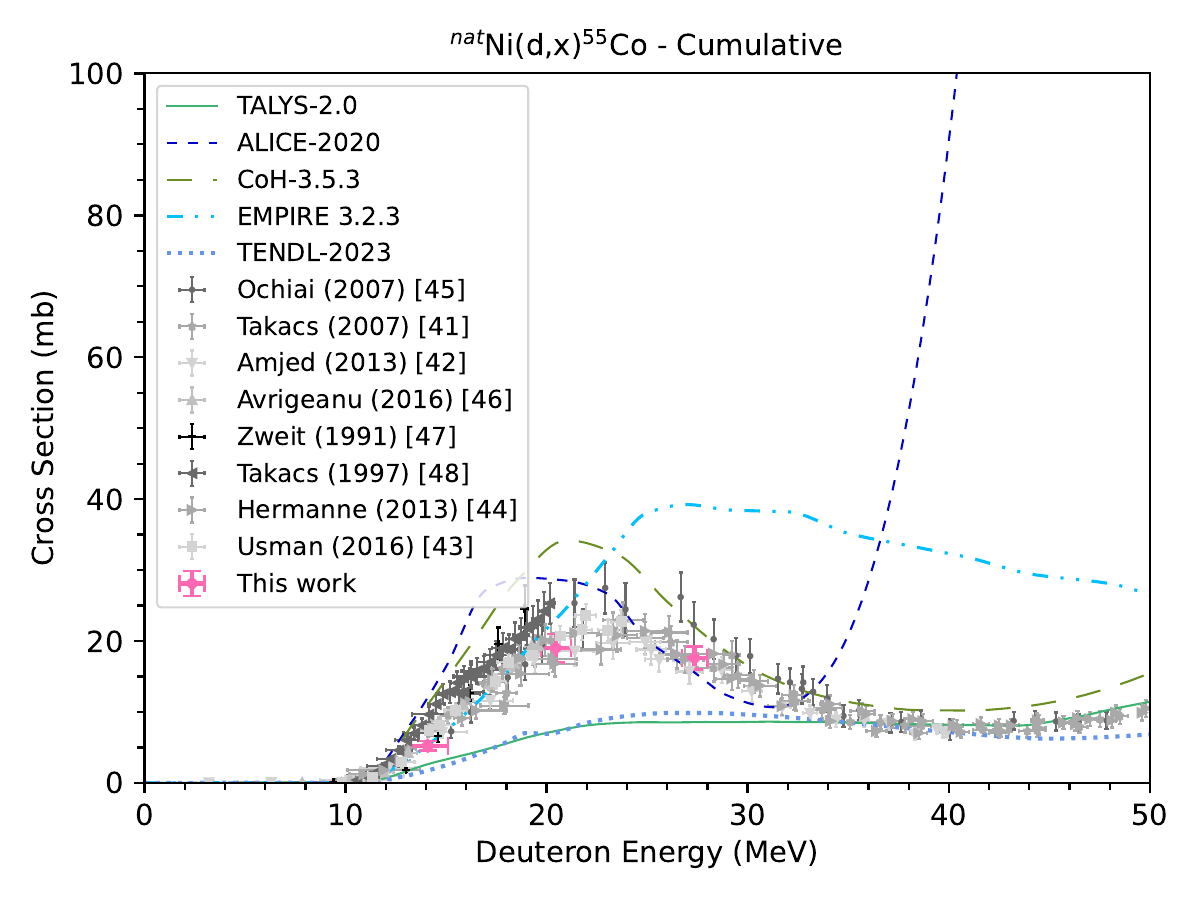}
    \caption{The excitation function of the cumulative production of $^\text{{nat}}$Ni($d$,$x$)$^{55}$Co.}
\end{figure}

\begin{figure}[h!]
    \centering
    \includegraphics[width=\linewidth]{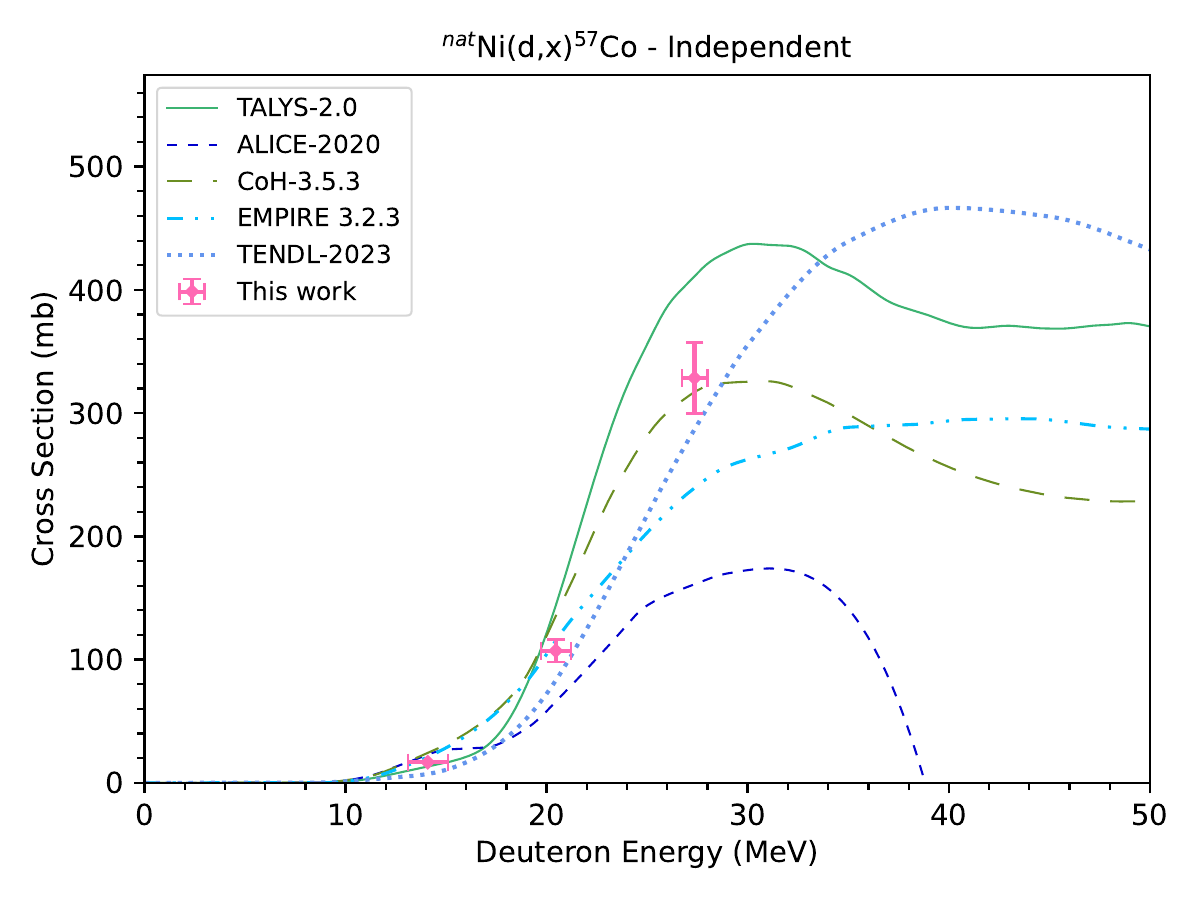}
    \caption{The excitation function of the independent production of $^\text{{nat}}$Ni($d$,$x$)$^{55}$Co.}
\end{figure}

\begin{figure}[h!]
    \centering
    \includegraphics[width=\linewidth]{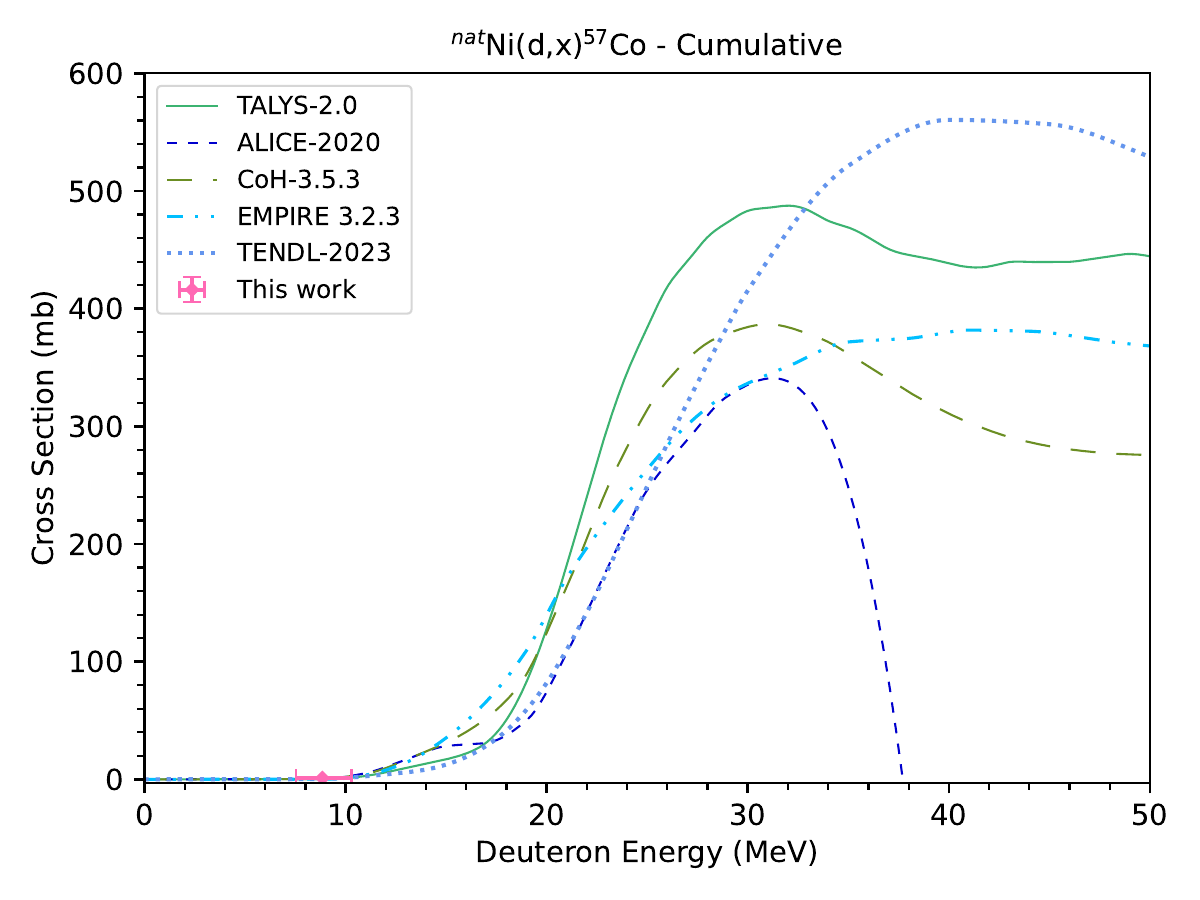}
    \caption{The excitation function of the cumulative production of $^\text{{nat}}$Ni($d$,$x$)$^{57}$Co.}
\end{figure}

\begin{figure}[h!]
    \centering
    \includegraphics[width=\linewidth]{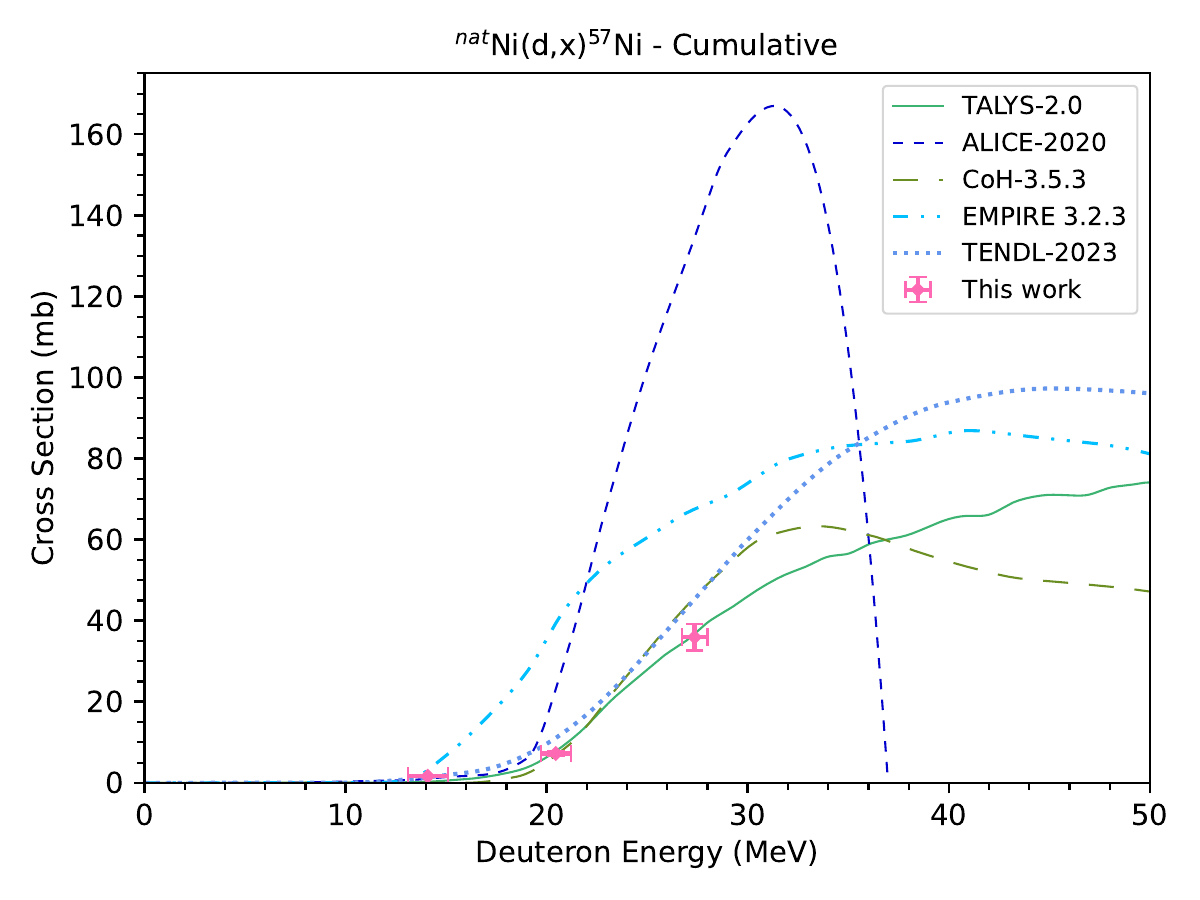}
    \caption{The excitation function of the cumulative production of $^\text{{nat}}$Ni($d$,$x$)$^{57}$Ni.}
\end{figure}

\begin{figure}[h!]
    \centering
    \includegraphics[width=\linewidth]{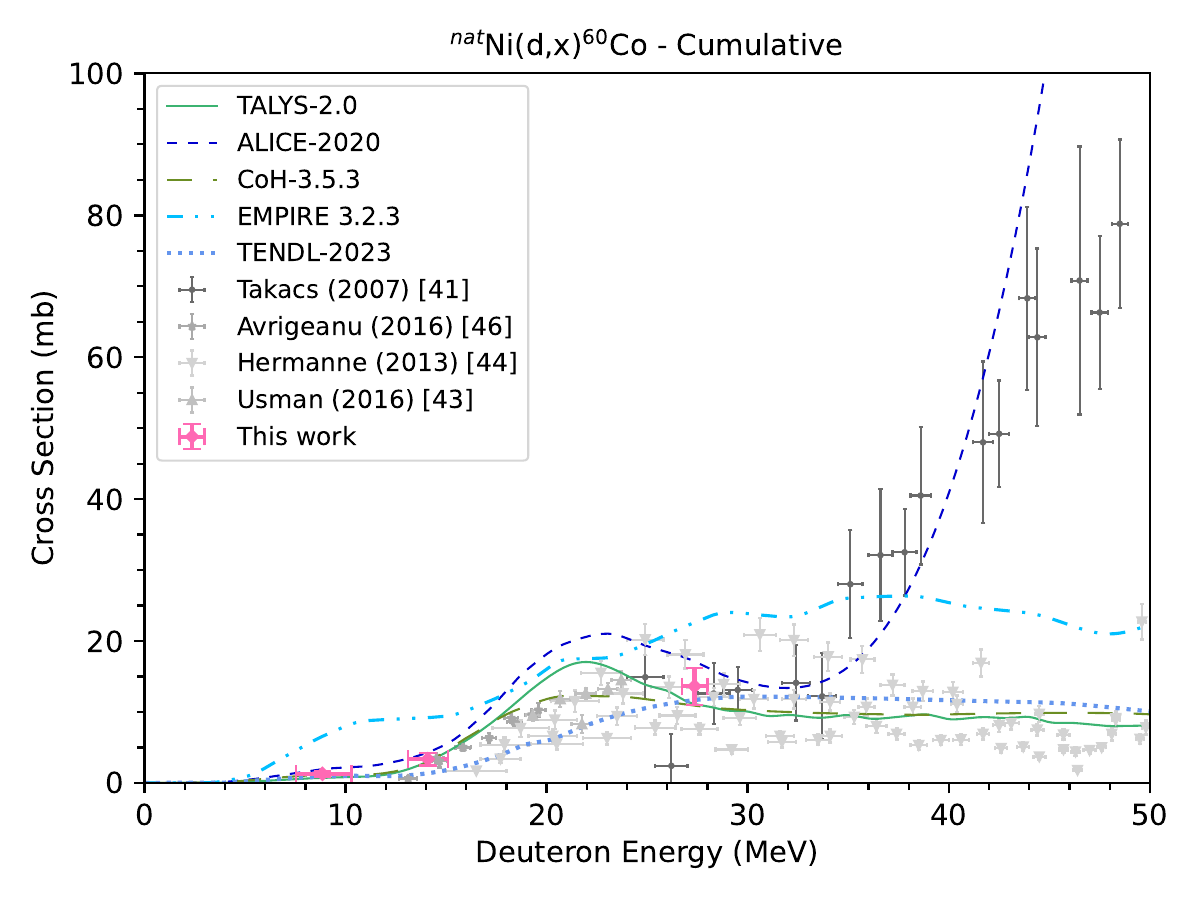}
    \caption{The excitation function of the cumulative production of $^\text{{nat}}$Ni($d$,$x$)$^{60}$Co.}
\end{figure}

\begin{figure}[h!]
    \centering
    \includegraphics[width=\linewidth]{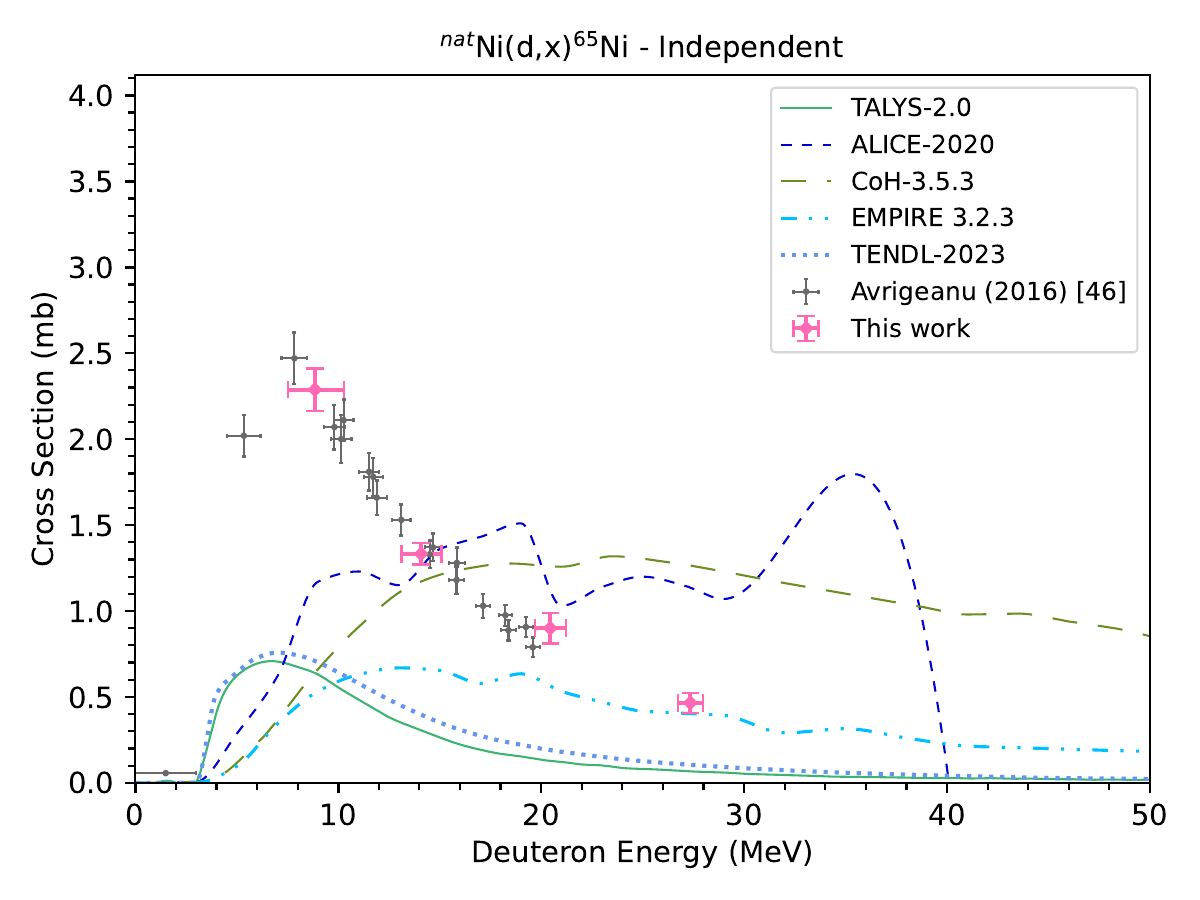}
    \caption{The excitation function of the independent production of $^\text{{nat}}$Ni($d$,$x$)$^{65}$Ni.}
\end{figure}

\subsubsection{$^\text{nat}$Ti($d$,$x$) reactions}
The measured cross sections presented in this section are compared to literature data \cite{Lebeda2015ExperimentalReactions, Gagnon2010Experimental51Ti, Takacs2007EvaluatedTitanium, Khandaker2013Excitation24MeV, Takacs1997ExcitationBeams, Duchemin2015Cross34MeV, Khandaker2014ActivationTitanium, Hermanne2000ExperimentalTi}, the $\textsc{TENDL}-2023$ data library \cite{TENDL} and the reaction modeling codes $\textsc{TALYS}-2.0$ \cite{TALYS}, $\textsc{ALICE}-2020$ \cite{ALICE}, $\textsc{CoH}-3.5.3$ \cite{CoH} and $\textsc{EMPIRE}-3.2.3$ \cite{EMPIRE}.

\begin{figure}[h!]
    \centering
    \includegraphics[width=\linewidth]{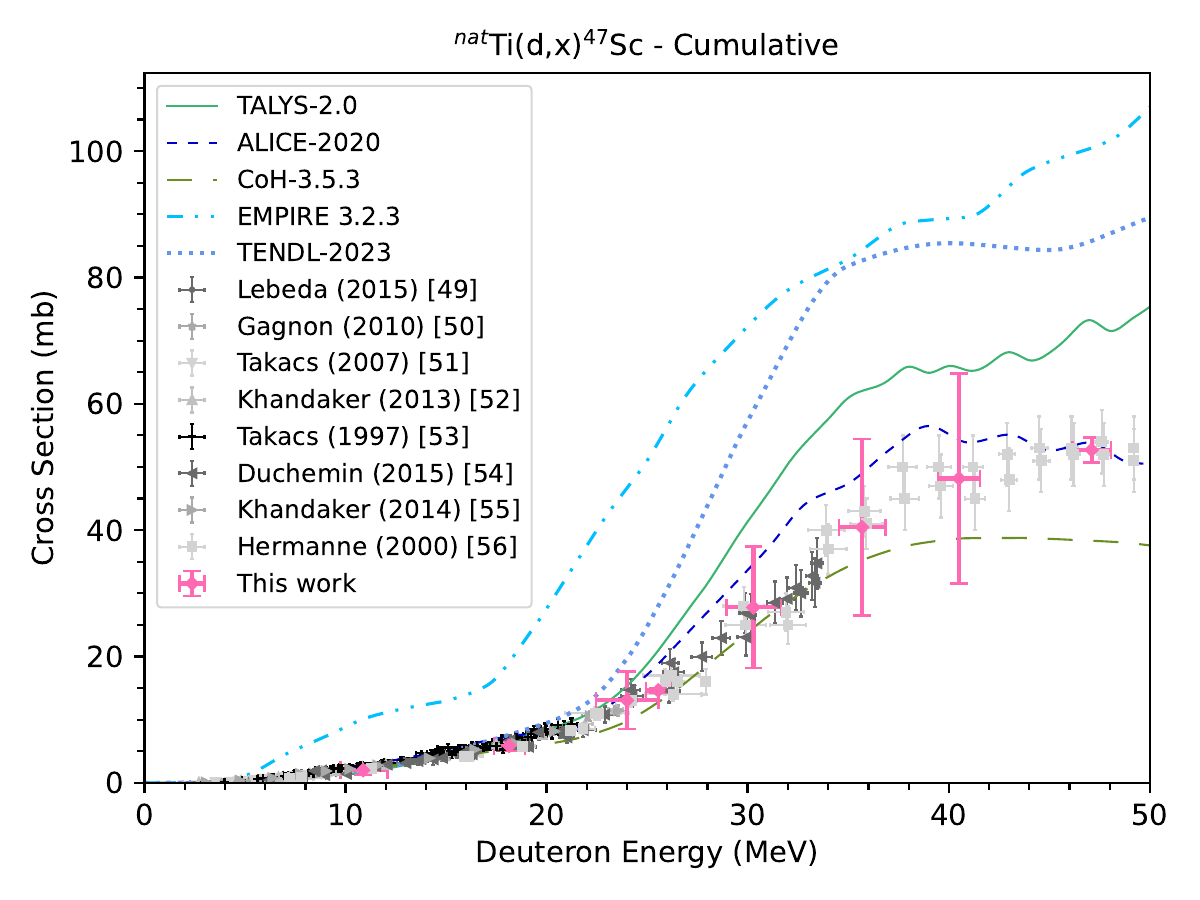}
    \caption{The excitation function of the cumulative production of $^\text{{nat}}$Ti($d$,$x$)$^{47}$Sc.}
\end{figure}

\begin{figure}[h!]
    \centering
    \includegraphics[width=\linewidth]{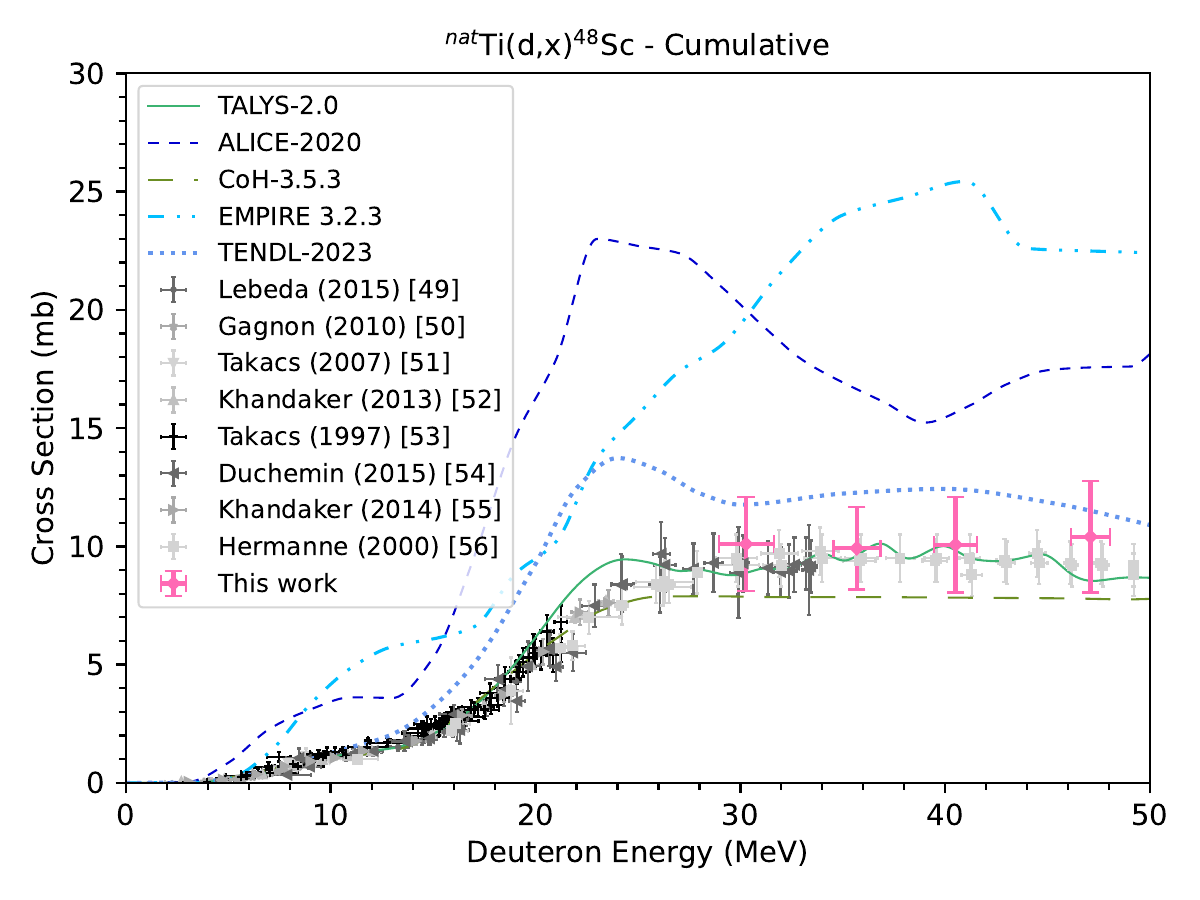}
    \caption{The excitation function of the cumulative production of $^\text{{nat}}$Ti($d$,$x$)$^{48}$Sc.}
\end{figure}

\subsubsection{$^\text{nat}$Fe($d$,$x$) reactions}
The measured cross sections presented in this section are compared to literature data \cite{Ochiai2007DeuteronIFMIF, Hermanne2000ExperimentalTi, Kiraly2009Evaluated10MeV, Zhao1995ExcitationIron, Khandaker2013Activation24MeV, Dmitriev1969MethodsIsotope., Clark1969ExcitationIron, Avrigeanu2014LowIsotopes, zhenlan1984excitation, Jung1992CrossDeuterons, Nakao2006MeasurementsMaterials, Sudar1994ExcitationMeV, Takacs1996StudyTechnique}, the $\textsc{TENDL}-2023$ data library \cite{TENDL} and the reaction modeling codes $\textsc{TALYS}-2.0$ \cite{TALYS}, $\textsc{ALICE}-2020$ \cite{ALICE}, $\textsc{CoH}-3.5.3$ \cite{CoH} and $\textsc{EMPIRE}-3.2.3$ \cite{EMPIRE}.

\begin{figure}[h!]
    \centering
    \includegraphics[width=\linewidth]{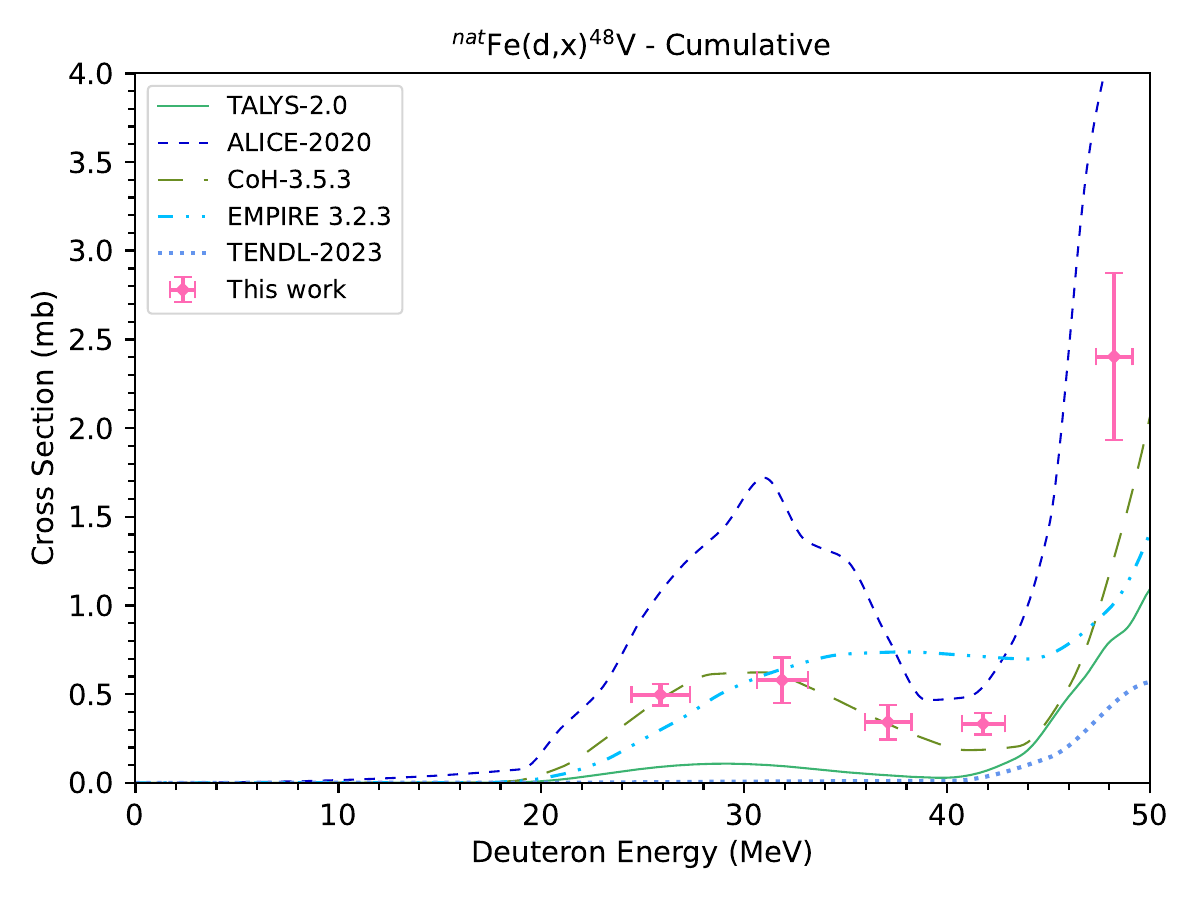}
    \caption{The excitation function of the cumulative production of $^\text{{nat}}$Fe($d$,$x$)$^{48}$V.}
\end{figure}

\begin{figure}[h!]
    \centering
    \includegraphics[width=\linewidth]{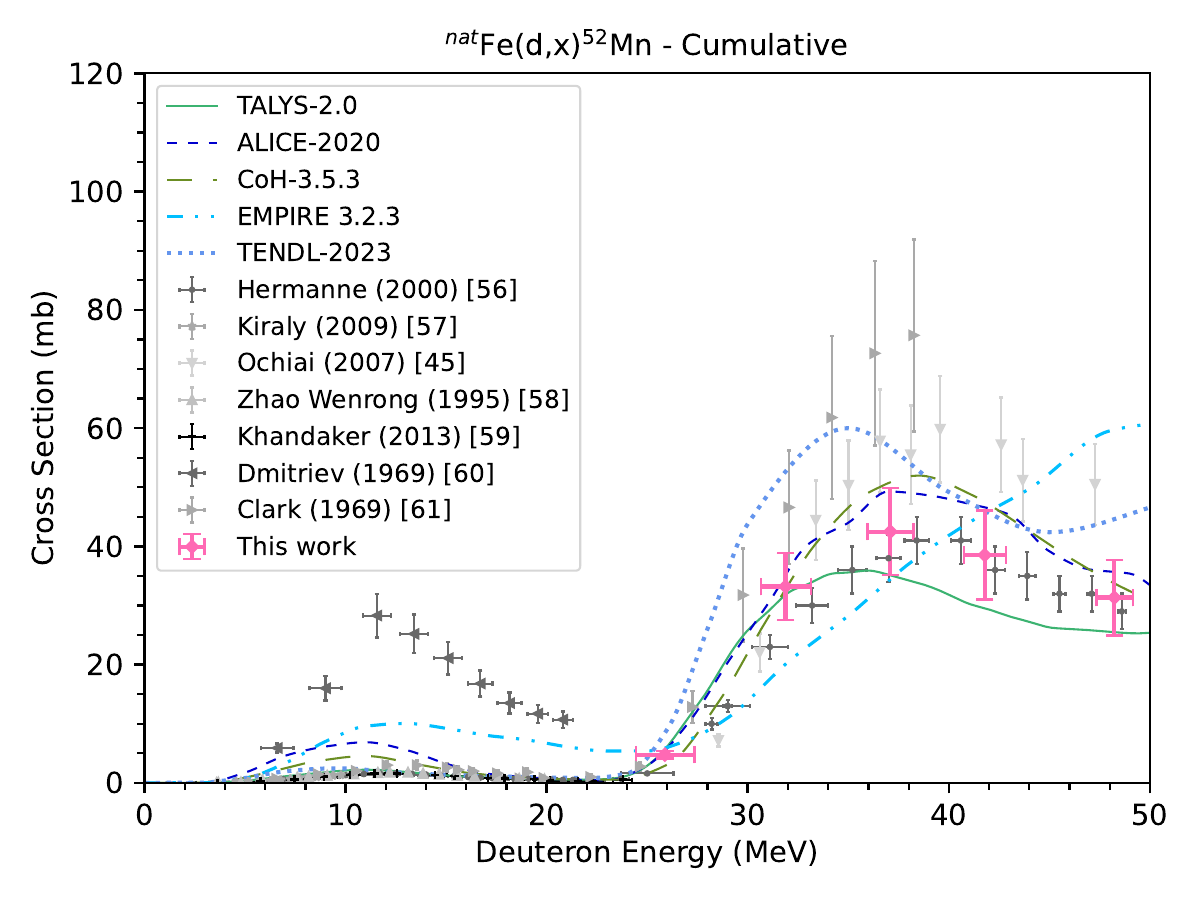}
    \caption{The excitation function of the cumulative production of $^\text{{nat}}$Fe($d$,$x$)$^{52}$Mn.}
\end{figure}

\begin{figure}[h!]
    \centering
    \includegraphics[width=\linewidth]{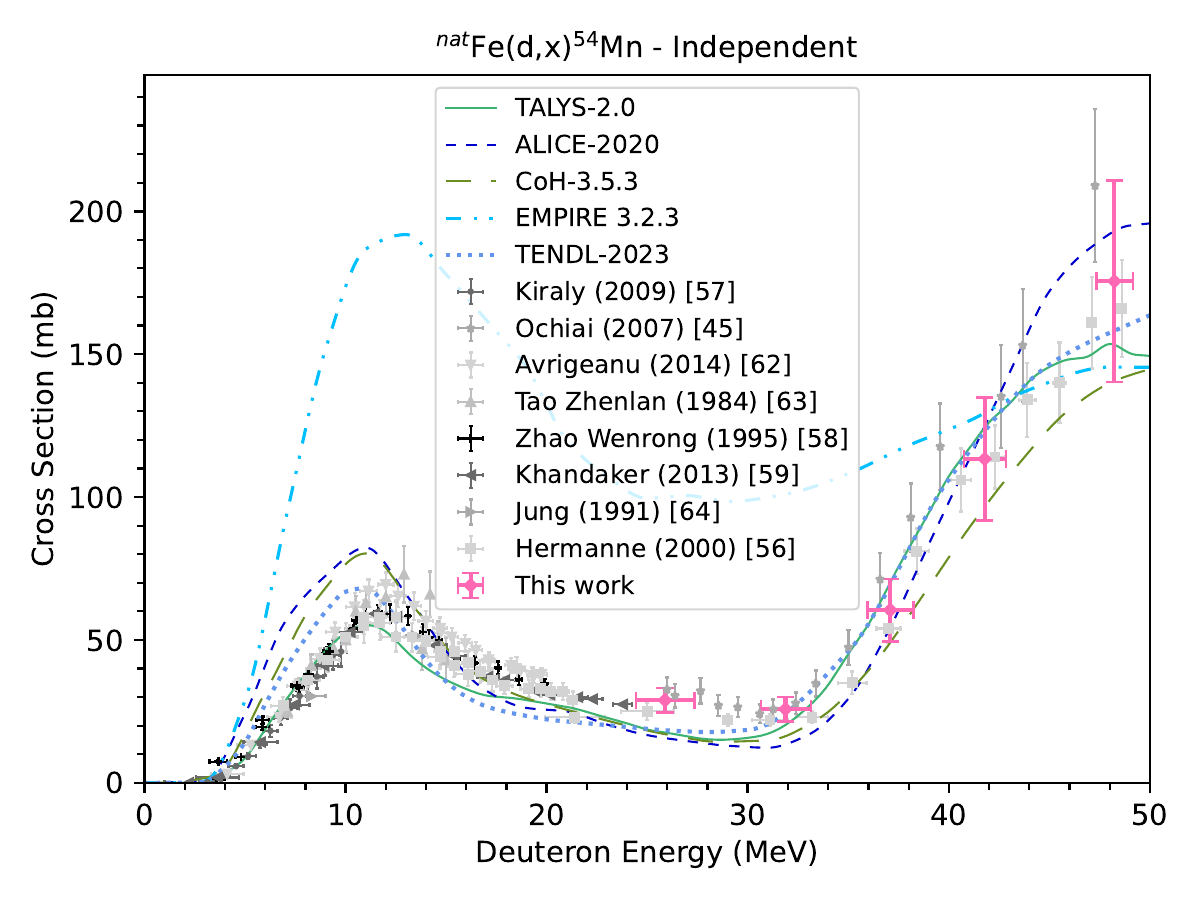}
    \caption{The excitation function of the independent production of $^\text{{nat}}$Fe($d$,$x$)$^{54}$Mn.}
\end{figure}

\begin{figure}[h!]
    \centering
    \includegraphics[width=\linewidth]{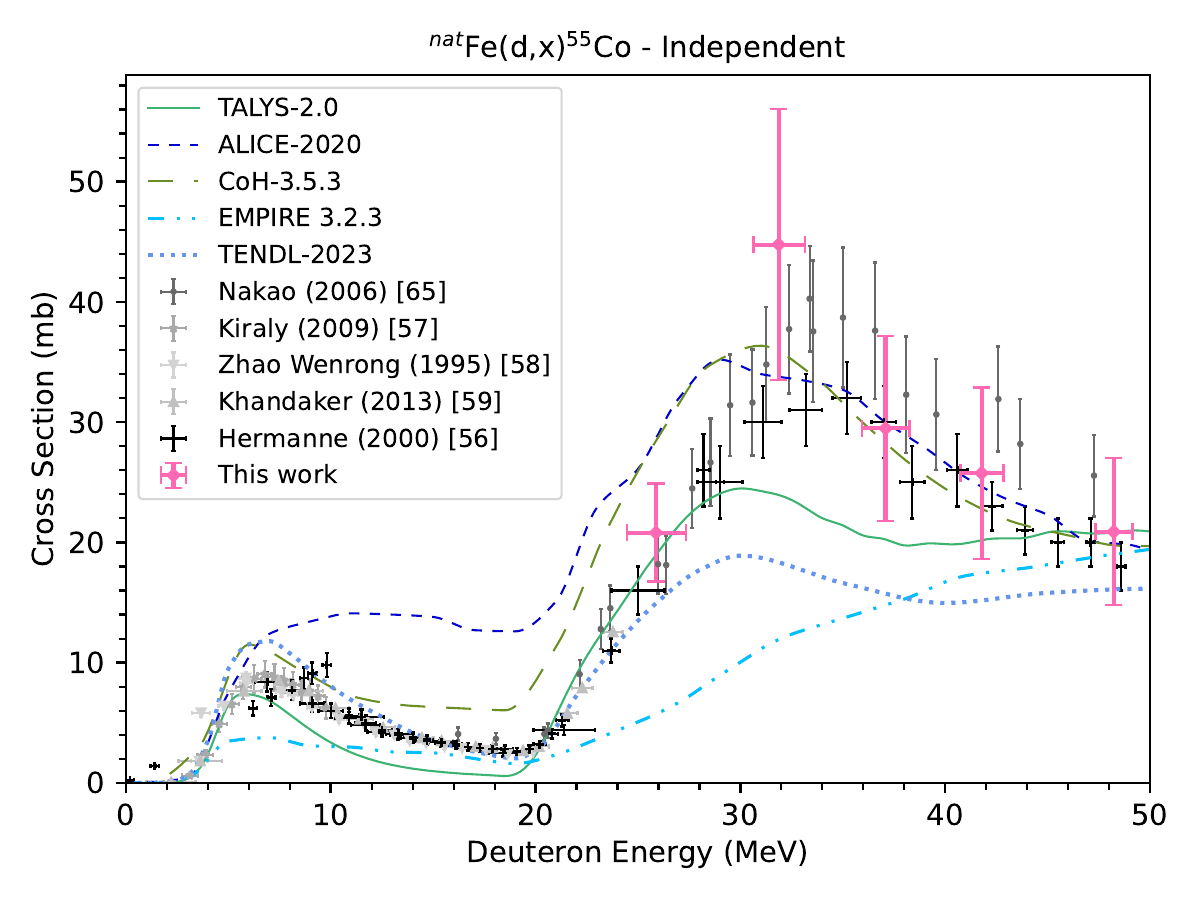}
    \caption{The excitation function of the independent production of $^\text{{nat}}$Fe($d$,$x$)$^{55}$Co.}
\end{figure}

\begin{figure}[h!]
    \centering
    \includegraphics[width=\linewidth]{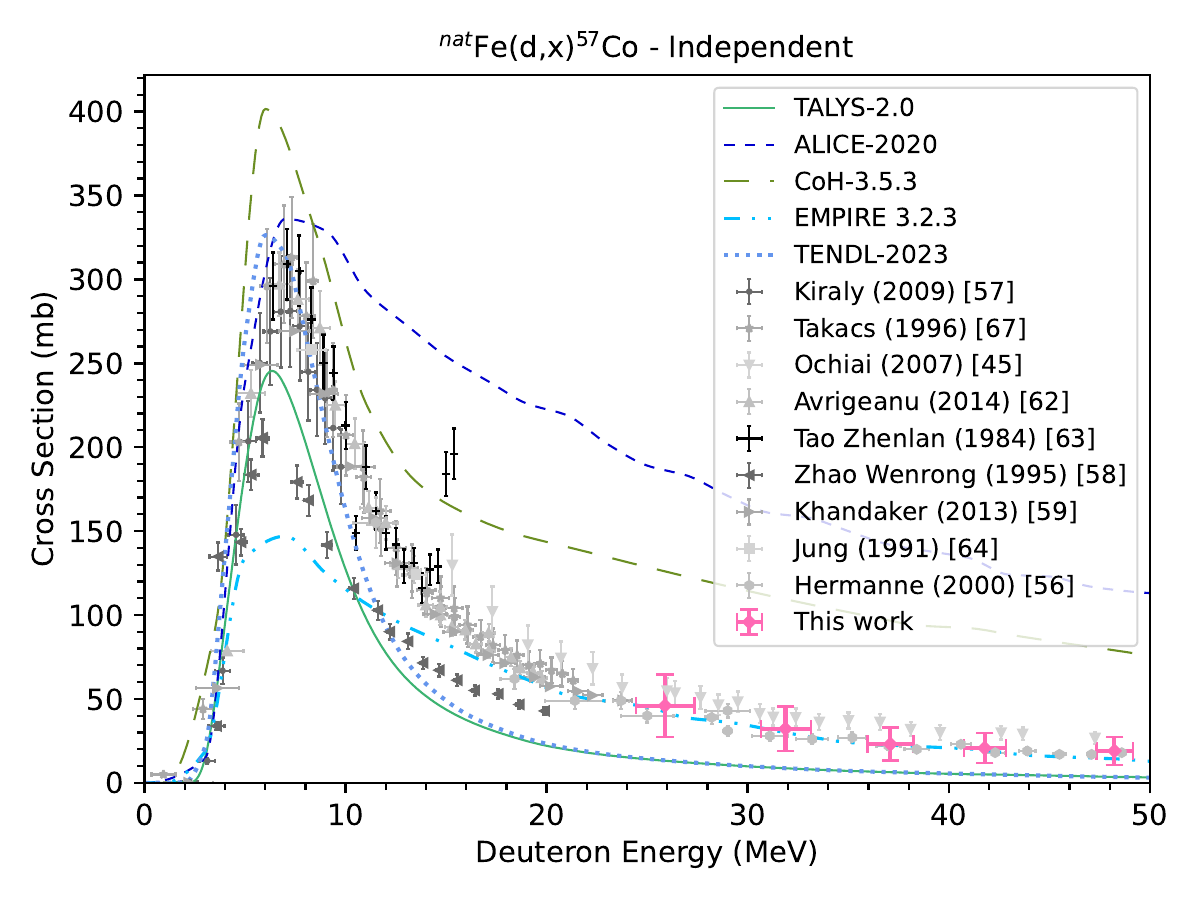}
    \caption{The excitation function of the independent production of $^\text{{nat}}$Fe($d$,$x$)$^{57}$Co.}
\end{figure}

\begin{figure}[h!]
    \centering
    \includegraphics[width=\linewidth]{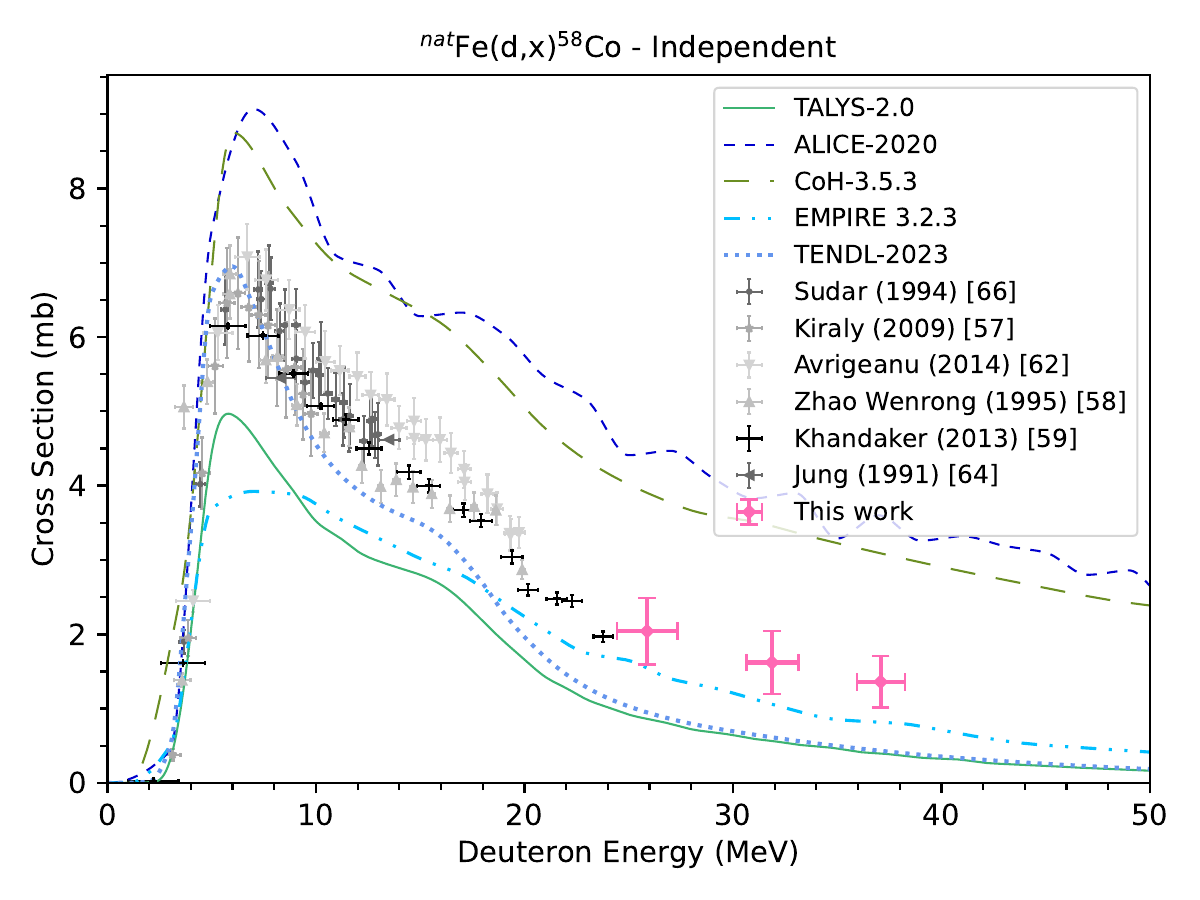}
    \caption{The excitation function of the independent production of $^\text{{nat}}$Fe($d$,$x$)$^{58}$Co.}
\end{figure}

\subsection{Monitor reaction measurements}
 Our measured monitor reactions are compared to the IAEA recommended values \cite{Hermanne2018ReferenceReactions} and previously measured data \cite{Ochiai2007DeuteronIFMIF, Takacs2007EvaluatedNickel, CLINE1971437, Amjed2013Activation40MeV, zhu1983measurements, Hermanne2013New50MeV, Jung1992CrossDeuterons, Zweit1991ExcitationTomography, Takacs1997ActivationPurpose, COETZEEPEISACH+1972+1+6, HERMANNE2007308, Gagnon2010Experimental51Ti, Takacs2007EvaluatedTitanium, Khandaker2013Excitation24MeV, HERMANNE201431, Duchemin2015Cross34MeV, Hermanne2000ExperimentalTi, Sudar1994ExcitationMeV, Kiraly2009Evaluated10MeV, takacs2001new, zhenlan1984excitation, Zhao1995ExcitationIron, Khandaker2013Activation24MeV} to demonstrate reproduction of well-established monitor reactions. We are not reporting these as new cs measurements.
 
\begin{figure}[h!]
    \centering
    \includegraphics[width=\linewidth]{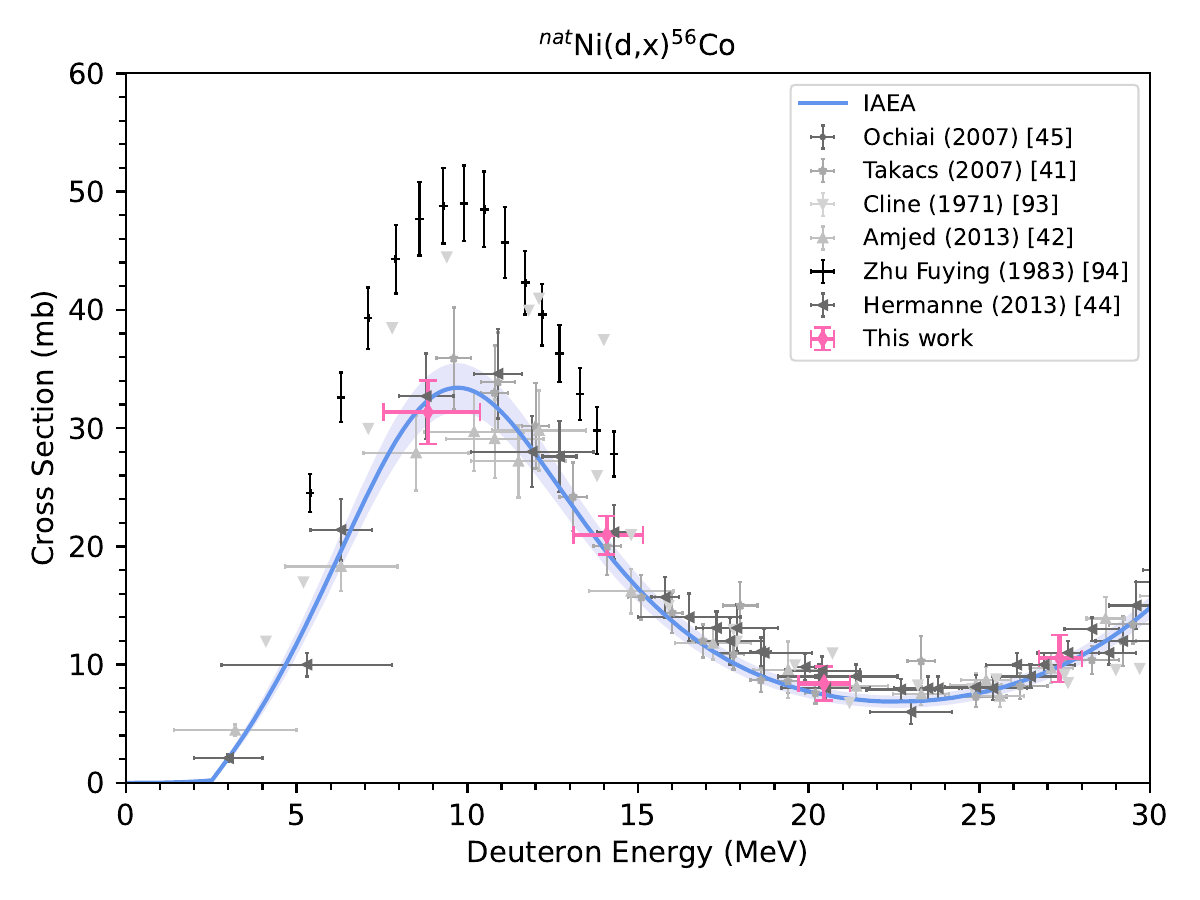}
    \caption{Our measured excitation function for the $^\text{{nat}}$Ni($d$,$x$)$^{56}$Co monitor reaction compared to the IAEA recommended values.}
    \label{fig:xs_mon_Ni_56Co}
\end{figure}

\begin{figure}[h!]
    \centering
    \includegraphics[width=\linewidth]{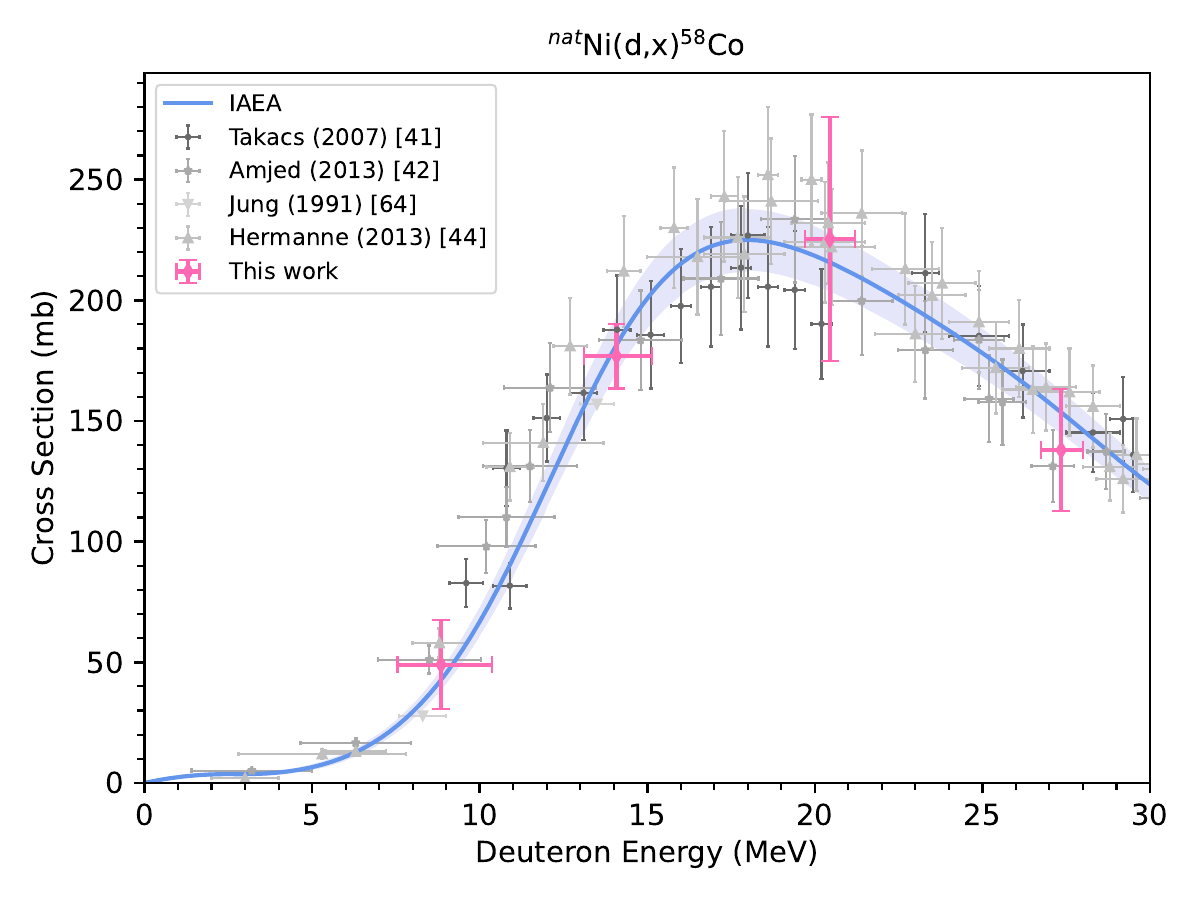}
    \caption{Our measured excitation function for the $^\text{{nat}}$Ni($d$,$x$)$^{58}$Co monitor reaction compared to the IAEA recommended values.}
    \label{fig:xs_mon_Ni_58Co}
\end{figure}

\begin{figure}[h!]
    \centering
    \includegraphics[width=\linewidth]{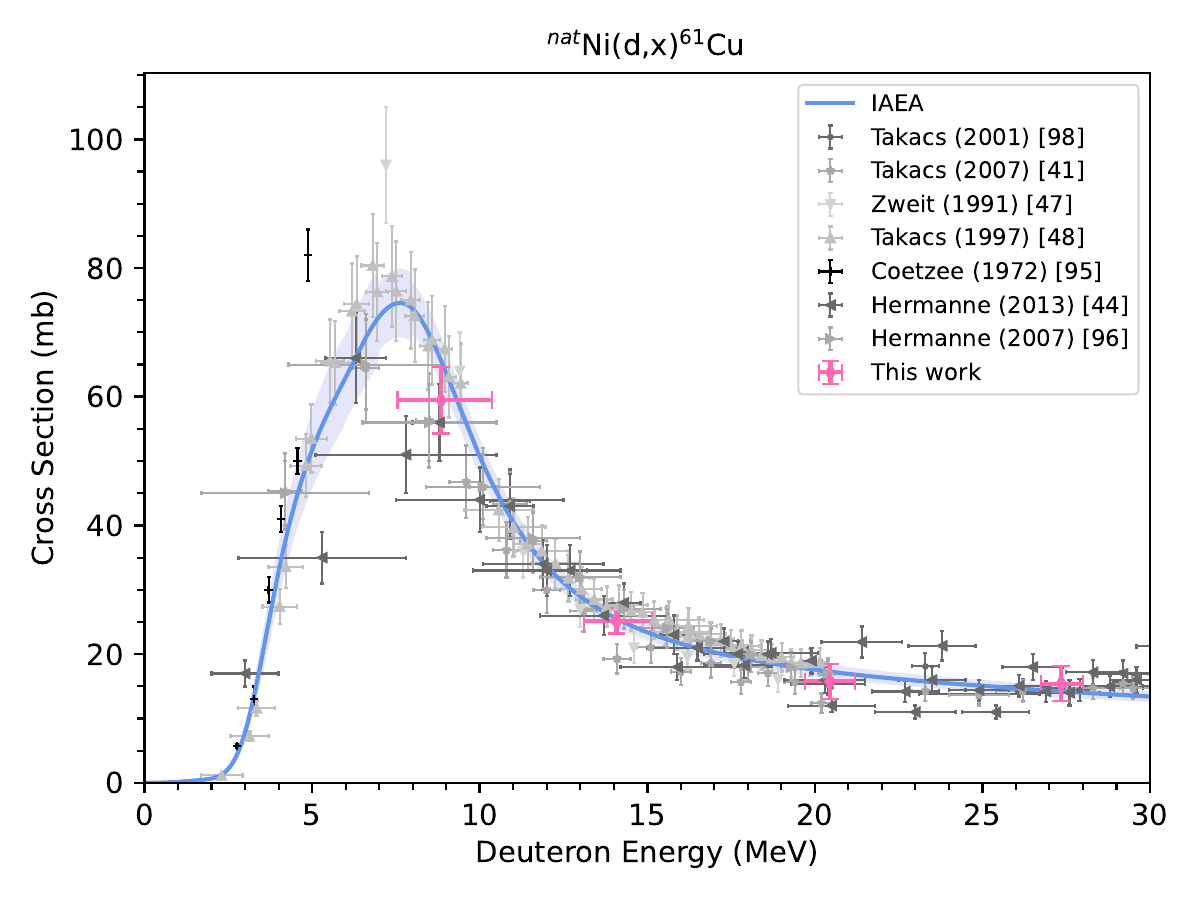}
    \caption{Our measured excitation function for the $^\text{{nat}}$Ni($d$,$x$)$^{61}$Cu monitor reaction compared to the IAEA recommended values.}
    \label{fig:xs_mon_Ni_61Cu}
\end{figure}

\begin{figure}[h!]
    \centering
    \includegraphics[width=\linewidth]{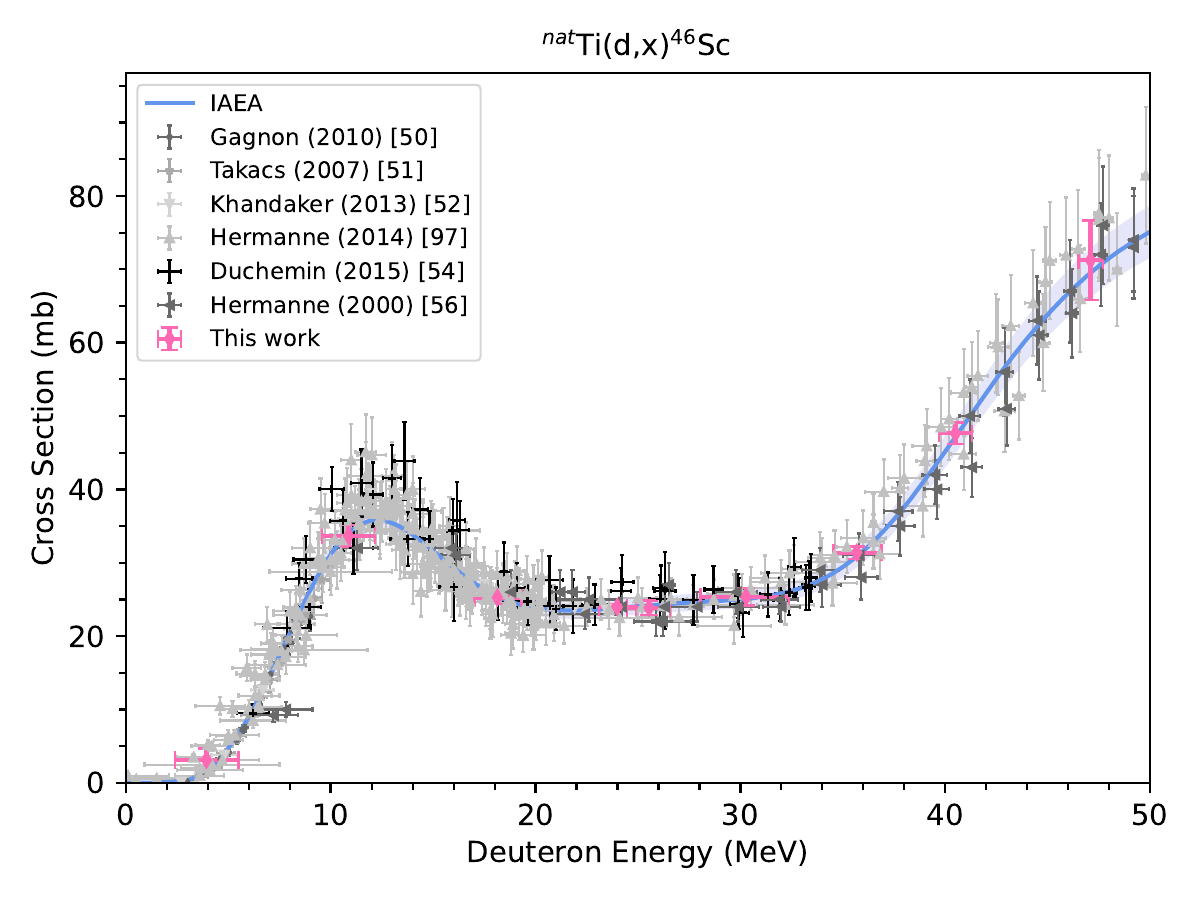}
    \caption{Our measured excitation function for the $^\text{{nat}}$Ti($d$,$x$)$^{46}$Sc monitor reaction compared to the IAEA recommended values.}
    \label{fig:xs_mon_Ti_46Sc}
\end{figure}

\begin{figure}[h!]
    \centering
    \includegraphics[width=\linewidth]{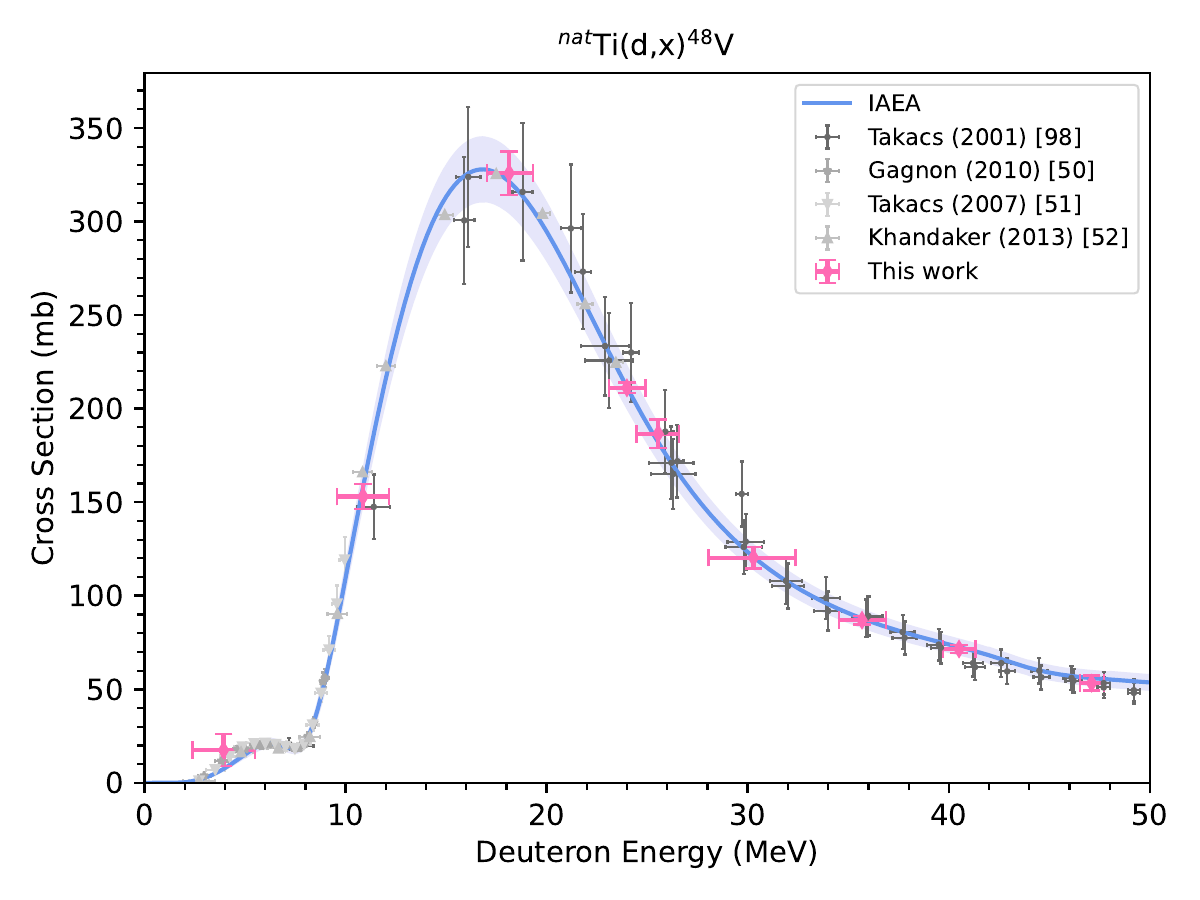}
    \caption{Our measured excitation function for the $^\text{{nat}}$Ti($d$,$x$)$^{48}$V monitor reaction compared to the IAEA recommended values.}
    \label{fig:xs_mon_Ti_48V}
\end{figure}

\begin{figure}[h!]
    \centering
    \includegraphics[width=\linewidth]{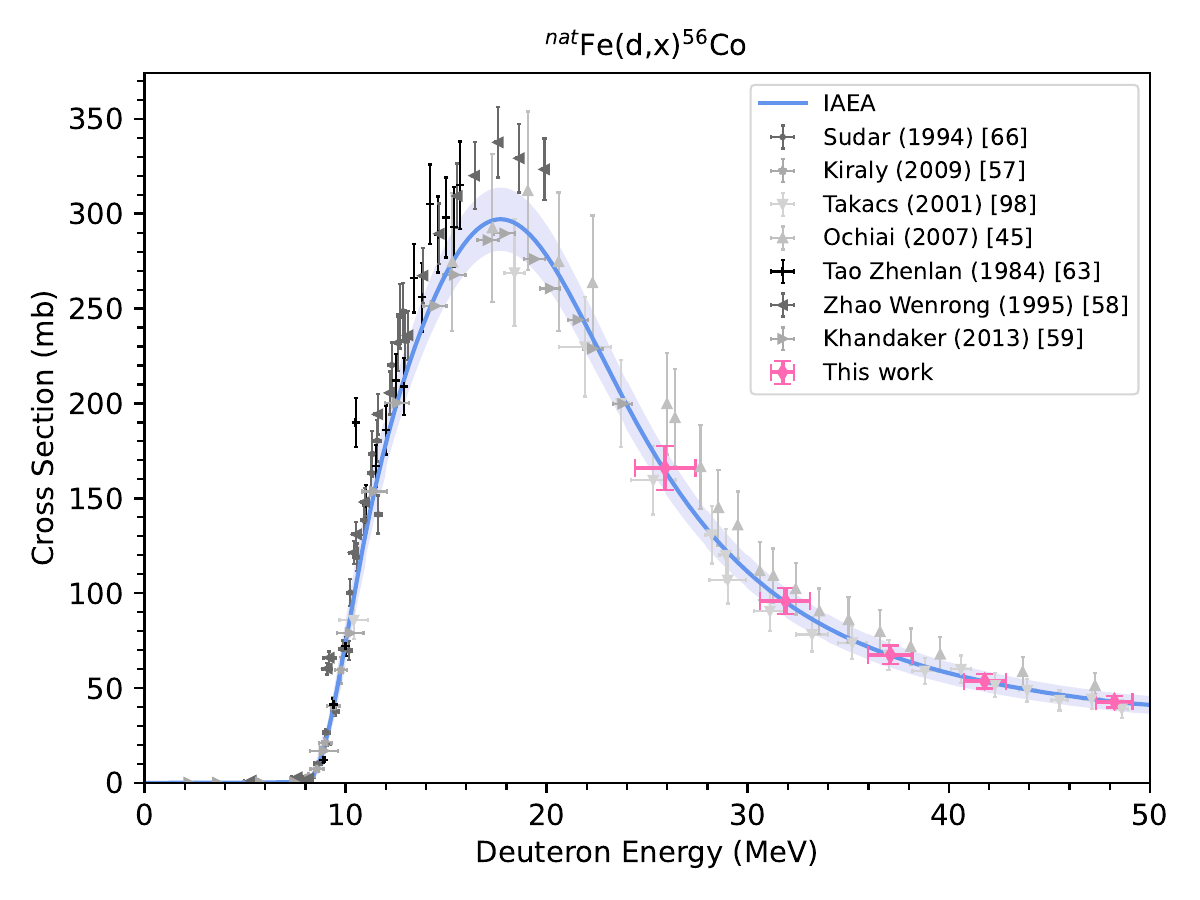}
    \caption{Our measured excitation function for the $^\text{{nat}}$Fe($d$,$x$)$^{56}$Co monitor reaction compared to the IAEA recommended values.}
    \label{fig:xs_mon_Fe_56Co}
\end{figure}

 \bibliographystyle{unsrt}

 \bibliography{references}
\end{document}